\documentclass[pdftex,twocolumn,epjc3]{svjour3}          

\RequirePackage[T1]{fontenc}

\smartqed  
\usepackage{graphicx}
\usepackage{ulem}
\usepackage{amsfonts,amstext,amssymb,amsmath,appendix}
\usepackage{latexsym}
\usepackage{lipsum}
\usepackage[dvipsnames]{xcolor}
\usepackage{braket}
\usepackage{bbold}
\usepackage{bm}
\usepackage{placeins}
\usepackage{enumerate}
\usepackage{standalone}
\usepackage{float}
\usepackage{xcolor}
\usepackage{mathtools}

\RequirePackage{graphicx}
\RequirePackage{flushend}
\RequirePackage[numbers,sort&compress]{natbib}
\usepackage{amstext,amssymb,amsmath,appendix}
\usepackage{latexsym}
\usepackage[dvipsnames]{xcolor}
\usepackage{braket}
\usepackage{bbold}
\usepackage{bm}
\usepackage{placeins}
\usepackage{enumerate}
\usepackage{standalone}
\usepackage{float}
\usepackage{xcolor}
\usepackage{mathtools}

\usepackage[colorlinks=true,
  linkcolor=blue,
  citecolor=blue,
  urlcolor =blue]{hyperref}
\newcommand{\be}{\begin{equation}}
\newcommand{\ee}{\end{equation}}
\newcommand*{\f}[2]{\frac{#1}{#2}}

\newcommand{\epr}{\ket{\mathrm{EPR}}}
\newcommand{\tfd}{\ket{\mathrm{TFD}}}
\newcommand{\figmanybodyteleportation}{\begin{figure}[t]
\centering
\includegraphics[width=1\linewidth]{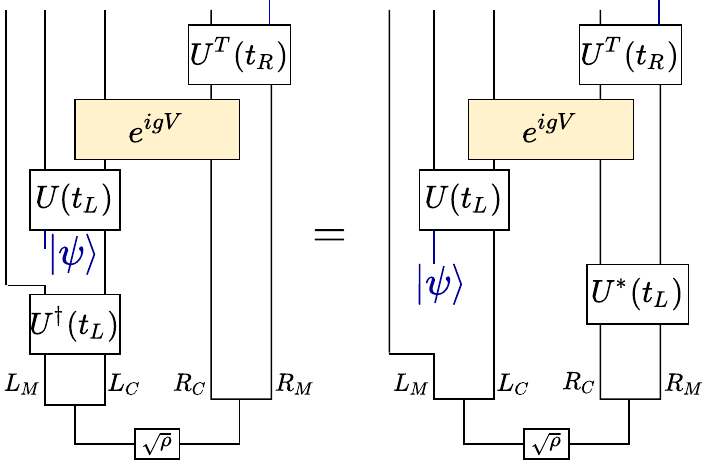}
\caption{\textit{Many-body generalization of the teleportation through wormhole:} Both the left and right systems are divided into message (labeled by a subscript $M$) and carrier  (labeled by subscript $C$) subsystems. Inspired from the gravity description, the left of  an initially prepared TFD at $t=0$ is evolved backward with $U^\dagger$ to reach $-t$.  At $-t$ an information,  shown with a state $|\psi\rangle$, is inserted. Then after forward evolution of left, a momentary coupling is introduced between the carrier subsystems on two sides. The right side is forward evolved with $U^T$, after which the teleported state can be read (see main text). The left and right circuits are exactly the same for evolution unitary described by the underlying one sided Hamiltonian $U=\exp(-iHt)$, on accounts of the identity \eqref{eq:shifting-tfd}.}
\label{fig:tw-many-body}
\end{figure}
}
\newcommand{\hpfigure}{\begin{figure*}[!ht]
    \centering
    \includegraphics[width=0.5\linewidth]{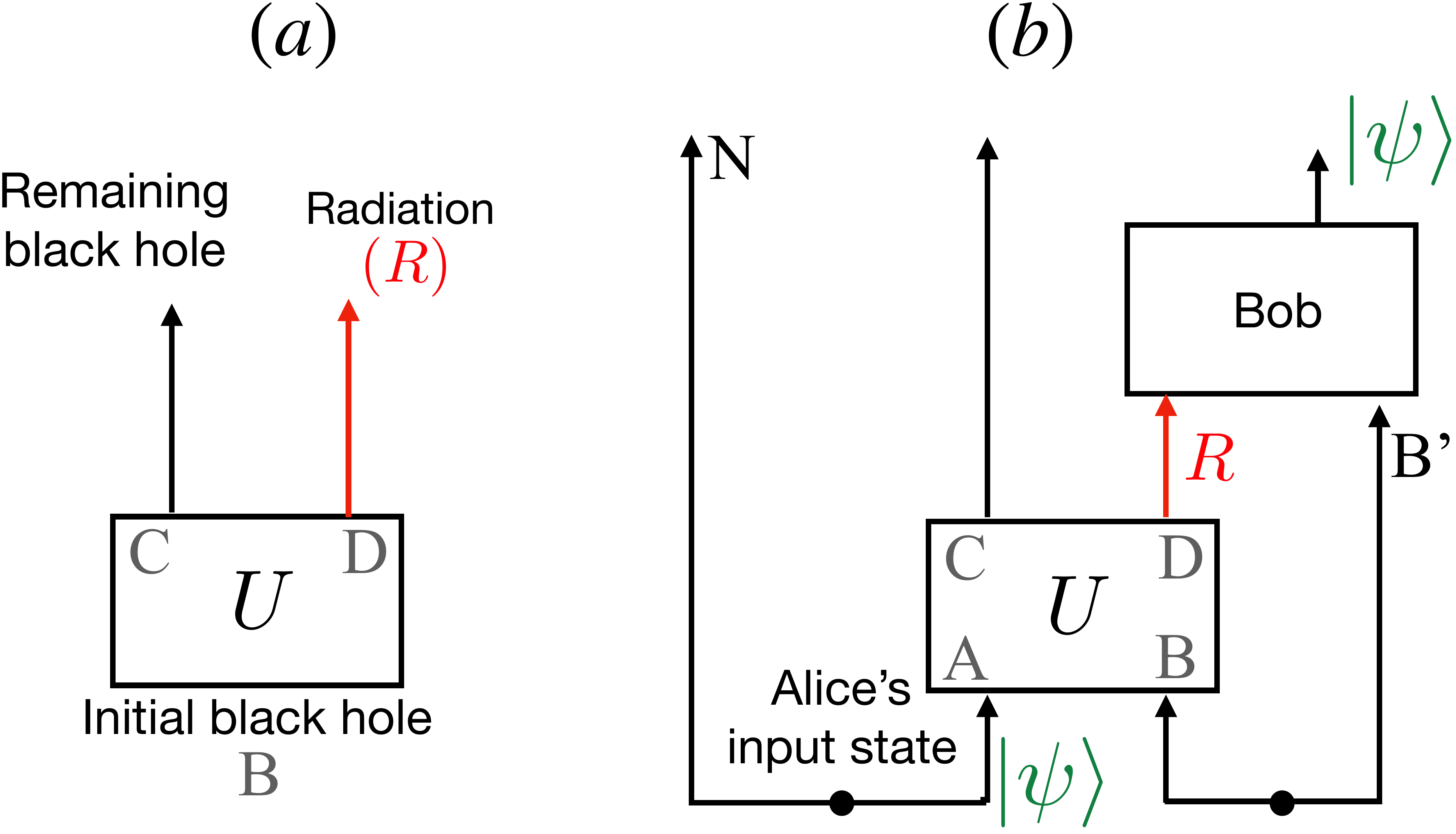}\hspace{1cm}
        \includegraphics[width=0.35\linewidth]{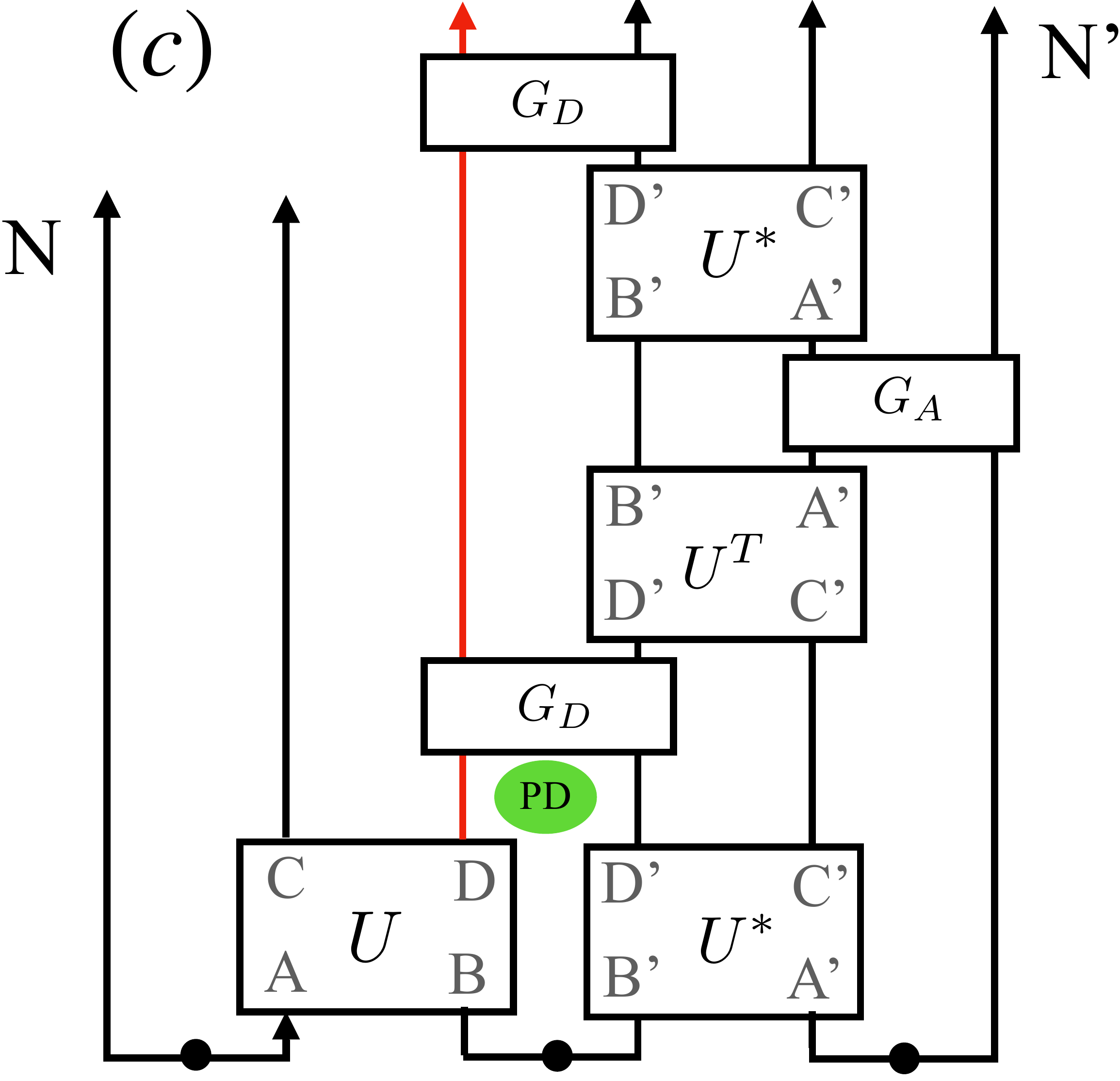}
    \caption{ \textit{Information recovery from a black hole} a) In Page's calculations, an initial black hole $B$ is evaporating radiation R with time. The growing size of the radiation should be compared to the forward time, here denoted with up arrow.  Assuming the Haar random dynamics to model black hole, Page showed that to learn about the black hole from the radiation one has to wait for the black hole to evaporate half of its entropy, and this time is of the order $\sim M^3$, here $M$ is black hole mass. b)  Hayden-Preskill protocol begins with a maximally entangled pair between BB' (black hole B and old radiation B'). The initial input from Alice A is  maximally entangled with  a system N. Bob collects  radiation  R and  the conditions on the recovery of Alice's input information are analyzed. c) In a further protocol, a quantum circuit  for any quantum unitary, Yoshida-Kitaev protocol has similar settings as the Hayden-Preskill, but  the information recovery procedure is made more concrete. Two possible  ways to recover information are discussed  $(i)$ a probabilistic protocol (denoted by PD with green oval here) and $(ii)$ a deterministic protocol. See text for details.}
    \label{fig:HP}
\end{figure*}
}
\newcommand{\otocmanybody}{\begin{figure}[!ht]
\includegraphics[width=\linewidth]{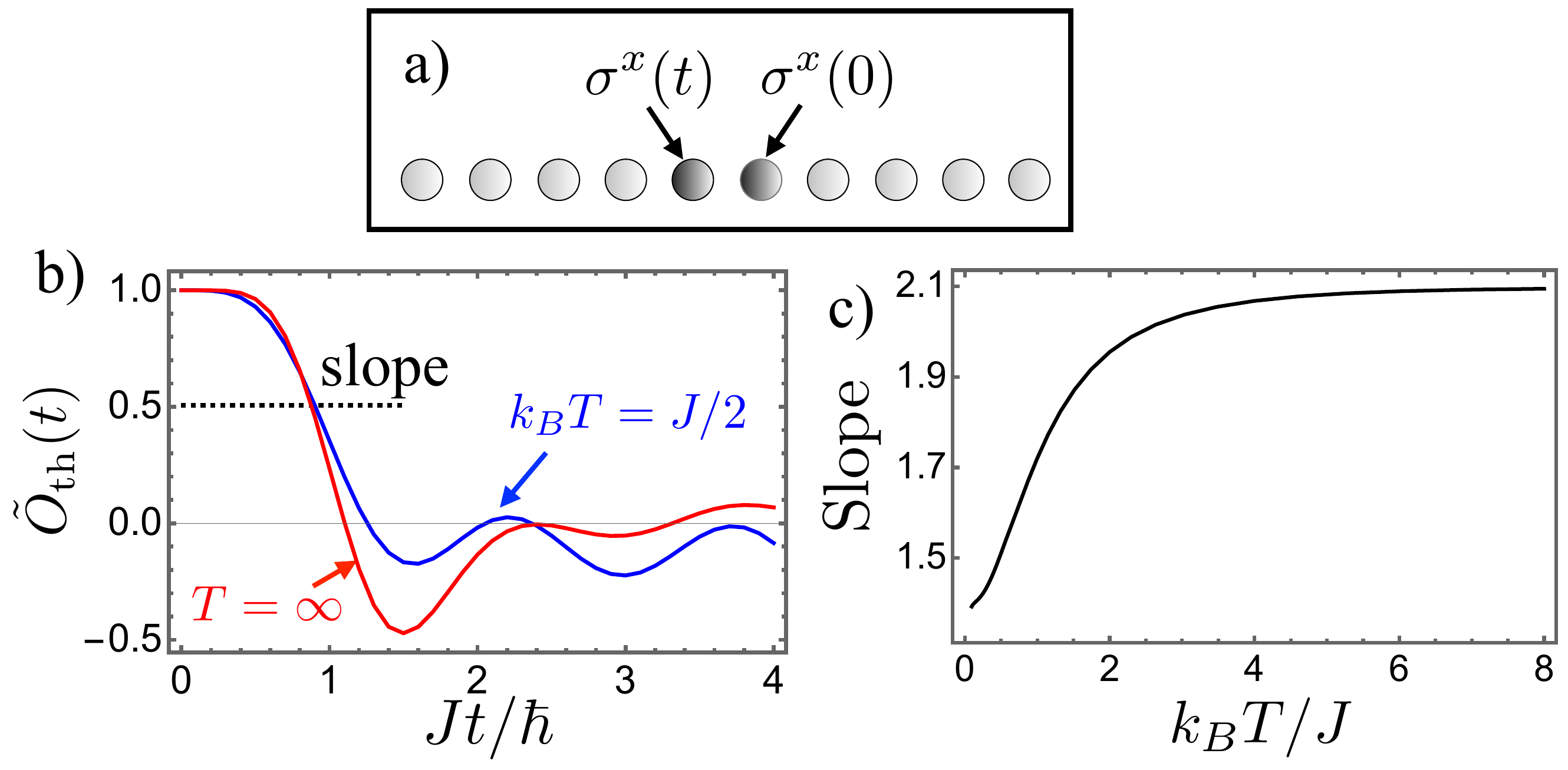}
\caption{\textit{Thermal OTOC in a many-body model given by Eq.~(\ref{eqn:H-tim}) for a system of size $N=10$ qubits.} a) The $W$ and $V$ are chosen to be Pauli operators at adjacent sites. b) The decay of the normalized OTOC $\tilde{O}_{\mathrm{th}}(t)=O_{\mathrm{th}}(t)/O_{\mathrm{th}}(0)$, is shown for different temperatures $T$.
c) The temperature dependence can be studied by inferring the slope when $\tilde{O}_{\mathrm{th}}(t)=0.5$. We see that the rate of decay increases with temperature, settling at a constant for large temperatures.}
\label{fig:otoc-tim}
\end{figure}
}
\newcommand{\figyao}{\begin{figure}[!ht]
\includegraphics[width=0.75\linewidth]{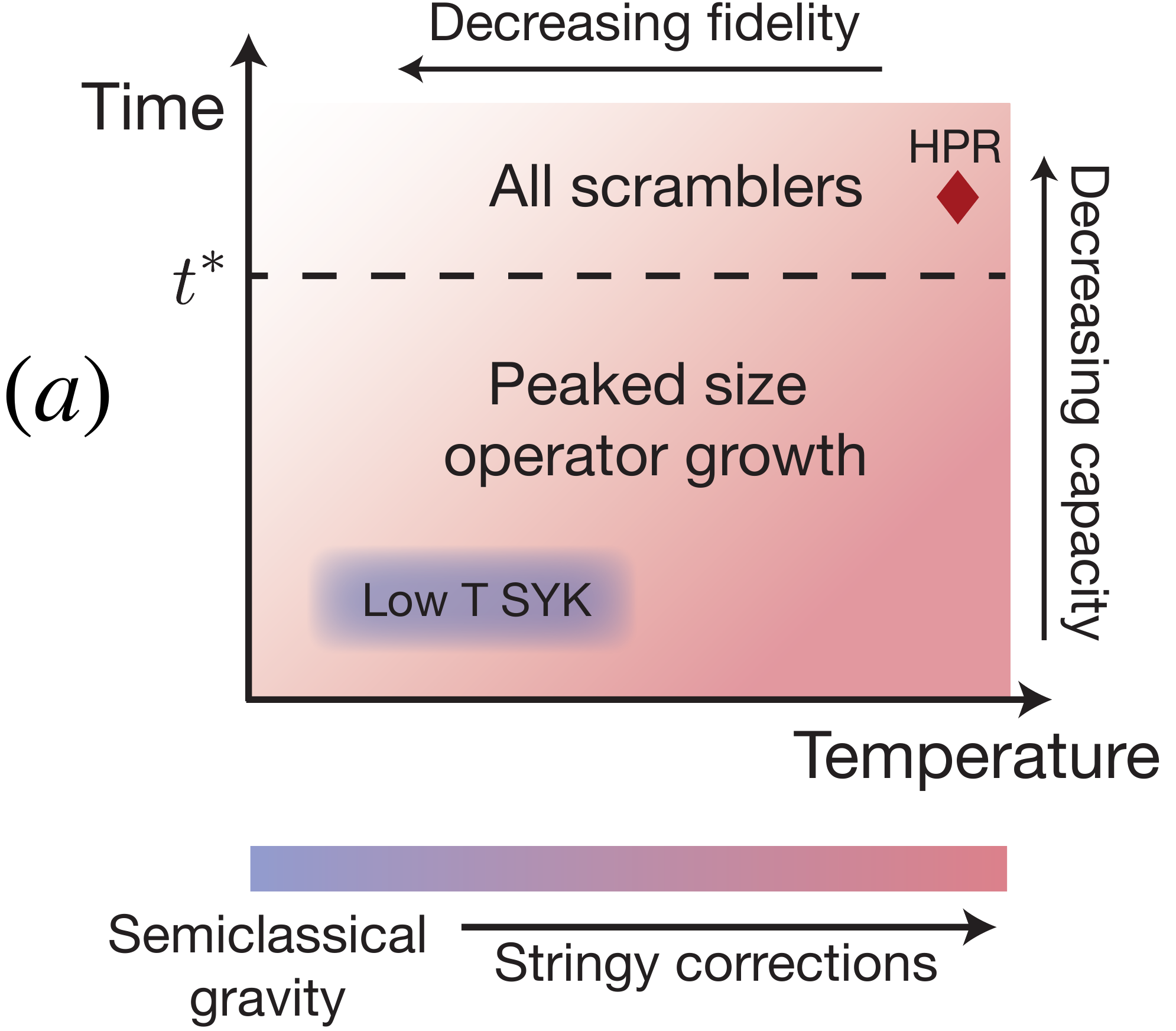} \vskip 0.5cm
\includegraphics[width=0.75\linewidth]{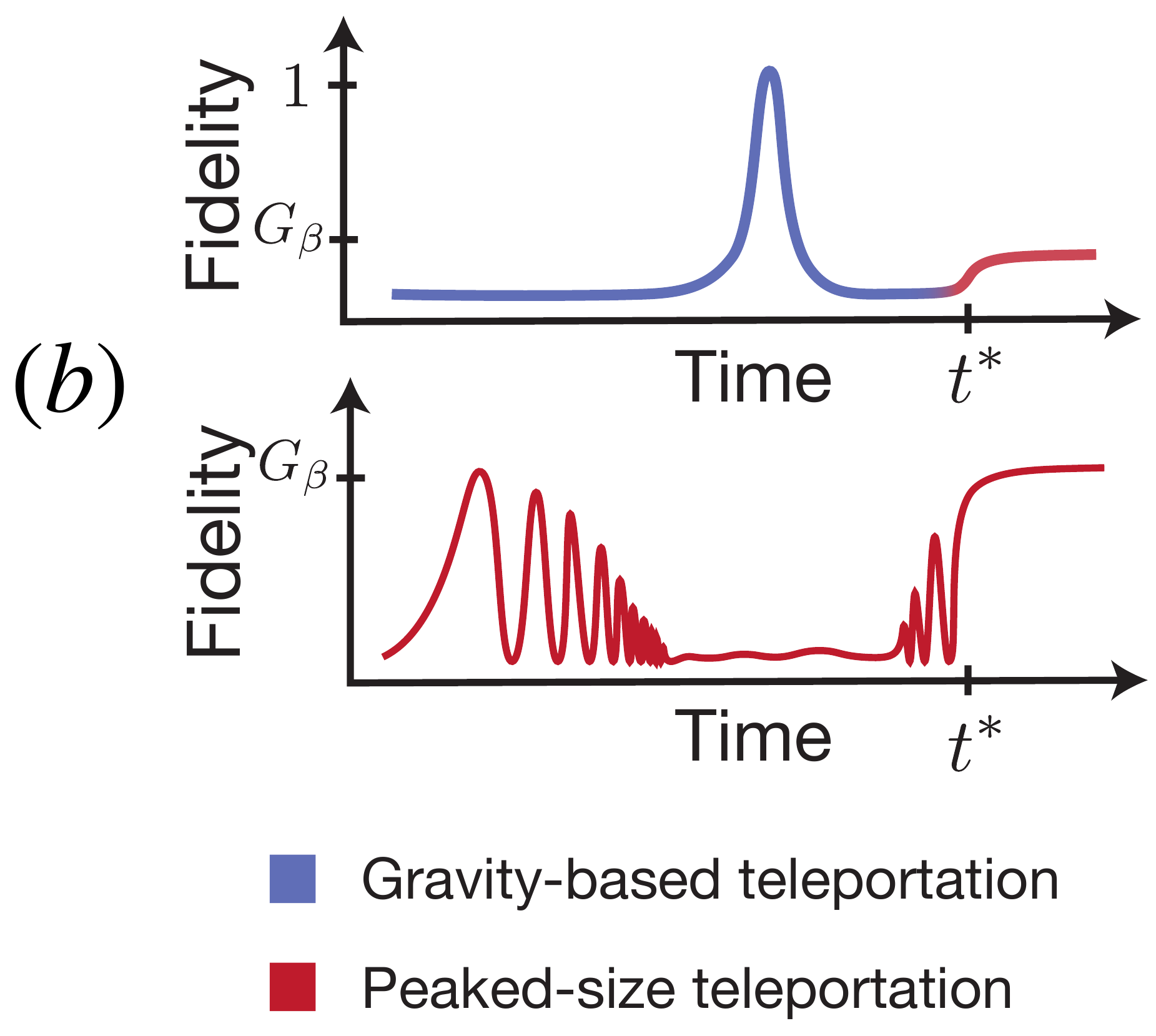}
\caption{\textit{Summary of the teleportation  for different unitary dynamics}. These plots are  taken from Ref. \cite{Schuster2021} with authors' consent. (a) The fidelity of teleportation decreases with temperature. The channel capacity: the number of teleported qubits, decreases at long times. (b) The fidelity features distinct behavior for holographic and other scramblers for $t<t^*$. For low temperature SYK, which is a model of black holes, the fidelity has a peak at $t=t_{\mathrm{scr}}$ while zero otherwise. Whereas for other scramblers it has a ripple like behavior. After $t>t^*$ we see a revival of fidelity for SYK saturating at $\propto G_\beta$. Thus, after $t^*$ all scramblers have fidelity $G_\beta$ and follow peak-size mechanism for teleportation.}
\label{fig:summary-yao}
\end{figure}
}

\journalname{Eur. Phys. J. C}
\begin{document}

\title{Quantum Information Scrambling: From Holography to Quantum Simulators\thanksref{t1}}

\author{Arpan Bhattacharyya\thanksref{e1,addr1}
        \and
        Lata Kh Joshi\thanksref{e2,addr2,addr3} 
        \and
        Bhuvanesh Sundar \thanksref{e3, addr4, addr5}
}

\thankstext[]{t1}{All authors  have contributed equally to this work. }
\thankstext{e1}{e-mail: abhattacharyya@iitgn.ac.in}
\thankstext{e2}{e-mail: lata.joshi@uibk.ac.at}
\thankstext{e3}{e-mail: bhuvanesh.sundar@colorado.edu}

\institute{Indian Institute of Technology, Gandhinagar, Gujarat-382355, India\label{addr1}
          \and
          Center for Quantum Physics, University of Innsbruck, Innsbruck A-6020, Austria\label{addr2}
          \and
          Institute for Quantum Optics and Quantum Information of the Austrian Academy of Sciences,  Innsbruck A-6020, Austria\label{addr3}
	\and
	JILA, Department of Physics, University of Colorado, Boulder, CO 80309, USA \label{addr4} \and Center for Theory of Quantum Matter, University of Colorado, Boulder, CO 80309, USA \label{addr5}
}

\date{Received: date / Accepted: date}

\maketitle
\begin{abstract}
In this review, we present the ongoing developments in bridging the gap between holography and experiments. To this end, we discuss information scrambling and models of quantum teleportation via Gao-Jafferis-Wall wormhole teleportation. We review the essential basics and summarize some of the recent works  that have so far been obtained in quantum simulators  towards a goal of realizing analogous models of holography in a lab. 

\end{abstract}
\section{Introduction}

Holographic correspondence  has been the most surprising and celebrated conjecture \cite{maldacena1999large,witten-adscft,Gubser:1998bc,AHARONY2000183} for almost three decades now. It connects special quantum field theories (called the boundary theory) to gravity living in  one extra dimension (called the bulk theory). Using the holographic toolbox, several advances have been made in the physics of strongly coupled quantum field theories-- the transport properties in hydrodynamics \cite{etabys1, etabys2, teaney2009,Son:2009tf, Landsteiner:2011cp,Buchel:2008vz, Brigante:2008gz,Bhattacharyya:2007vjd,Rangamani:2009xk, banerjee2011holographic}, renormalization group flow \cite{Zamolodchikov:1986gt,deHaro:2000vlm,Freedman:1999gp,Barnes:2004jj, Intriligator:2003jj,Myers2011,Myers:2010xs,Cardy:1988cwa, Komargodski:2011vj,Luty:2012ww,Elvang:2012st, Bhattacharyya:2012tc, Elvang:2012yc, Casini:2012ei}, and entanglement entropy \cite{RT-EE,Hubeny:2007xt,Lewkowycz:2013nqa,Fursaev:2013fta,Camps:2013zua,Dong:2013qoa,Bhattacharyya:2013jma,Bhattacharyya:2013gra,Bhattacharyya:2014yga,Miao:2014nxa,Bhattacharyya:2013sia,Rangamani:2016dms}, to name a few.
At a more microscopic level, the relations established between the geometry and quantum entanglement through the entanglement entropy proposal from Ryu and Takayanagi \cite{RT-EE}, ER=EPR \cite{Maldacena:2013xja, Susskind:2014yaa} have been suggestive of the fact that the gravity is an emergent phenomenon  \cite{Swingle:2009bg,Nozaki:2012zj,deBoer:2016pqk,Pastawski:2015qua,Hayden:2016cfa,Czech:2015kbp,Bhattacharyya:2016hbx,Bhattacharyya:2017aly,Yang:2018iki,Chen:2021ipv,Caputa:2017urj,Erdmenger:2020vmo,Brown:2019whu,Chen:2021lnq,error-review-2021}\footnote{References in all of these cases are by no means exhaustive. Readers are encouraged to consult the references and the citations of the papers mentioned in the main text.}. 

On the other side of the duality are gravity and black holes. The duality has also helped to advance us to understand the quantum nature of black holes \cite{Harlow:2014yka} through quantum information processing in the boundary quantum systems. In recent years, the simplicity and analytic amenability of the duality between Sachdev-Ye-Kitaev (SYK) model and nearly Anti-de Sitter spacetime \cite{Sachdev:1992fk, Sachdev:2010um, Kitaev-talks:2015, maldacena2016conformal, Mandal2017, Gaikwad2020} has served as a guiding lamppost for many developments in our understanding of black holes. This refers to, but is not limited to, quantum chaotic properties of black holes \cite{Shenker2014b,Shenker2015a, maldacena2016bound, Cotler2017a,saad2019semiclassical}, and recent progress towards the black hole information paradox \cite{polchinski-BHinfo,Almheiri-review}.

Towards the information content of the Hawking radiation, Hayden and Preskill \cite{Hayden2007} proposed a fascinating thought experiment wherein information thrown into an \textit{old} black hole can be recovered quickly having observed only a few quanta of Hawking radiation. This proposal was later made concrete for generic quantum systems by providing mechanisms for decoding the intended information \cite{Yoshida2017}. At a first thought, one can visualize decoding of information in a quantum circuit as a form of teleportation of information from the input to the output. Whether or when the above is true constitutes some parts of this review. It has been recently argued that the Hayden-Preskill inspired information decoding circuits for generic quantum channels are actually similar (and same in some limits)  to the circuits inspired by teleportation through a wormhole \cite{Brown, Nezami2021, Schuster2021}.

In the first part of this review we discuss these concepts and provide a summary of recent developments on wormhole teleportation inspired  quantum circuits. We begin with holographic dictionary connecting eternal black holes to thermofield double state (TFD) \cite{Maldacena:2001kr} where the two asymptotic regions of left and right black holes are causally disconnected. What it means is that any perturbation on one side can not travel to the other, thus such two-sided {wormholes} are not traversable (see more on wormholes in \cite{kundu-wormhole}). 
Traversable wormholes have fascinated researchers for long \cite{wormhole_1}, however it is also known that we need to violate null energy conditions, or inject negative energy, in order to achieve traversability \cite{wormholes_2, wormholes_3, wormholes_4}. To this end, Gao, Jafferis, and Wall \cite{Gao:2016bin}, followed by \cite{Maldacena2017}, put forth a seminal work where a coupling between the two-sided geometry was proposed, that renders the wormhole traversable. 

Remarkably, the Hayden-Preskill and Gao-Jafferis-Wall protocols are quite generally applicable for quantum many-body systems, and can be realized in the lab using programmable quantum devices. This is possible due to tremendous experimental advances in noisy intermediate scale quantum (NISQ) devices \cite{Preskill2018quantumcomputingin, bharti2021noisy}, which provide a powerful toolset for analog and universal digital quantum simulation. 

In the second part of our review, we describe how these protocols can be implemented in a lab with quantum simulators. Geared towards the goal of observing quantum gravity in a lab, in the holographic language, one requires initially a bridge to translate the tools of holography in terms of many-body dynamics,  see also \cite{liu2020quantum} for a review of the connection between holography and quantum many-body dynamics. Quantum simulators provide unique opportunities to study the time evolution of many-body systems in highly controlled laboratory settings. In this direction, we describe two out of many quantum simulation platforms -- based on trapped ions~\cite{blatt2012quantum,monroe2021programmable} and Rydberg atoms~\cite{browaeys2020many}. We emphasis that while an observation of models marking dynamics dual to black holes is still far away, 
the preparation and benchmarking steps provide promising directions for future experiments. For example, this refers to protocols  \cite{PhysRevLett.123.220502, cottrell2019build, sundar-otoc} and preparation of TFD states \cite{zhu2020generation, wu19, su21, mar19}, observation of Hayden-Preskill variant of quantum teleportation \cite{Landsman2019}, and theoretical proposals \cite{yoshida-yao, daug2019detection2, vermersch2019probing, sundar-otoc, extra1,extra2,extra3,extra4,extra5,extra6,extra7} and experimental observation of out-of-time-ordered correlators (OTOC) \cite{li2017measuring, wei2018exploring, nie2019detecting, Joshi2020,pegahan2021energy,garttner2017measuring, Landsman2019,braumuller2021probing, alaina}  in small-scale quantum simulators.

\paragraph*{Overview:$\ $} This review is organized as follows: In Section \ref{sec:holographic}, we review some basics of the holographic correspondence. To be concrete, we present the example of duality between eternal black holes and TFD states and discuss how the wormholes are made traversable by introducing double trace deformation. In Section \ref{sec:info-spreading}, we discuss and set up basic notations regarding the information spread in quantum systems. We describe that the spread of initial information and the measures of it are the central mechanisms to understand teleportation in quantum circuits. In this section we also review the Hayden-Preskill protocol, and its variant generically applicable to quantum dynamics. In Section \ref{sec:many-body-teleportation}, we discuss the circuits, motivated from the wormhole teleportation, as teleportation circuits for many-body dynamics. We present a mechanism of transfer based on operator size and summarize the recent results. In Section \ref{sec:Qsim-platforms}, we describe in detail two platforms for quantum simulation,  and present realization of many-body models in Section~\ref{sec:many-body-realization}. We then present the measurement protocols, directly accessible in experiments, to measure OTOC and perform many-body teleportation in Section \ref{sec:MeasProt}. We conclude in Section \ref{sec:conclusion} with some additional remarks and future prospects.

\section{AdS$_{d+1}$/CFT$_{d}$ and Wormhole}
\label{sec:holographic}
 AdS/CFT correspondence can be embodied in various avatars, but we will only briefly review some aspects of it which will be relevant for the rest of the review. Essentially, the AdS/CFT  duality links two different theories: a conformal field theory (CFT) which is strongly coupled (typically a large N gauge theory) and a weakly coupled gravity theory defined on the background of Anti-de Sitter (AdS) spacetime which is a spacetime with a negative curvature \cite{maldacena1999large, Gubser:1998bc,witten-adscft}. $d+1$ dimensional AdS spacetime represents the maximally symmetric solution for the Einstein field equation with  a negative cosmological constant $\Lambda=-\frac{d(d-1)}{2\, L^2},$ where $L$ is the AdS radius. The most well-understood example of this duality comes from the String theory. It has been demonstrated in \cite{maldacena1999large,Gubser:1998bc,witten-adscft,AHARONY2000183},  that there exists an equivalence between a strongly coupled $\mathcal{N} = 4$ supersymmetric $SU(N)$ Yang-Mills (SYM) theory and Type $IIB$ String theory on $AdS_5 \times S^5$ in the large $N$ limit, where $N$ is the rank of the gauge group. In this context, one first starts with a stack of $N$ number of $D3$-branes. The low energy dynamics of it is described by $\mathcal{N}=4$ SYM with a Gauge group $ SU(N)$ with the 't-Hooft coupling $\lambda=g_{YM}^2 N$,
 where $g_{YM}$ denotes the Yang-Mills coupling.
 We can analyze this theory perturbatively when $\lambda \ll 1.$ On the other hand, we can have a 10-dimensional metric solution emerging from the low energy description of Type $IIB$ String theory, 
 \begin{align}
     \begin{split}\label{eq1}
     ds^2=&\alpha^\prime\Big[\frac{r^2}{\sqrt{4\pi g_sN}}(-dt^2+dx_1^2+dx_2^2+dx_3^2)\\
     &+\sqrt{4\pi g_s N} \frac{dr}{r^2}+\sqrt{4\pi g_sN}\, d\Omega_5^2\Big], 
     \end{split}
     \end{align}
where $g_s$ is the string coupling.
     We work in $\alpha'\rightarrow 0$ limit where $\alpha'$ is the string tension. In this limit we can effectively neglect any stringy effect and hence work in the supergravity limit (which is essentially a Type $IIB$ Supergravity theory for this case).  In the AdS/CFT duality the couplings on the two sides are related by 
     \be
     \lambda=g_{YM}^2 N= 2\pi\,g_s N~.
     \ee
     We can identify $L^2=\alpha'\sqrt{2\, g_{YM}^2 N}\,$. After this, we can easily see from \eqref{eq1} that the spacetime described by \eqref{eq1} is nothing but $AdS_5\times S^5.$ The supergravity limit necessarily implies, 
    \begin{equation} \label{eq2}
        \Big(\frac{L}{l_s}\Big)^4=2\, g_{YM}^2 N \gg  1,
    \end{equation}
    where we have used the fact that the string length $l_s=\sqrt{\alpha'}.$ This equation simply tells us that,  classical gravity description is valid when the AdS length scale is much
bigger than the string length and from our previous identification the ’t-Hooft
coupling becomes very large. We have classical gravity (and weakly coupled) description in the bulk in the limit described in (\ref{eq2}) and it is equivalent to a strongly coupled gauge theory at the boundary for which standard perturbation theory will not work anymore. Also, the Newton's constant (10-dimensional) which is the coupling for the gravity theory can be shown, 
\begin{equation}
      8\,\pi\, G_{10}= (2\pi)^7\, \alpha'^4 g_s^2.
\end{equation}
From this, it is evident that the gravity theory is weakly coupled. Then the 5-dimensional Newton's constant $G_5$ can be related to $G_{10}$ by simply dividing it by the volume of unit 5-sphere \cite{Ammon:2015wua}. \par
Although this conjecture has not been proven yet, it passes several essential checks, such as matching the spectrum of chiral operators and correlation functions. One obtains a precise dictionary between field theory correlators and correlators of fields living inside the AdS spacetime \cite{maldacena1999large,Gubser:1998bc,witten-adscft,Ammon:2015wua,Penedones:2016voo,DHoker:2002nbb}\footnote{More details about the dictionary are given in the \ref{appendA}.}. Holography is being used to study hydrodynamic transport coefficients,  phase transitions in condensed matter systems, some aspects of QCD, open quantum systems, quantum chaos, black hole information paradox etc \cite{Rangamani:2016dms,Rangamani:2009xk,Jahnke:2018off,Sachdev:2010ch, Nishioka:2018khk, Bhattacharyya:2015nvf,Natsuume:2014sfa,Erdmenger:2007cm,Jana:2020vyx, Liu:2018crr,Beisert:2010jr}\footnote{ We encourage interested readers to consult \cite{Ammon:2015wua} and the references therein  for a comprehensive review of AdS/CFT and its applications.}.\par
Although evidence supports the holographic principle, it is still not clear how gravity emerges from field theory. In recent times, tools of information-theoretic quantities, e.g. entanglement, have provided a more profound insight into the inner workings of AdS/CFT, see a recent review \cite{kibe2021holographic} on bulk emergence and quantum error correction in holography. Following holography, a plethora of interesting studies have resulted \cite{Rangamani:2016dms,Bhattacharyya:2015nvf} and several setups based on quantum information scrambling have been proposed to test certain predictions coming from holography. In the rest of this review, we will discuss some of them.  Also, we will work in natural units where we will set $c=\hbar=k_B=1.$

\subsection{ER=EPR and Wormholes}
We know that quantum mechanics allows for Einstein-Podolsky-Rosen (EPR) correlations \cite{Einstein:1935rr}, which basically stem from the underlying entanglement structure of the wavefunction describing the system. On the other hand, one can find solutions in general relativity that can connect far away points of spacetime via wormholes \cite{Misner:1957mt} which are called Einstein-Rosen bridges (ER) \cite{Einstein:1935tc}. These two phenomena seem to challenge the notion of locality \cite{Einstein:1935rr}. The locality plays an important role in physics, primarily because we cannot send a signal faster than light. From the point of view of spacetime, all points of spacetime are not causally connected.  Maldacena and Susskind later proposed in \cite{Maldacena:2013xja,Susskind:2014yaa} that these two effects are related. In the context of AdS/CFT duality, two entangled copies of a conformal field theory having EPR-type correlation have a bulk dual that connects them through a wormhole.  In particular, two black holes that are spatially far away but have EPR correlation between their microstates described by CFT are actually connected through an ER bridge. To elaborate a little bit more, let us take an analogy from quantum mechanics. Let us consider two CFTs on two spatially disconnected regions $A$ and $B$, and consider the following wavefunction,
\begin{equation}\label{eq3}
|\psi\rangle=|\psi_A\rangle \otimes |\psi_B \rangle,
\end{equation}
where $|\psi_A\rangle$ and $|\psi_B\rangle$ are the wavefunctions of the two non-interacting CFTs at $A$ and $B.$ From (\ref{eq3}), it is evident that $|\psi\rangle$ does not have any entanglement as it is a direct product state. This can be confirmed by computing von-Neumann entropy by tracing out either $A$ or $B$. This state corresponds to two disconnected geometries in the context of holography \cite{VanRaamsdonk:2010pw}. 
\begin{figure}[!ht]
\centering
\includegraphics[width=0.65\linewidth]{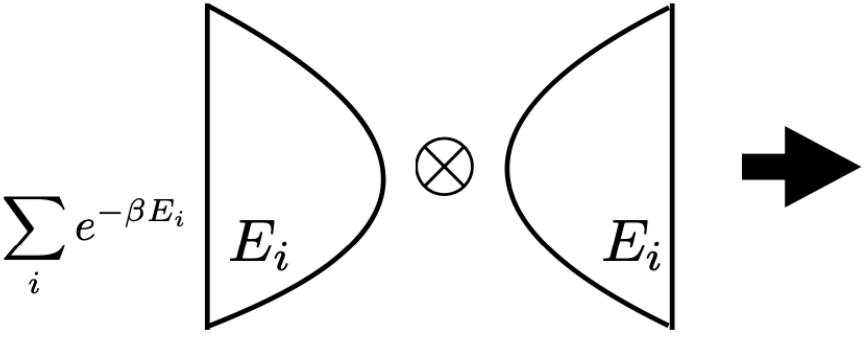}
\includegraphics[width=0.3\linewidth]{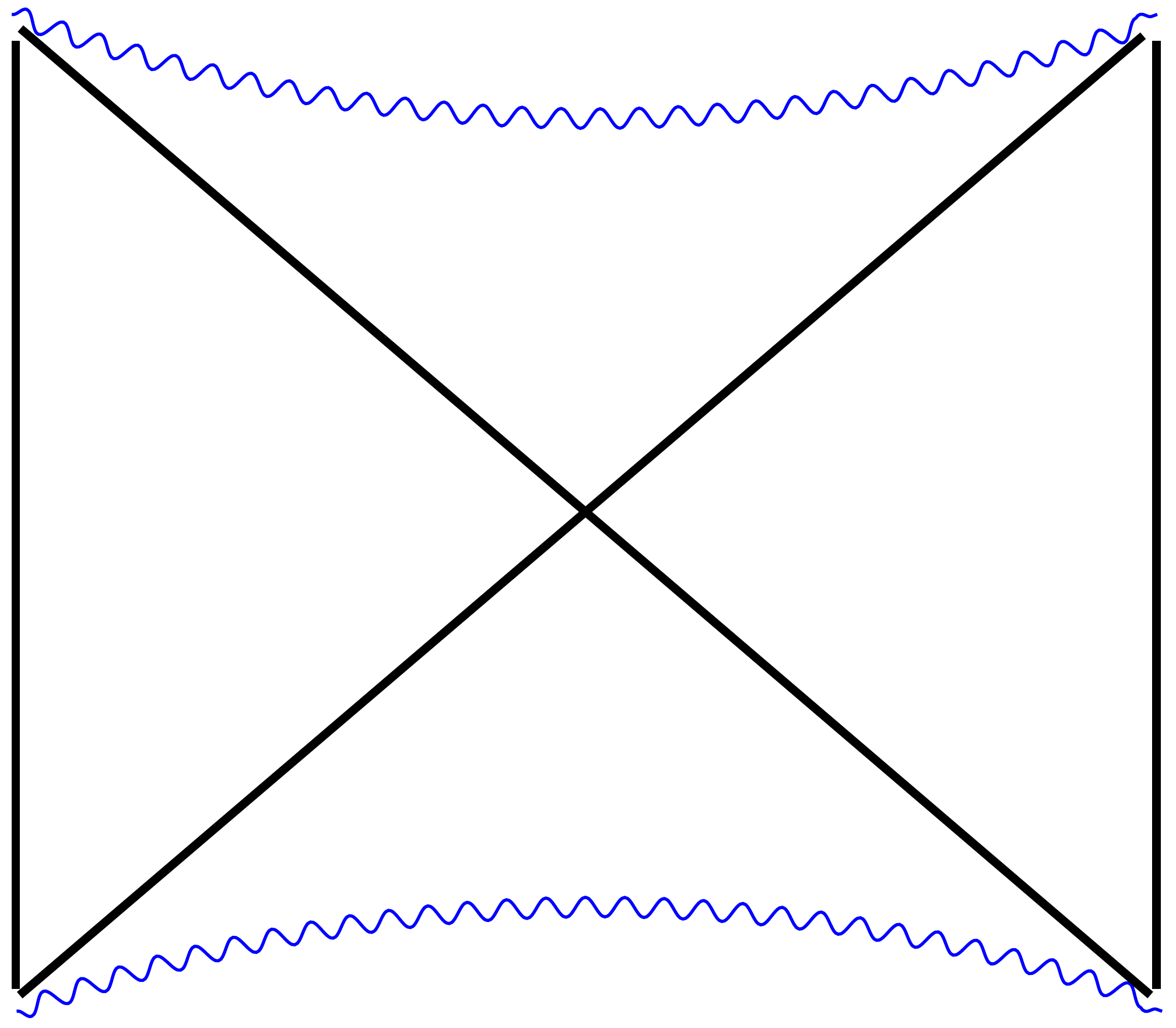}
\caption{Following, Ref.~\cite{VanRaamsdonk:2010pw}, the figure depicts that the entanglement product of two disconnected CFT corresponds to a connected geometry, Penrose diagram of which is shown in the right,  explained later in Fig.~\ref{et}.} \label{Fig1}
\end{figure}

Now following \cite{VanRaamsdonk:2010pw}, we can consider two CFTs placed on $S^d$ and let us denote the $i^{\rm th}$ energy eigenstate of each CFT by $E_i.$ Then let us consider the following wavefunction (up to some normalization),
\begin{equation} \label{eq4}
      |\psi\rangle \propto \sum_{i=1}^{n} \, e^{-\frac{\beta E_i}{2}}|E_i\rangle \otimes |E_i\rangle.
\end{equation}
This is basically a sum of the product state $|E_i\rangle\otimes |E_i\rangle.$ From (\ref{eq4}) it is evident that this state does contain some amount of entanglement, which can be estimated by computing von-Neumann entropy by tracing out one of the CFTs. This is a particular example of the so-called ``thermofield double" state. In the context of holography, this can be shown to be dual to a Euclidean ``eternal black hole" geometry \cite{Maldacena:2001kr} as shown in the Fig.~\ref{Fig1}, which is basically a two-sided Euclidean black hole. So the quantum superposition of two states of two classically disconnected CFTs corresponds to a classically connected geometry (for our case, the two sides are connected by ER bridge). Next, we briefly discuss the geometry of this two-sided black hole.\\
\begin{figure}[!ht]
\centering
\includegraphics[width=1\linewidth]{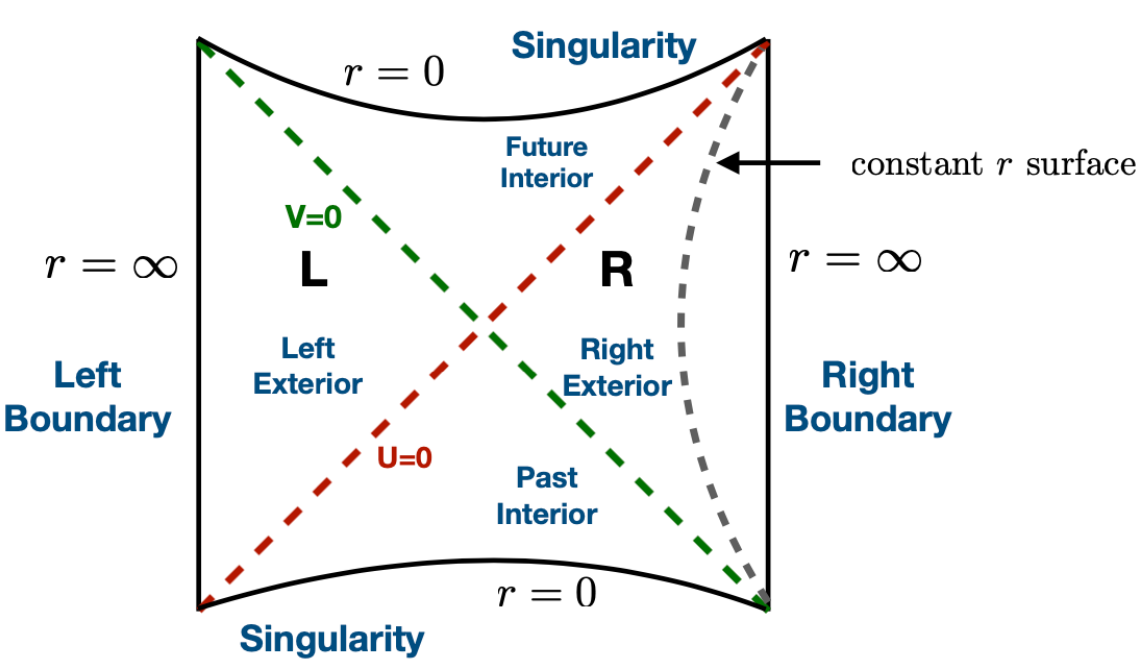}
\caption{\textit{Penrose diagram representing an eternal Schwarzschild-AdS black hole \cite{Harlow:2014yka}.} Also shown are the left and right boundaries where the CFT lies and which the system is dual to. The diagonal lines represent the left and right black hole’s horizons. $r=0 $ corresponds to the singularity of the spacetime. The original exterior region is the right one ($R$) and the new exterior is the left one ($L$). No radial null geodesic can escape the future interior into one of the exterior and no null geodesic can connect the left and right exterior.} \label{et}
\end{figure}

\textit{Eternal black hole:} We consider the eternal AdS black hole with two asymptotic regions. Its Penrose diagram is depicted in Fig.~\ref{et}. An eternal black hole consists of two causally disconnected black holes that share a common time \cite{Maldacena:2001kr}. The separated spaces have non-interacting degrees of freedom, but the two black holes are highly entangled \cite{Harlow:2014yka}, and they form a wormhole that connects both of them \cite{Harlow:2014yka}. To elaborate a little more, let us consider the example of Euclidean non-rotating Ba\~{n}ados-Teitelboim-Zanelli (BTZ) metric\footnote{For the Euclidean case, we have analytically continued the Lorentzian time: $t \rightarrow i\, \tau.$}, 
\begin{align}
      ds^2= f(r) d\tau^2+ \frac{dr^2}{f(r)}+r^2 d\phi^2,
\end{align}
where, $ f(r)=\frac{r^2-r_+^2}{L^2}$, $L$ is the AdS radius, $r=r_+$ is the horizon where the $f(r)$ vanishes.
The period of the $\tau$ coordinate $\beta=\frac{2\pi\, L^2}{r_+}$ is identified with the inverse of the temperature $T$  of the black hole. The period of $\phi$ is $2\pi.$ Together $\tau$ and $\phi$ provide the coordinates for the space on which the dual CFT is defined.  The metric becomes ill-defined at $r=r_+$ but this is just a coordinate singularity. One can define the following coordinate transformation,
\begin{align}
     U=-e^{-\kappa\, u}, V= e^{\kappa\, v}, 
\end{align}
where $\kappa=\frac{r_+}{l^2}$ is the surface gravity and $u,v=i\,\tau \pm r_*,$ with, 
\begin{align} \nonumber
      r_*=-\int_r^{\infty}\frac{dr'}{f(r')}=\frac{L^2}{2 r_+}\log\Big[\frac{r-r_+}{r+r+}\Big].
\end{align}
This is nothing but a Kruskal transformation \cite{Harlow:2014yka}. The metric becomes,
\begin{align} \label{eq5}
      ds^2=-\frac{4\, L^2\, dU\, dV}{(1+ U\, V)^2}+\frac{r_+^2\, (1- U\, V)^2}{(1+U\, V)^2}d\phi^2.
\end{align}
$U=0$ and $V=0$ are the two horizons. From (\ref{eq5}), it is evident that the metric is well defined even when either $U=0$ or $V=0.$ 
While doing the coordinate transformation, we implicitly assumed that $r > r_+,$ thereby making $U$ negative and $V$ positive. Similarly, for the region $r < r_+$ we can perform the same type of coordinate transformation only with the difference that for that case, $U$ will be positive and $V$ will be negative. Then we again end up with the same form of the metric as shown in (\ref{eq5}). Finally, the Penrose diagram for the spacetime looks like as shown in Fig.~\ref{et} \footnote{Note that, to draw the Penrose diagram, we need to do further a conformal compactification of the metric defined in (\ref{eq5}). This can be done by using a particular coordinate transformation and then throwing out an overall conformal factor. Interested readers are referred to \cite{Harlow:2014yka, Banados:1992wn} for more details. Also, we have ignored the angular coordinate. Each of the points on Fig.~\ref{et} corresponds to a $S^1.$}. The spacetime now has four regions, as shown in the Fig.~\ref{et}. The two singularities occur at $U\, V=1\,$ $ (r=0)$, and the $U\, V=-1\,\,$ $(r=\infty)$ corresponds to the two asymptotic AdS boundaries. Combining all four regions, we can now interpret the full two-sided Euclidean BTZ space as a wormhole connecting the two asymptotically-AdS spaces. The wormhole is non-traversable in the sense that no signal can be sent from the region- $L$ to the region $R$ as shown in the Fig.~\ref{et}, but two people, Alice and Bob, will be able to jump from these two sides and reach the middle point (the bifurcation point where $U=0$ and $V=0$ line intersect as shown in the Fig.~\ref{et}) and exchange notes. Although we have used mainly the BTZ metric, all these analyses can be extended to higher dimensions.\\

\textit{Thermofield double state:} As we know that the\\ AdS/CFT is a two-way street, we briefly now discuss the dual of this geometry. Within the context of holography, each geometry corresponds to a certain state of the dual field theory. From the boundary point of view, the CFT lives on a space described by two coordinates, both of which are periodic. The space looks like a product of two spheres: $S_{\beta}^{1} \times S^{d-1}.$  $S_{\beta}$ is coming from the $\tau$ coordinate, and $S^{d-1}$ is coming from the rest of the angular coordinates. For (eternal) BTZ, we have $d=2$, and for a constant time slice, the boundary will be the sum of two disconnected spheres $ S^{1}+ S^{1}$. Then the Euclidean time direction then connects these two spheres.  Then following \cite{Maldacena:2001kr}, we can write down the dual state as,
\begin{equation} \label{tfd}
|\psi\rangle=\frac{1}{\sqrt{Z (\beta)}}\sum_j e^{-\beta E_j/2} | E_j\rangle_L \otimes | E_{j}^*\rangle_R~,
\end{equation}
where $| E_{i}\rangle$ denotes the energy eigenstate of the CFT placed on the sphere, $L$ and $R$ indicate the two asymptotic regions, the sum over $i$ goes over all the eigenstates\footnote{At this point, we are still in the field theory limit. Hence this sum goes up to $\infty$ as we are dealing with infinite-dimensional Hilbert space.} and $Z(\beta)$ is the thermal partition function for one copy of the CFT. The star denotes the CPT conjugation. 
From (\ref{tfd}) it is evident that this state is an entangled state defined on a Hilbert space of the form $\mathcal{H}=\mathcal{H}_{L}\otimes \mathcal{H}_{R}.$ In general finite dimensional quantum systems, TFD is a useful way to purify a given thermal state, we discuss this aspect in the next Section \ref{sec:info-spreading}.

Due to the presence of the factor $e^{-\beta E_j/2}$, one can easily see (by computing the von-Neumann entropy by tracing one of the subsystems, either $L$ or $R$) that $\ket{\psi}$ possesses non-vanishing entanglement. From the wavefunction $\ket{\psi}$ in (\ref{tfd}), we can compute the thermal expectation value of any operator in the following way, 
\begin{equation}
      \langle \psi |O_L|\psi\rangle= Tr (\rho_L^{\beta} O_L),
    \label{eq:exp-tfd}
\end{equation}
where, $O_L$\footnote{ This should be read as $O_L\otimes I_R,$ $I_R$ is the identity operator acting on the right region.} is an operator which acts on the left asymptotic boundary. Then one can trace over the right region, and effectively the expectation value of this operator will be given by tracing over the reduced density matrix of the left region ($\rho_L^{\beta}$) times the operator $O_L.$ The reduced density matrix $\rho_L^{\beta}$ comes from the fact that we have traced out the right region entirely. The subscript $\beta$ denotes the fact that it is a thermal density matrix that arises due to the entanglement between the two copies. Similarly, one can compute higher point correlation functions also. On the dual side, one can use the standard techniques of holography to compute these correlators. Following \cite{Maldacena:2001kr,Keski-Vakkuri:1998gmz} we will below quote the result for two-point functions of two spinless primary operators of scaling dimension $\Delta$\footnote{In the context of AdS/CFT, this corresponds to scalar fields on AdS with a certain mass.} acting on $L$ and $R$ boundaries (both at $t=0$) respectively\footnote{Following \cite{Keski-Vakkuri:1998gmz}, one can compute this correlator by using the standard holographic method. One first computes the bulk to boundary propagator using the method of image; hence one has to shift the $\phi$ coordinate by the factor of $2\pi n$ and then sum over all the values of $n.$}.
\begin{align}
    \begin{split}\label{eq6}
&\langle \psi |O_R (0,\phi_R) O_L (0,\phi_L)|\psi\rangle  \sim \\& \sum_{n=-\infty}^{\infty}\frac{1}{\Big[1+\cosh\Big(\frac{2\pi(\phi_R-\phi_L)+2\pi\, n}{\beta}\Big)\Big]^{2\Delta}}.
\end{split}
\end{align}
From (\ref{eq6}), it is evident that we indeed get non-vanishing correlations between two operators acting on two disconnected CFTs. This is because the underlying geometry and the dual state have some entanglement, although the two boundary regions are causally disconnected. This provides evidence to the ER=EPR conjecture discussed previously.

\subsection{Teleportation through traversable Wormholes} \label{sec:gjw-section}
The rest of the review will mainly focus on quantum information spreading and its implications for holography. Particularly, we will focus on the teleportation of quantum information and the corresponding holographic model. This provides us with an interesting playground to test some of the predictions from holography in the experimental setting. It is evident from our previous discussion that wormholes provide an ideal setting for quantum teleportation \cite{Susskind:2017nto} because they have EPR-like correlations. However, the wormhole that we have discussed previously is not traversable \cite{Misner:1957mt,Harlow:2014yka}. 
\begin{figure}[!ht]
\centering
\includegraphics[width=1\linewidth]{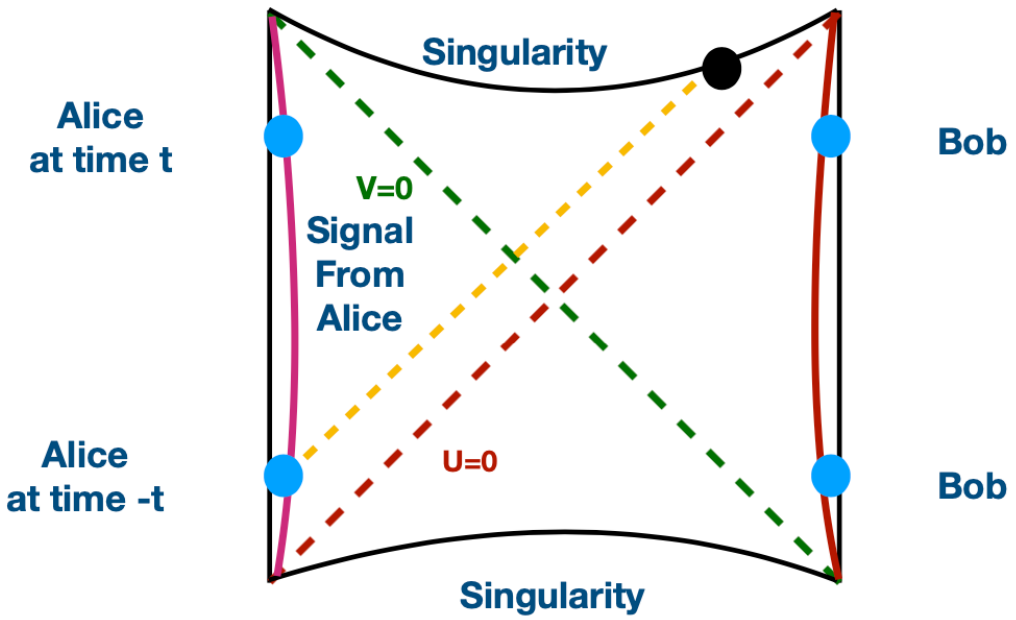}
\caption{Alice and Bob are accelerating near the left and right boundary. Alice sends a signal at time $-t$ at the left boundary, (shown by yellow dashes). Instead of reaching Bob, it will be lost into the singularity as no light like trajectory can escape into one of exterior region from the other, bypassing the future interior. 
}. 
\label{et1}
\end{figure}

As shown in the Fig.~\ref{et1}, Alice sends a signal from the left boundary at some time $-t.$ She is accelerating near the left boundary, as shown by the hyperbolic trajectory. She hopes that Bob, who is accelerating near the right boundary, will receive the signal. But as evident from the diagram, as the signal moves at the speed of light, it will always hit the singularity and Bob will never receive it. So we cannot send a signal through this non-traversable wormhole even if it possesses EPR-like correlation.\par 
For teleportation, we need a traversable wormhole \cite{Morris:1988cz}. The exact protocols for quantum teleportation through a traversable wormhole will be reviewed in detail in the later sections. In this section, we briefly discuss the argument put forward in \cite{Gao:2016bin,Maldacena2017} to make a traversable wormhole.  It is well known that in general relativity, the traversable wormhole only occurs when the stress tensor for the matter sector violates the null energy condition \cite{Visser:1995cc,Gao:2016bin,Maldacena:2020sxe,Ahn:2020csv}. In the context of AdS/CFT, there is a precise protocol to achieve this, and we will discuss this in the context of the eternal AdS black hole following Ref.~\cite{Gao:2016bin}.  We first deform the system by adding a relevant double trace deformation. So the change in the action (boundary CFT action) is given by\footnote{Time runs in the opposite direction for two exterior wedges of Eternal black hole geometry. Hence the $\mathcal{O}_R$ and $\mathcal{O}_L$ are inserted at $t$ and$-t$ respectively.},
\begin{equation}
    \delta S= \int dt\, d^{d-1}x\, h(t,x) \mathcal{O}_R(t,x) \mathcal{O}_L(-t,x),
    \label{eq:original-coupling}
\end{equation}
where $\mathcal{O}_L$ and $\mathcal{O}_R$ are scalar operators with scaling dimension less than $\frac{d}{2}$ and acting on the left and right boundary, respectively. For the case of the eternal BTZ black hole \cite{Banados:1992wn}, $d+1=3$ and $x$ will be the azimuthal coordinate $\phi.$ By AdS/CFT dictionary, these two operators will be dual to a scalar field $\varphi$ with certain mass propagating inside the bulk spacetime. Also, we remember that time runs in the opposite direction in left and right wedges of this eternal geometry. The function $h(t,x)$ is turned on only after a certain time, which is referred to as ``turn-on" time. The integral over time makes sure that we do not get contribution from very high energy states. In the path-integral, this will have the contribution of the form $\sim e^{i\,\delta S}.$ In the subsequent section, we will ignore this time integral following \cite{Maldacena2017} then we will have the contribution to the path integral simply as $\sim e^{i\,\tilde g \mathcal{O}_L(0)\, \mathcal{O}_R(0)},$ where $\tilde g$ is an overall coupling constant. $\mathcal{O}_L(0)$ and $\mathcal{O}_R(0)$ are inserted at the two asymptotic boundaries at $t=0.$\par One can further compute the stress-energy tensor of this scalar field $\varphi$ in the bulk spacetime
\begin{equation}
    T_{\mu\nu}=\partial_{\mu}\varphi\partial_{\nu}\varphi-\frac{1}{2}g_{\mu\nu}(\partial \varphi)^2-\frac{1}{2} m^2\varphi^2,\end{equation}
where $m$ is the mass of the scalar field. From this, one can compute the 1-loop expectation value of this stress tensor. Following \cite{Gao:2016bin}, we get,
\begin{align}
    \begin{split}
    \langle T_{\mu\nu}\rangle =\lim_{x\rightarrow x'}\Big[&\partial_{\mu}\partial'_{\nu} G(x,x')-\frac{1}{2}g_{\mu\nu}\partial_{\alpha}\partial'^{\alpha}G(x,x')\\&-\frac{1}{2}g_{\mu\nu} m^2 G(x,x')\Big].
\end{split}
\end{align}
One uses point splitting method to compute this stress-tensor and one has to normalize it to get a finite result. $G(x,x')$ is a two-point function of the scalar field. One such two-point function when there is no double trace deformation is shown in (\ref{eq6}). But in the presence of this deformation it will get modified. A detailed calculation of it is given in \cite{Gao:2016bin}. Now as mentioned earlier to make the wormhole traversable we need to break the null energy condition. In this case, we have to violate the average null energy condition \cite{Gao:2016bin}. Let $k^\mu$ be the tangent vector of the null geodesic passing through the wormhole and let $\lambda$ be the affine parameter, then average null energy condition (ANEC) is,
\begin{equation}    \int_{-\infty}^{\infty} \langle T_{\mu\nu}\rangle\, k^{\mu}k^{\nu}d\lambda \geq 0.
\end{equation}
In our Kruskal coordinate, 
$\partial_{U}$ is the tangent vector to the infinite null geodesic along the horizon $V=0$ and we can choose $U$ as the affine parameter. So the violation of ANEC implies,
\begin{equation}
      \int dU \langle T_{UU}\rangle < 0. 
\end{equation}
 \begin{figure*}[t]
\centering
\includegraphics[width=0.9\linewidth]{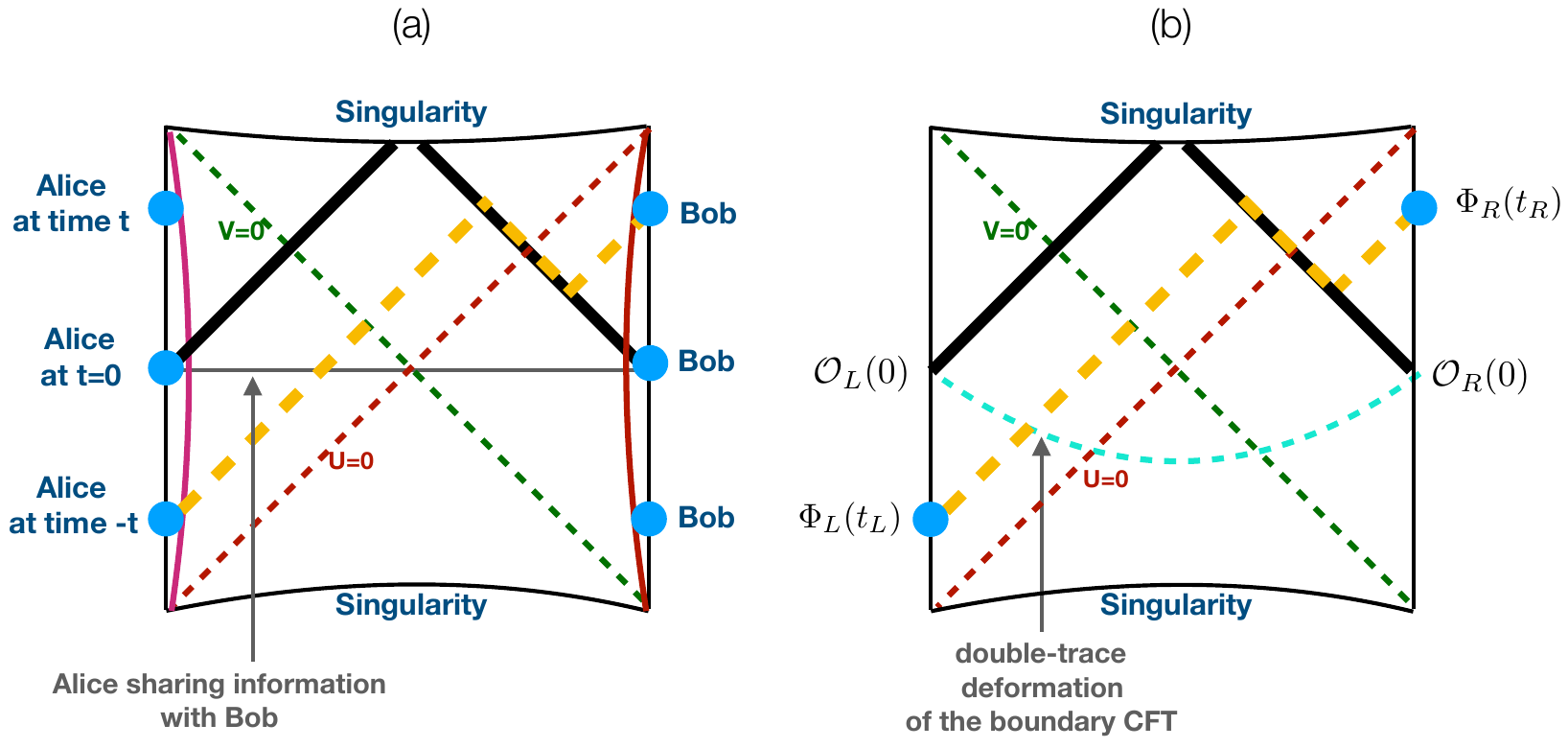}
\caption{In (a), Alice sends a signal at time $-t,$ (shown by yellow dashes) then she measures a part of the Hawking radiation and exchanges information with Bob at time $t=0$ (shown with gray line). This helps Bob send a negative energy shock (shown with solid black). Because of this, the signal reaches Bob at time $t$ due to a Shapiro time advance. This is the essence of quantum teleportation \cite{Maldacena2017, Nariman,youtube}. In (b), following \cite{Maldacena2017}, the same scenario is depicted in terms of operators, the message $\Phi_L(t_L)$ sent by Alice from the left boundary at time $t_L$, experiences the negative energy shock generated due to the double trace coupling $\mathcal{O}_L \mathcal{O}_R$ at $t=0$, and finally reaches to Bob $\Phi_R(t_R)$ at the right boundary at time $t_R.$ This diagram is motivated from \cite{Gao:2016bin, Maldacena2017}.}
\label{et2}
\end{figure*}

Now this will back react to the geometry, and for a small spherically symmetric perturbation from the relevant component of the linearized Einstein equation, one can find that at $V=0$ \cite{Gao:2016bin},
 \begin{align}
     \begin{split}\label{eq8}
 \frac{(d-1)}{4}\Big[&\Big(\frac{(d-2)}{r_h^2}+\frac{d}{L^2}\Big)\Big(\delta g_{UU}+\partial_U(U \delta g_{UU})\Big)\\&- \frac{2}{r_h^2}\partial_{U}^2 \delta g_{\phi\phi}\Big]=8\,\pi\, G_N\, \langle T_{UU}\rangle,
 \end{split}
 \end{align}
 where $r_h$ is the black hole horizon radius and $\phi$ denotes the azimuthal angle. $\delta g_{UU}$ is the linearized fluctuation of the metric. For the BTZ, $d+1=3$ and $r_h=r_+$ which follows from (\ref{eq5}). Again following \cite{Gao:2016bin}, we can argue that perturbations will reach a stationary state  with respect to the Killing symmetry $U \partial_U$ after the scrambling time as the deformation is small. Also, $T_{UU}$ will be decaying faster than $\frac{1}{U^2}$ and all other terms in the equation (\ref{eq8}). Then we integrate (\ref{eq8}) and drop all the total derivative terms as at the end points as they will vanish. Then we get,
 \begin{align} 
 \begin{split}\label{eq9}
   & \frac{(d-1)}{4}\Big(\frac{(d-2)}{r_h^2}+\frac{d}{L^2}\Big)\int dU\, \delta g_{UU}\\& = 8\,\pi\, G_N\, \int dU\, \langle T_{UU}\rangle .
 \end{split}
 \end{align}
This equation relates the integral of $\langle T_{UU} \rangle$ to the integral of $\delta g_{UU}.$ We also know that up to linear order in perturbation,
 \begin{equation}
       V(U)=-\frac{1}{2 g^0_{UV}}\int_{-\infty}^{U} dU\, \delta g_{UU}.
 \end{equation}
 
Note that,  $g^{0}_{UV},$ the original $U V$ component of metric is negative on $V=0$ slice. Now we can impose the ANEC condition. If ANEC violates, then from (\ref{eq8}) the integral over $\delta g_{UU}$ is also negative (note that the prefactor $\frac{(d-1)}{4}\Big(\frac{(d-2)}{r_h^2}+\frac{d}{L^2}\Big)$ in (\ref{eq8}) is positive for $d \geq 2$.). Following \cite{Gao:2016bin}, we can conclude that whenever ANEC violates, $V(+\infty)\rightarrow 0,$ so that a light ray from the left boundary will reach the right boundary after a finite time. Furthermore, one can also calculate the deviation of this light ray from the horizon ($\Delta V$) by computing the Shapiro time delay (in our case it is actually a time advance\footnote{Here we have a shockwave backreacting on the geometry, thereby generating this time advance. This has also been used in other contexts, for example, to discuss causality constraints \cite{Camanho:2014apa}.}!) and we can show that it is proportional to $h(t,x)$ \cite{Gao:2016bin} as defined in (\ref{eq:original-coupling}). Again for more details interested readers are referred to \cite{Gao:2016bin}.
 \par
Before we end the section, we give an intuitive picture of the exchange of classical information in the quantum teleportation protocol realized in the bulk dual through the classical coupling introduced to the system. We briefly sketch the argument provided in \cite{Maldacena2017, youtube,Nariman}.
As shown in Fig.~\ref{et2}, Alice first sends her message into the left horizon while accelerating near the left boundary (in our context, this message can be a scalar field propagating towards the black hole horizon). At time t= 0, she measures a part of the Hawking radiation emitted from the black hole. Remember, the Hawking radiation is generated due to vacuum fluctuations. Suppose that Alice measures the positive Hawking radiation energy, which corresponds to the positively charged particle of the Hawking pair created near the horizon. She then sends the result of her measurement to Bob, who is accelerating near the right horizon. So a classical communication takes place. Based on the result of Alice's measurement, Bob now has a sense of what the positive energy particle is, and then he can measure the Hawking radiation to identify the negative energy particle. This is possible since Alice and Bob share an entangled state (in our case, it corresponds to a thermofield double state). Then Bob can throw a negative energy pulse into the horizon from the right boundary as shown in the Fig.~\ref{et2}. This negative energy pulse causes the singularity to recede and help the signal from Alice to speed up. Specifically, signals (in our case, a scalar field propagating across the bulk) get delayed ( or advanced in this case) due to the negative energy shock. In general relativity, this well-known effect is known as the Shapiro time delay \cite{Shapiro:1964uw}. This delay (or the advancement) happens due to the double trace coupling $\mathcal{O}_L \mathcal{O}_R$ turned on for certain time interval results in the ANEC violation. So finally, the signal speeds up, and instead of hitting the singularity, it reaches Bob!\par
So far, in the present section, we have discussed the teleportation through a wormhole from the point of view of the bulk gravity. The coupling and the teleportation in the gravity have a straightforward representation in the boundary theory described by a TFD, wherein coupling the two Hamiltonians,  information is teleported from one Hilbert space to another. In quantum simulators, which can realize very general states and engineer interesting evolutions, one can ask  the question of the generality of such a gravity-inspired teleportation scheme. We review some recent developments in understanding the underlying mechanism of teleportation and their applicability in general many-body models in Section \ref{sec:many-body-teleportation}. In the next section, Section \ref{sec:info-spreading}, we first set up some useful notations and summarize important results on quantum information scrambling, which makes the basis for the following sections. 
\section{Quantum information spreading}
\label{sec:info-spreading}
Consider a Heisenberg operator $W$  evolving under a local Hamiltonian, $H$ acting on a lattice, such that $W(t)= e^{iHt}~W ~e^{-iHt}$. As a function of time, this operator can be written using the Baker–Campbell–Hausdorff formula as 
\begin{align}
W(t)&= W+i t[H, W]-\f{1}{2!} t^2 [H, [H, W]] \nonumber \\ &- \frac{1}{3!} it^3 [H, [H, [H, W]]] + \cdots
\label{eq:bch}
\end{align}
Thus, as time grows, the operator $W$ contains sums of many products of local operators. For example, if we consider a local Hamiltonian with interactions only on neighboring sites, the operator $W$ will spread to farther and farther sites as the time evolves. This is referred to as \textit{quantum information spreading}, and has been a central goal in various studies in recent years, involving the operator growth and the study of out-of-time ordered (OTOC) correlators (more details follow). Before continuing further towards operator growth and spreading, it will be useful to introduce some notations and diagrammatic representations which we use in several places. For the   diagrammatic notations we follow Ref.~\cite{hosur2016chaos}. 
\subsection{Operator-State correspondence}
\begin{figure}[!ht]
\includegraphics[width=\linewidth]{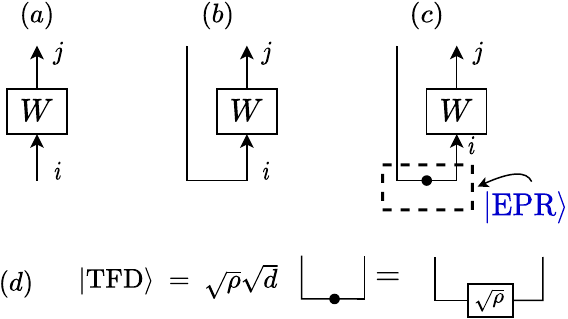}
\caption{\textit{Operator-state correspondence in diagrammatic form.} a) An operator $W$ is represented by an ingoing and an outgoing index. b) In the state representation both the ingoing and outgoing index are treated similarly and each of them denote a basis state in the two copies of the Hilbert space. c) The state $\ket{W}$ is related to the EPR, by the relation Eq.~\eqref{eq:stateW}, The dashed box denotes the EPR state \eqref{eq:epr}. d) The TFD denotes finite temperature generalization of EPR, where the density matrix $\rho$ is the density matrix in either the left or the right Hilbert space, $\rho=e^{-\beta H}/\mathrm{Tr}(e^{-\beta H})$.
}
\label{fig:op-state}
\end{figure}

An operator $W$, in a Hilbert space, can be expressed as 
\begin{equation}
   W=\sum_{i, j=1 }^d W_{ij} \ket{i}\bra{j}~,
\label{eq:operatorW}
\end{equation}
where $\ket i$, $\ket j$ denote the basis elements of the Hilbert space whose regularized dimension is $d$, and thus $i, j =1, 2,\cdots, d$. The coefficients $W_{ij}=\langle i| W |j \rangle$ denote the elements in the matrix representation of $W$ in this basis. In Fig. \ref{fig:op-state}(a) this operator is represented with an input leg $i$ and an output leg $j$. 

The operator-state correspondence relates an operator of the above form to a state in the doubled Hilbert space, $\mathcal{H}\otimes \mathcal{H}$, given as 
\begin{equation}
\ket{W}=\frac{1}{\sqrt{{\rm Tr}(W^\dagger W)}}\sum_{i, j=1 }^d W_{ij} \ket{i}\otimes\ket{j^*}~.
\label{eq:stateW}
\end{equation}
The basis states with a star $\ket{j^*}$ are the time reversed (or equivalently complex conjugated) states. These are related to $\ket j$ with an anti-unitary operator $\ket{j^*}=\Theta \ket{j}$. The prefactor $1/\sqrt{{\rm Tr}(W^\dagger W)}$ is the normalization constant. The above map from an operator in a single Hilbert space to a state in a doubled Hilbert space is also known as the `\textit{purification}', since the state $\ket W$ is a pure state, i.e. $\mathrm{Tr}((\ket{W}\bra{W})^2)=1$. We denote this state by a bent input line, as shown in Fig.~\ref{fig:op-state}(b).

An example of a pure state in the doubled Hilbert state is the EPR state. In its most simple form it can be understood as the product of $N$ Bell pairs, $\epr=(\ket{\Phi^+})^{\otimes N}$, where $\ket{\Phi^+}=(\ket{00}+\ket{11})/\sqrt{2}$  is a maximally entangled state between a pair consisting of one qubit from each Hilbert space, here $(0,1)$ are the computational basis or the qubit basis.   This definition can be rewritten using the basis elements of each Hilbert space as
\be
\epr=\f{1}{\sqrt{d}}\sum_j {\ket{j}\otimes \ket{j^*}}~.
\label{eq:epr}
\ee
Comparing with Eq.~\eqref{eq:stateW}, we note that the EPR state is a purification of the identity operator $\mathbb{1}$, which is  also the density matrix for a state at infinite temperature $\rho_{\infty}=\mathbb{1}/d$. Therefore, the EPR state denotes an infinite-temperature state. In what follows we denote the EPR state with a notation shown in the dashed box in Fig.~\ref{fig:op-state}(c). 
Using this definition, we can further write the state $\ket W$ in Eq.~\eqref{eq:stateW} as
\be
\ket{W} = \sqrt{\frac{ d }{{\rm Tr}(W^\dagger W)}} ( W\otimes \mathbb{1})\epr~.
\ee
The EPR state has a special property, often termed as operator shifting, i.e., an operator acting on the left is the same as the operator transpose acting on the right,
\be 
(O_L\otimes \mathbb 1)\epr=(\mathbb 1\otimes O_R^T)\epr~,
\label{eq:shifting-epr}
\ee
where the subscripts $L$ and $R$  denote the two copies, as in the case of the asymptotic region of holography. These subscripts  label the side an operator $O$ acts on. This identity is a direct consequence of the definition Eq.~\eqref{eq:operatorW}, which implies $W^T=\sum_{i,j}W_{ij}|j^*\rangle\langle i^*|$, combined with the definition of EPR. 
We can now revisit the  finite temperature generalization of the EPR, i.e., the thermofield double states (TFD).

\paragraph*{Thermofield Double States (TFD):$\ $} In the context of CFTs, we listed the TFD state, in the previous section, as the holographic dual to an eternal black hole. On the doubled Hilbert space  $\mathcal{H}_L\otimes \mathcal{H}_R$, with finite dimensional Hilbert spaces, the TFD state at temperature $T \equiv 1/(k_B\beta)$ is an entangled state on $2N$ qubits, defined as
\begin{equation}
\tfd = \frac{1}{ \sqrt{Z} }\sum_{i=1}^d e^{-\beta E_i/2}\ket{E_i}_L\otimes \ket{E_i^*}_R ,
\label{eq:tfd}
\end{equation}
where $Z=\mathrm{Tr}[\mathrm{exp}(-\beta H)]$.   The sum in the TFD runs over the eigenstates $\ket{E}$ of $H$, labeled by $i$, with respective eigenvalues $E$, i.e., $H\ket E=E \ket E$. The time reversed state ${E^*_i}$ satisfy, ${H^*}\ket {E^*}= E\ket {E^*}$. 
There have been many interesting works using the TFD state, in particular, in black holes \cite{ISRAEL1976107}, quantum field theory \cite{1975-tfd}, and more recently in connections with holography \cite{Maldacena:2001kr, Maldacena2017, maldacena2018eternal} and others. Some of the main properties that make it a valuable subject are:
\begin{itemize}
    \item {It is a pure state. Constructing a density matrix $\rho_{\mathrm{TFD}}=\tfd \bra{\mathrm{TFD}}$, one notes that $\mathrm{Tr}(\rho_{\mathrm{TFD}}^2)=1$.}
\item {By tracing one part of the system, we obtain,\\ $\mathrm{Tr}_R(\tfd \bra{{\rm TFD}})=\rho_{L}$ where $\rho_L$ is a the thermal density matrix on the left system with Hamiltonian $H_L$, $\rho_L=\mathrm{exp}(-\beta H_L)/Z$.}
\item {Since the state is defined on a product Hilbert space, expectation values of operators on one Hilbert space stay as thermal expectation values in that system. For example, for an operator in the left system,  $\langle \mathrm{TFD} |O_L| \mathrm{TFD}\rangle =\mathrm{Tr}(\rho_L O_L)$, as already mentioned in the previous section Eq.~\eqref{eq:exp-tfd}.}
\end{itemize}
In Fig.~\ref{fig:op-state}(d), the TFD is written in terms of the EPR state such that, 
\be
\tfd = \sqrt{d}\sqrt{{\rho_L}}\epr=\sqrt{d}\sqrt{\rho^*_R}\epr~.
\label{eq:tfd-epr}
\ee
Similar to the relation \eqref{eq:shifting-epr}, for the TFD state we find, 
\begin{align}
(O_L\otimes\mathbb 1)\tfd &= (O_L\otimes \mathbb 1)\sqrt{d}\sqrt{\rho^*_R}\epr~,\nonumber\\
&=\sqrt{d}\sqrt{\rho^*_R}( \mathbb 1 \otimes O_R^T)\epr~,\nonumber\\
\text{and}~~(\mathbb 1 \otimes O_R)\tfd &= \sqrt{d}\sqrt{\rho_L}(O_L^T\otimes \mathbb 1)\epr~.
\label{eq:shifting-tfd}
\end{align}
These relations will be useful in next sections where we discuss the measures of the information scrambling and the many-body teleportation circuit. 
For this purpose, in the next subsection we return to  quantifying information scrambling using the out-of-time-ordered correlators.  

\subsection{Out of time ordered correlators}\label{subsec:otoc}

To quantify the spread of information in a quantum system we  can ask the question in terms of commutators representing the information and a probe. The effects of an initial perturbation, say $W$, on a later measurement of another operator $V$ can be understood by computing the commutator $[W(0), V(t)]$. Even if the operators $W$ and $V$ at $t=0$ commute, after the time-evolution following \eqref{eq:bch} the operators need  not commute.  As an observable, it is meaningful to consider
\begin{equation}
C(t)=\langle [W(0), V(t)]^\dagger[W(0), V(t)]\rangle~,
\label{eq:coft}
\end{equation}
where the angle brackets denote expectation value in a state $\rho$, $\langle C(t)\rangle = \mathrm{Tr}(\rho C(t))$. Thus, $C(t)$ for initially commuting operators grows in magnitude with time. When expanded, $C(t)$ contains time-ordered and out-of-time-ordered correlators (OTOCs). One of the four terms in the expansion consists  of the composite operator  
\be
O(t)=W^\dagger (0)V^\dagger(t)W(0)V(t)~.
\label{eq:op-O}
\ee
The expectation value of this operator in some state, $\langle O(t)\rangle$, denotes a correlation function between two operators $W$ and $V$, where the times appear out of order. 

Lately in connection with quantum chaos, OTOC has acquired  a lot of interest in condensed matter systems, black holes, SYK, many-body quantum systems \cite{Shenker2014b,hosur2016chaos, Roberts2015, maldacena2016bound, Hashimoto2017,Jahnke:2018off, HydroOTOCEE,GalitskiLiao, vermersch2019probing,monika-fast-tree} to cite a few. Some intuition for this connection is often given as follows. In a classical system characterized by position ($x$) and momentum ($p$), the change in the position due to changes in initial conditions can be denoted by $\delta x(t)/\delta x(0)$. The classically chaotic systems are known to display butterfly effect, wherein $\delta x(t)/\delta x(0) \sim \mathrm{exp}(\lambda t)$, i.e., nearby trajectories differ exponentially at a later time-- the exponent $\lambda$ is known as the Lyapunov exponent. Quantum mechanically, such deterministic information about the coordinates of a system or particle is not possible and therefore effects of initial perturbations are studied through the real observable $C(t)$. A quantum butterfly effect is often stated as the scenario when the $C(t)$ becomes as large as 2$\langle W^\dagger W \rangle \langle V V^\dagger\rangle$ at late times \cite{Roberts2015, maldacena2016bound}, which implies that at these times, the OTOC decay to zero. This time is known as the scrambling time $t_{\mathrm{scr}}$, which is when the initial local information is spread to all the degrees of freedom.  

Writing analytically an expression for OTOC depends on the underlying evolution operator $U$ and may not always be possible. However for systems evolving under Haar random unitaries\footnote{We assume familiarity with Haar measure. }, it can be shown that after long time $t>t_{\mathrm{scr}}$, and for large systems, the OTOC between general operators takes the following form,
\begin{align}
&\langle W(t)Y(0)Z(t)X(0)  \rangle \approx\nonumber\\ &\langle W Z\rangle \langle Y\rangle \langle X\rangle+\langle W \rangle \langle Z\rangle \langle YX\rangle-\langle  Z\rangle\langle  W\rangle \langle Y\rangle \langle X\rangle
    \label{eq:otoc-haar}
\end{align}
This result has been obtained and used in Ref. \cite{Yoshida2017}  to derive important bounds on the success fidelities in the teleportation protocol, which we will quote in this review.

\subsubsection{Thermal OTOC}
We represent the operator $O(t)$ in Eq.~\eqref{eq:op-O} in the state representation, following Eq.~\eqref{eq:stateW}, as
\be
\ket{O(t)} = (O(t)\otimes \mathbb{1}) \epr~,
\ee
up to a normalization constant. Projecting this into the EPR state will give us the OTOC in the infinite temperature state,
\begin{eqnarray}
\includegraphics[width=0.9\linewidth]{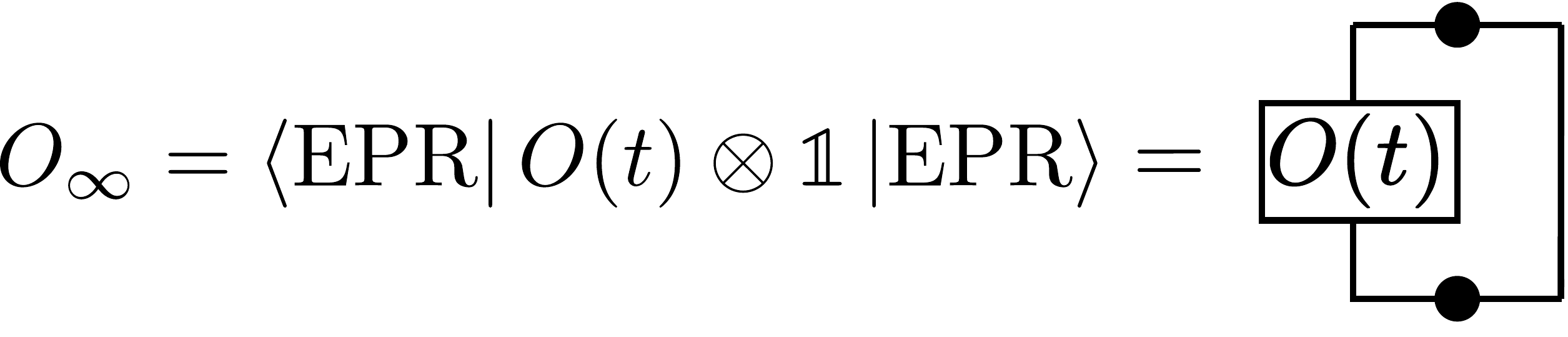}
\end{eqnarray}

An operator acting on one side of the EPR can be shifted to another following Eq.\eqref{eq:shifting-epr}, 
using this property we can rewrite,
\begin{eqnarray}
O_\infty(t)&&=\bra{\mathrm{EPR}}W^\dagger V^\dagger(t)W\otimes [V(t)]^T\ket{\mathrm{EPR}}\nonumber\\ &&
\includegraphics[]{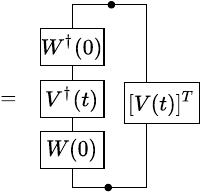}
\end{eqnarray}
Of particular interest is  the spread of information in  a thermal background. Therefore, we proceed to generalize the above definition to include finite temperatures by considering TFD instead of the EPR, which leads to
\begin{eqnarray}
O_{\mathrm{th}}(\beta, t)&&=\bra{\mathrm{TFD}}W^\dagger V^\dagger(t)W\otimes [V(t)]^T\ket{\mathrm{TFD}}\nonumber\\ &&
\includegraphics[scale=0.75]{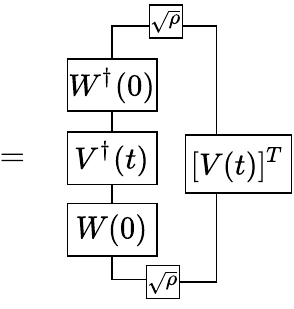}\nonumber\\
&&= \frac{ \mathrm{Tr}\left( e^{-\beta H/2} W^\dagger V^\dagger(t) W e^{-\beta H/2} V(t) \right)}{ Z }~,
\label{eq:otoc-def}
\end{eqnarray}
where $Z=\mathrm{Tr}(\mathrm{exp}(-\beta H))$ is the thermal partition function, and we have used Eq.~\eqref{eq:shifting-tfd} in the last line. As should be noted, the OTOC here also depends on the parameter $\beta$ besides time, where $\beta$ is the inverse temperature of one half of the TFD state \eqref{eq:tfd}.

We remark that, the above definition of the thermal  OTOC  is one of the different regularizations\footnote{For example, following the traditional definition, the expectation value of an operator $O$ in a thermal state is given by
\begin{equation*}
\includegraphics[]{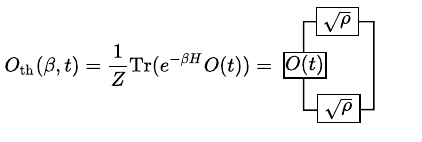}
\end{equation*}
which corresponds to a different thermal OTOC.} often considered to introduce finite temperatures \cite{lantagne2020diagnosing}. In particular, in the seminal work proposing bound on the  growth of $C(t)$ \cite{maldacena2016bound} the finite temperature OTOC is of the form of $W^\dagger \rho^{1/4} V^\dagger(t)\rho^{1/4}W \rho^{1/4}V(t)\rho^{1/4}$. However, we work with the form \eqref{eq:otoc-def} of the thermal OTOC for two reasons. Firstly  because of its accessibility in the experiments \cite{sundar-otoc,alaina}, where one only needs to perform local measurements of operator $V^\dagger$ and $V^T$ on a prepared state in the two copies, for detailed measurement protocol see Section \ref{sec:MeasProt}. And secondly because, as we will see in Section \ref{sec:many-body-teleportation}, we note that an averaged form of this thermal OTOC is related  to the operator size  which is central in the teleportation mechanism in many-body systems.

\subsubsection{Illustration in many-body dynamics}
 To gain intuition about the properties of the OTOC and its dependence on the temperature, let us take an example. We consider the transverse field Ising Hamiltonian in presence of longitudinal fields, on a lattice of $N$ spin-$1/2$s, 
\begin{equation}
\label{eqn:H-tim}
H = J \sum_{i=1}^{N-1} \sigma^x_i\sigma^x_{i+1} + \sum_{i=1}^N (b\sigma^z_i+ h \sigma^x_i),
\end{equation}
where $\sigma^a, ~a\in(x, y,z)$ is the Pauli operator. The coefficient $J$ denotes the  interaction strength between neighboring spins, and $b$, $h$ are transverse and longitudinal field strengths respectively. For concreteness, we choose $b=J$ and $h=J/2$. 
\otocmanybody

In Fig.~\ref{fig:otoc-tim} we plot the numerically calculated OTOC in this model. As shown in \ref{fig:otoc-tim}(a) we chose the operators $V$ and $W$ as the Pauli operators $\sigma^x$ on adjacent qubits. The initial time dependence of the OTOC depends on the spatial positioning of the operators, here we have chosen the operators in the middle of the 1D lattice chain separated by unit lattice distance, as seen in panel $(a)$.
For generic operators we can chose to normalize such that $\tilde{O}_{\mathrm{th}}(t=0)=1$. We plot $\tilde{O}_{\mathrm{th}}(t)=O_{\mathrm{th}}(t)/O_{\mathrm{th}}(t=0)$ in Fig.~\ref{fig:otoc-tim}(b), and note that it decays from an initial value $1$.  Upon subsequent time-evolution the correlations between $W$ and $V$ decay finally reaching late time thermal expectation value. The late-time ($Jt/\hbar \sim 4$) behavior for the OTOCs in Fig.~\ref{fig:otoc-tim}(b) are also affected by finite size effects.

We have also presented the behavior at different temperatures. It is best seen by plotting the slope of the OTOC when it becomes half of its initial value, i.e., the slope when $\tilde{O}_{\mathrm{th}}(t)=0.5$. The numerically computed slope, $(d\tilde{O}_{\mathrm{th}}(t)/dt)\big|_{\tilde{O}_{\mathrm{th}}= 0.5}$, is presented as a function of temperature in Fig.~\ref{fig:otoc-tim}(c).

The decay of the OTOC as  discussed above  with an example of a local Hamiltonian with $N=10$ sites, is a generic feature of OTOC, expected to hold in all systems which scramble information. In Section \ref{sec:Qsim-platforms}, we discuss in detail the experimental platforms, which can realize the Hamiltonian \eqref{eqn:H-tim} as well as measure the OTOC, protocols discussed in Section \ref{sec:MeasProt}. 
\hpfigure

Quantum information scrambling has been central in the studies of the quantum nature of black holes. In this direction, we next briefly recapitulate the Hayden-Preskill recovery protocol \cite{Hayden2007} for information sent into the black holes and its generalization \cite{Yoshida2017} to general quantum channels.

\subsection{Hayden-Preskill recovery protocol}
\label{sec:HPR-YK}
According to the original calculation using Schwarzschild black hole, the Hawking radiation contains information only about the macroscopic details, like the mass (equivalently temperature) of the black hole. Since then the questions about the information content of the black hole interior have been explored in many directions \cite{Almheiri-review}, in particular revolving around the question of  how can the thermal radiation reveal any information about the formation of a black hole? While this can be a difficult problem, the black hole thermodynamics suggests that, on average, black holes show similar thermodynamic properties as generally expected in unitary quantum mechanics. For example, they have a finite entropy $S$, proportional to the horizon area,  using which a Hilbert space with dimension $d=\mathrm{exp}(S)$ is associated with the black holes. Page considered black holes as quantum objects whose dynamics in the long time limit can be mimicked by Haar random unitaries \cite{Page1993_1, Page1993_2}. 

Let us denote the initial state of the black hole by a random pure state $\ket{\Psi}$, and consider it evaporating with time. In Fig.~\ref{fig:HP}(a) such a set up is schematically drawn, where $U$ denotes Haar random unitary describing black hole internal dynamics. At initial time we have a pure  black hole which with time evaporates into radiation R, here the upward direction denotes time, which should be thought of as the growing size of the radiation subspace. Associating dimensions $d_R, d_C$ to the Hilbert spaces of emitted radiation (R) and remaining black hole C, it holds that, $d_R d_C=d$. The density matrix describing the radiation should be 
\be
\rho_{\mathrm{rad}}=\mathrm{Tr}_{\mathrm{C}}(\ket{\Psi}\bra{\Psi}).
\label{eq:radstate}
\ee
For a small amount of radiation, $d_R\ll d_C$, Page showed that $\rho_{\mathrm{rad}}=\mathbb{1}/d_R$.  Thus, the radiation remains maximally mixed and  one can not access information of the black hole just by looking at the radiation itself. However as the black hole evaporates half of its entropy away, and a point of $d_R=d_C$ is reached, the radiation is maximally entangled with the remaining black hole. After this point we have $d_R>d_C$ and the correlations between remaining black hole and the radiation are sufficient to learn about the information from the black hole. However, in Page's setting, to reach this half way point, one has to wait a time which scales as the cube of the black hole mass ($\sim M^3$), which is impractical for all purposes. 

The problem that Hayden and Preskill discussed \cite{Hayden2007} in the context of the information in a black hole is as follows.  They consider an old black hole which has evaporated at least up to half of its entropy. Bob has been collecting all this radiation and Hayden-Preskill protocol begins with considering maximal entanglement between black hole B and B' which is the radiation collected by Bob, see schematics in Fig.~\ref{fig:HP}(b). Furthermore, Bob has access to all the future radiation.  Alice (A) wants to hide her quantum information ($\psi$) by throwing it into the black hole. 

The information recovery problem can be further simplified by considering the black hole as a quantum object of $N$ qubits, such that $d=2^N$, and the dynamics given by a unitary $U(d)$ from the circular unitary ensemble of dimension $d$, which makes a unitary group over Haar measure. We think of Alice's state to be composed of $k$ qubits, then questions that Hayden-Preskill answered are $(i)$ How many qubits does Bob need to collect to recover Alice's state and $(ii)$ how long does he need to do so?

 The analysis of the problem further reduces if one considers Alice to be in maximal entanglement with $N$.  The information content of Alice's diary will be entirely in the radiation R and information theoretically it will be possible for Bob to learn about $\psi$ only when the black hole evaporates to a point after which there is no entanglement between  N and the remaining black hole C. This translate to the case when the combined density matrix of the NC system separates out as
\be
   \rho_{\rm NC}=\rho_{\rm N}\otimes \rho_{\rm {C}}~,
\label{eq:HPdecouple}
\ee 
where we find the density matrix in a system by partial tracing every other system, as also done in \eqref{eq:radstate}. Without going into more details, we summarize the answers to the above questions here, as they will be directly related to the theme of this review. 

\textit{(i) How many qubits does Bob need?} To answer this we need to find whether and when Eq. \eqref{eq:HPdecouple} holds. Ref. \cite{Hayden2007} used the notion of $L_1$ norm $||\cdots||$ of states, which is to say that any states closer in the $L_1$ norm are indistinguishable in measurements. Assuming Haar random evolution of black holes, they showed that,
\be
\int dU ||\rho_{\rm NC}-\rho_{\rm N}\otimes \rho_{\rm C} ||^2 \leq 2^{2k-2s}~,
\label{eq:L1norm}
\ee
where $dU$ is the Haar measure and $s$ is the number of qubits collected by Bob. Clearly, when $s>k$, the condition $\eqref{eq:HPdecouple}$ holds up to some tolerance. So Bob needs to only collect a little more than the qubits thrown in by Alice. 

\textit{(ii) How long does it take?} The time needed was shown to be $t_{\rm scr}$ plus the time needed to radiate $s$ qubits.

Even though these are answers to some basic questions, how the information recovery, called also the decoding, is manifested was presented in a variation of the HP protocol applicable to generic quantum channels by  Yoshida and Kitaev \cite{Yoshida2017}, shown in Fig.~\ref{fig:HP}(c).  The information recovery protocol assumes that the dynamics is sufficiently mixing, i.e., any initial local information spreads to all degrees of freedom, referred to as \textit{maximally mixing}. 

Drawn in  Fig.~\ref{fig:HP}(c), The Yoshida-Kitaev protocol begins with the black hole unitary $U$ and Bob's unitary $U^*$. Alice's input is at A, and the black hole B is in maximally mixed state with B' which is a subsystem of the system Bob possesses.  There is a reference system N maximally mixed with A and another maximally mixed pair of A' and N' with Bob. The protocol has two ways of decoding the information, both the algorithms work as long as the dimension of the D subsystem is $d_D \geq d_A^2$. In the derivation of this bound an averaged definition of the OTOC of the form \eqref{eq:otoc-haar} is used, we refer the readers to the interesting and detailed calculations in Ref.~\cite{Yoshida2017}. 

\paragraph{Probabilistic decoding:}
In the probabilistic decoding, after the input of initial information both the systems ABCD and A'B'C'D' are forward evolved with $U$ and $U^*$ respectively. After which the probabilistic decoding is performed (labeled with green oval with PD in the Fig.~\ref{fig:HP}(c)). This involves projecting the combined system DD' onto EPR pair, while leaving the C' and N' as they are. Recall  that only D, D', C', N' are in Bob's possession, so all decoding operations can only be performed in these subsystems. The EPR projector is taken to be 
\begin{align}
P_{[DD'] (C'N')}&=[\epr_{DD'}\langle\mathrm{EPR}|_{DD'}]\otimes (\mathbb{1}_{C'N'})=\nonumber\\
&\hskip 1cm \includegraphics[width=0.40\linewidth]{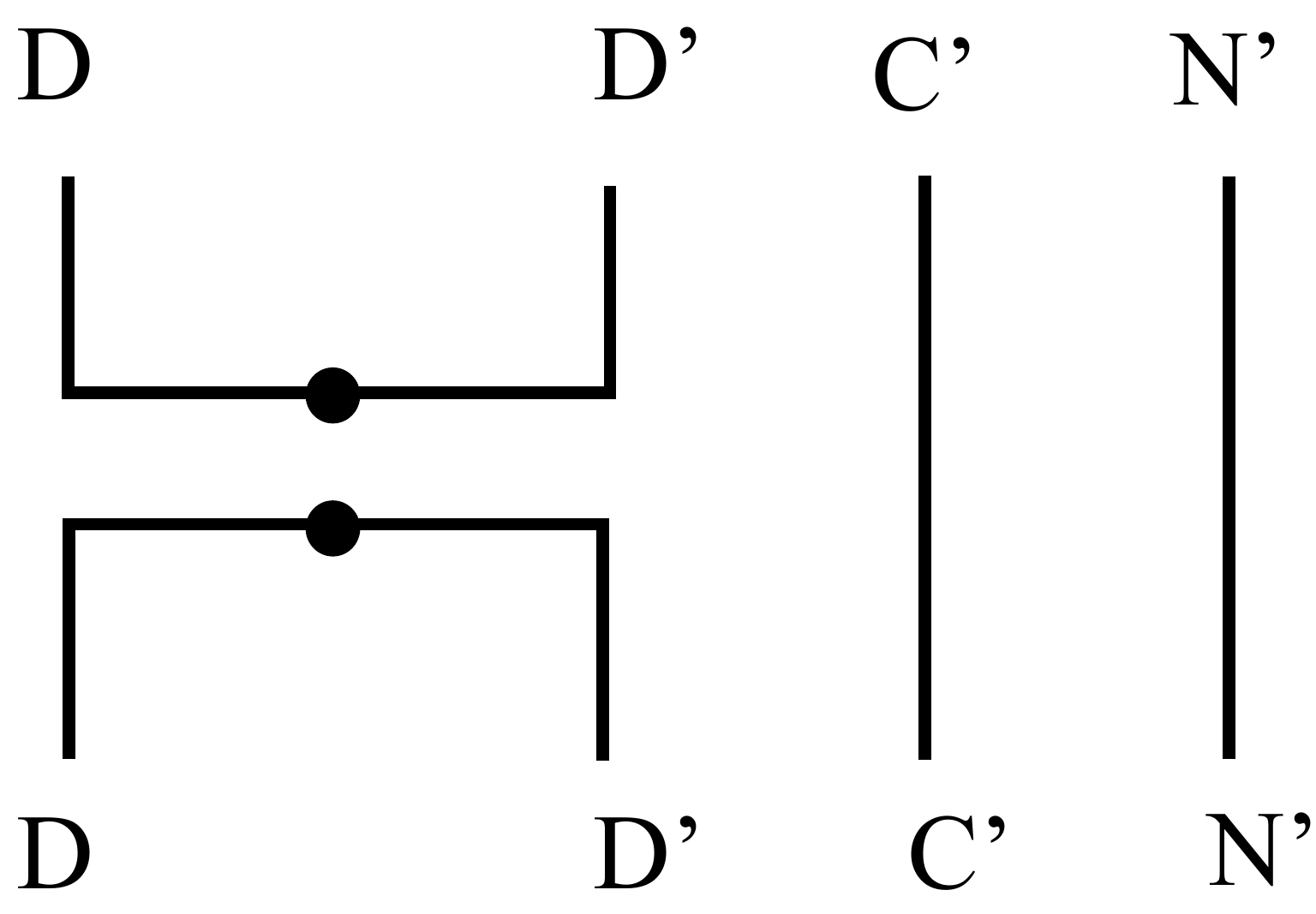}
\end{align}
We have used in the above definition in the subscript of $P$ the bracket $[..]$ to denote the pair which is projected on EPR and $(..)$ for the subsystems where no operations are performed. The projection to EPR pair succeeds with the probability given in terms of an averaged OTOC and is $ \ge 1/d_A^2$. From this, it can be shown that  if DD' is projected to EPR with a probability of $1$,  then the N and N' should make an EPR pair. And thus from N' Bob can read the initial input state. 

\paragraph{Deterministic decoding:}
Success probability in the probabilistic decoder goes down as $1/d_A^2$ as the size of the subsystem A increases. The success probability can be  boosted with a Grover variant. The idea is similar to that of the Grover's algorithm -- the probability of measuring a target solution can be improved by repeated applications of the Grover oracle and Grover diffusion. Instead of doing EPR projective measurements after initial evolutions with $U$ and $U^*$, we instate the Grover's iterations, see Fig.~\ref{fig:HP}(c). One iteration involves evolving DD' with $G_D$ (defined below), followed by evolving D'C' with $U^T$, A'N' with $G_A$ and A'B' with $U^*$, where,
\be
G_D=1-2P_{[DD'](C'N')}~~\text{and}~~ G_A= 2P_{[A'N'](B'D)}-1
\ee
The first operation $G_D$ is the Grover's oracle and the second operation $U^TG_AU^*$ is the Grover's diffusion operator \cite{grover1997quantum}. With this decoder, the probability for successfully decoding the initial input  after $m$ Grover steps is $\sin^2\left( (m+1/2)\theta \right)$, where $\theta = 2 \arcsin( 1/d_A )$. The probability approaches 1 when $m \sim \pi d_A/4$ for $d_A \gg 1$~\footnote{For $d_A = 2$, the probability of decoding is $1$ at $m=1$.}.

Both the probabilistic and deterministic protocols have been demonstrated in an experiment based on trapped ions, we present the setup in Section~\ref{sec:Qsim-platforms} and results in Section~\ref{sec:MeasProt}. 
In the next section we discuss the recent developments connecting many-body quantum teleportation to  wormhole teleportation \cite{Brown, Gao2019, Nezami2021, Schuster2021}. For late times and a single-qubit input bit, we find that the Yoshida-Kitaev circuit is the same as the single-qubit teleportation circuit.

\section{Wormhole teleportation and many-body quantum teleportation}

\label{sec:many-body-teleportation} Motivated from the gravity calculations in the Section \ref{sec:gjw-section} describing the negative energy shock wave in eternal black hole, Fig.~\ref{et2}, we can devise a quantum circuit, designed for a many-body system  on a lattice shown in Fig.~\ref{fig:tw-many-body}. We provide description of this protocol in the next subsection. Here, we present the mechanism behind the teleportation in terms of the growth of the initially inserted operator and identify the criteria for a successful many-body teleportation using this wormhole teleportation inspired circuit. In subsequent subsection \ref{sec:many-body-tw-illustration} we  illustrate teleportation of a single-qubit. We then provide a summary of results from the existing literature in Section \ref{sec:summary-TW}. We end this section in \ref{sec:size-winding} by discussing a different origin of the teleportation, known as the \textit{size-winding}.

\subsection{Description of the protocol}
As first steps, we make a one-to-one map of the wormhole teleportation as in Section \ref{sec:gjw-section} to obtain a circuit for many-body dynamics, as considered in \cite{Brown, Nezami2021}.
The circuit consists of the following steps. The description is easier to follow when  we divide the left and right systems into message and carrier subsystems labeled by the subscripts $M$ and $C$ respectively. For an $N$ qubit system, the message to be teleported is inserted in the message subsystem $L_M$, which is composed of $m$ qubits on the left and received at the message subsystem $R_M$ at the right, also of the same size $m$.  Thus, the carrier subsystem contains $K=N-m$ qubits. 

\figmanybodyteleportation

The circuit begins with a TFD state in the product Hilbert state $\mathcal{H}_L\otimes \mathcal{H}_R$ at time $t=0$. This corresponds to a non-traversable eternal black hole (as discussed in Section \ref{sec:holographic}). We consider scrambling and thermalizing unitary dynamics in the two sides of the TFD where the forward time evolution in the left is governed by $U_L=U=\mathrm{exp}(-iH t)$ and that on the right is by $U_R=U^T=\exp(-iH^Tt)$. The left side of the TFD is evolved with the adjoint unitary $U^\dagger$ to reach a time $-t$, at which point a message, to be teleported, is inserted as a state $\ket\psi$. This can be done by performing a swap between $\ket\psi$ and the state of the message subsystem. Next, this left system is forward evolved with $U$, which results in the scrambling of the input information, to reach time $t=0$. At this state, a momentary coupling $\mathrm{exp}(i g V)$ is applied between the left carrier $(L_C)$ and the right carrier ($R_C$) subsystems. This is similar to the Gao-Jafferis-Wall coupling introduced for wormhole teleportation in Eq.~\eqref{eq:original-coupling} but  now adapted for a lattice model \cite{Brown}. The right system is then forward evolved, after which if the teleportation is successful, the initial state should be teleported \cite{Brown, Nezami2021}. 
The coupling at $t=0$ is,
\begin{align}
&G=e^{igV}\nonumber\\
\text{where,\quad} &V= \frac{1}{K} \sum_{j=1}^{K} O_{j, L}(0) O_{j,R}(0)~,
\label{eq:gjw-coupling}
\end{align}
where $g$ denotes the coupling strength, and $K=N-m$ is the number of qubits in the carrier subsystems. The above operation can be seen either as quantum gates between the two sides or simply as communicating the results $o_{j,L}$ of the measurement of an operator $O_{j,L}$ on the left, followed by doing a conditioned operation on the right carrier by  \cite{Maldacena2017},
\begin{equation}
\text{Operator on the } R \text{ system}=e^{i g \sum_j o_{j,L} O_{j,R} }~,
\end{equation}
similar to the wormhole discussion in Fig.~\ref{et2}. It should be noted that the above teleportation is different than the conventional quantum teleportation, where the measurement of the initial quantum information is classically sent to a decoder. In the above teleportation, the information is first scrambled and then the results of the classical measurements are used to perform quantum operations on the right carrier subsystem.

In recent works, the above circuit, though inspired from wormholes, is found to be teleporting initial information not only for the gravity models but also for models far from it; high temperature SYK \cite{Gao2019}, spin models and random unitary channels \cite{Nezami2021, Schuster2021}.  
So there seems to be a unified underlying mechanism assisting the  teleportation. As we explain below, this mechanism is based on the  growth of the operators under scrambling dynamics, see for the generic notion of operator growth the Ref.~\cite{Qi2019a}.

\subsubsection{Mechanism of teleportation: Operator size}
Let us use the Pauli basis to expand operators. The Pauli basis for $N$ qubits is formed by taking tensor product of $N$  single-qubit Pauli operators,\\ $\mathcal P=  (\{\mathbb{1}, \sigma^x, \sigma^y, \sigma^z\}^{\otimes N})$. The circuit shows a state $\ket\psi$ insertion at the time $-t$, by removing the qubits in the message subsystem $L_M$. This should be viewed as an operator $Q_L$ acting on the qubits in $L_M$ such that
\be
  Q_L=  |\psi\rangle \langle\phi|~,
\label{eq:operatorQ}
\ee
where $|\phi\rangle$ denotes the state of the subsystem $L_M$ at $-t$. The coupling in the teleportation circuit acts on a state, $Q_L(-t)\rho^{1/2}$, at time $t=0$.
Since an operator applied at $-t$ can be related to a state insertion in the above fashion, from here on we have used the words \textit{operator} and \textit{state} synonymously to talk about the inserted message. We have also dropped the minus sign in front of the time, for brevity. However, we keep in mind that an operator with a subscript  $L$ is inserted at $-t$, and use it explicitly whenever it is not obvious. We begin by expanding this operator in Pauli basis, 
\begin{eqnarray}
   Q_L(t)\rho^{1/2}&&=\frac{1}{\sqrt{d}}\sum_{P\in \mathcal P} c_{P}(t) P~,
\label{eq:op-pauli}
\end{eqnarray}
where the coefficients $c_P$ are such that $\sum_P |c_P|^2=1$. For a Pauli string $P$, the size $|P|$ of the string is defined as the number of non-identity operators in the string. As is evident from the basis set $\mathcal P$, many Pauli strings can share the same size, and they will enter in the operator in Eq.~\eqref{eq:op-pauli}, with some coefficient $c_P$. Thus, there will be distribution of sizes, which is defined  for a size $l$ as, 
\begin{equation}
   q(l)=\sum_{|P|=l} |c_P(t)|^2
\label{eq:size-dist}~.
\end{equation}
Summing over all possible  sizes the distribution follows $\sum_l q(l)=1$, which is simply the sum of the probabilities to find the operator in Eq.~\eqref{eq:op-pauli} in one of the $P$ strings.

At this point, we take a slight detour to learn a trick used to obtain the size of the Pauli string. We discuss it here for bosonic operators and closely follow Ref. \cite{Schuster2021}. The size of an operator can be found by considering EPR projectors in the doubled Hilbert space. To see this, first let us consider an EPR projector for a single-qubit in the doubled Hilbert space of $N$ qubits, 
\begin{eqnarray}
P_{\mathrm{EPR}, i} &&=\mathbb{1}\otimes \cdots \left(\epr \bra{\mathrm{EPR}}\right)_i \otimes \cdots \mathbb{1}\nonumber\\
&&=\mathbb{1}\otimes \cdots\frac{1}{4}\sum_{P_i}P_{i,L} P_{i,R}^*\otimes\cdots \mathbb{1}
\label{eq:single-epr}
\end{eqnarray}
where $P_i \in \{\mathbb{1}, \sigma^x, \sigma^y, \sigma^z\}$. Next, we note that the expectation value of a Pauli string $P$ in a single-qubit EPR state $_i \langle\mathrm{EPR}| P| \mathrm{EPR}\rangle_i$ gives the trace of the Pauli $P_i$ at this qubit i.e. $_i \langle\mathrm{EPR}| P| \mathrm{EPR}\rangle_i$= $\mathrm{tr}(P_i)/2$, which is $=\delta_{P_i, \mathbb{1}}$. Thus, the above projector acts on $P\epr$ as, 
\begin{eqnarray}
  P_{\mathrm{EPR},i} (P\epr) =  \delta_{P_i, \mathbb 1}(P\epr)~.
\label{eq:single-epr-projector-eigenstate}
\end{eqnarray}
Thus, the eigenvalue of the single-qubit EPR projector at $i^{\rm th}$ qubit index is non-zero only when there is an identity at the  $i^{\rm th}$ site in the string $P$. This property can be utilized to count the number of identities or vice-versa to count the size of the string by considering a sum of all such single-qubit EPR projectors, i.e., we can define a counting operator, 
\be
   \widetilde{V}=\frac{1}{N}\sum_{i=1}^{N}P_{\mathrm{EPR},i}
\label{eq:counting}
\ee
which follows [directly from Eq.~\eqref{eq:single-epr-projector-eigenstate}],
\begin{eqnarray}
\widetilde{V}(P\epr)=\frac{N-|P|}{N} (P\epr)~,
\end{eqnarray}
by counting the identities in the Pauli string, and in return  giving the size $|P|$ of the string $P$. For the states of the form of \eqref{eq:op-pauli}, which are linear in $P$, we note that, 
\begin{align}
\widetilde{V}(Q_L(t)\tfd)&=\widetilde{V}\left(\sum_Pc_P(t)P\right)\epr\nonumber\\
&=\sum_P\frac{N-|P|}{N} c_P(t) (P\epr)~.
\label{eq:action-Vtilde}
\end{align}
which immediately leads to, 
\begin{align}
\langle Q_L(t)\mathrm{TFD}|\widetilde{V}|Q_L(t)\mathrm{TFD}\rangle=  \sum_P\left(1-\frac{|P|}{N}\right)|c_P(t)|^2
\label{eq:average-size}
\end{align}
Thereby the expectation value of $\widetilde{V}$ in the state, just before the coupling is inserted in Fig.~\ref{fig:tw-many-body}, gives an average of the operator size.

We return to our discussion regarding the effect of the coupling $G$ in \eqref{eq:gjw-coupling}. From the form of the operator $\widetilde{V}$, by now, it should be clear where we are headed to with this discussion. The coupling \eqref{eq:gjw-coupling}, central in the teleportation protocol, is of the same form as the operator $\widetilde{V}$ and thus measures the average size of the operators that have acted before $t=0$. The effects of this coupling can be further simplified.

First, note that the counting operator in Eq. \eqref{eq:counting} is generic. For a non-trivial coupling we should remove the trivial identity operation. That would result in considering, in single-qubit EPR projector, in Eq.~\eqref{eq:single-epr}, a sum over $P_i$ restricted with $P_i\neq \mathbb 1$. Such that,

\begin{align}
\widetilde{V}_{P_i\neq \mathbb 1}&=\frac{1}{N}\sum_{i=1}^N\left(\frac{1}{3}\sum_{(P_i,{P_i\neq \mathbb 1})} P_{i,L}P^*_{i,R}\right)\nonumber\\
&=\frac{4}{3}\widetilde{V}-\frac{\mathbb 1}{3}\ ,
\end{align}
thus, the eigenvalue of $\widetilde{V}_{P_i\neq \mathbb 1}$ on the state $(P|\mathrm{EPR}\rangle)$ is $[(N-4|P|/3)/N]$.
Next, note that we have assumed the dynamics to be scrambling and thermalizing, in this case, after we have inserted $Q_L$ and let the system scramble for time $t$, it is sufficient to just consider $1$ out of the $3$ non-trivial $P_i$. This assumption is justified if we have taken $t\geq t_{\mathrm{scr}}$, since then the initial information has spread equally to all sites, and all $3$ non-trivial Pauli operators will probe the operator size similarly. Thus the coupling $V$ at $t=0$, without loss of generality, becomes much simpler, written as  \cite{Brown}, 
\begin{equation}
   V = \frac{1}{K} \sum_{i=1}^{K}\sigma^z_{i,L} \sigma^z_{i,R}~,
\end{equation} 
The coupling contains the operator $V$ between $K$ carrier qubits only. We focus, in this work, on $m\ll N$, strictly $m=1$.  In this case the  average size distribution $\sum_P |c_P|^2 |P|/N$ in \eqref{eq:average-size} which uses all $N$ qubits can be regarded as the same as the  average size distribution $\sum_{P_c} |c_{P_c}|^2|P_c|/K$ for $K=N-m$ qubits, where $P_c$  is the Pauli string only on the $N-m$ carrier qubits. 

Continuing the same calculation as presented above for a generic $\widetilde V$, we find the expectation value of $V$ in the state prepared before $t=0$ to be, 
\begin{align}
\langle V \rangle_Q&= \langle\mathrm{TFD}| Q_L^\dagger(t) V Q_L(t) |\mathrm{TFD}\rangle\nonumber\\
&= \left(1-\frac{4}{3}|\varrho_{\epsilon_-}|\right)~,
\label{eq:phase-size}\end{align}
where  $|\varrho(\epsilon_-)| =\sum_{P_c}|c_{P_c}|^2|P_c|/K\approx \sum_{P}|c_{P}|^2|P|/N$ is  the average size over $K$ qubits for the state that existed just before $t=0$ (hence the use of $\epsilon_-$), i.e, $\varrho({\epsilon_-})=Q_L(t)\rho^{1/2}$.

We can now ask what are the effects of the coupling $G=\mathrm{exp}(i g V)$? In the large number of carrier qubits $K$ we use the property of factorization such that, any expectation value of the form, $\langle B | G|B\rangle \approx \exp(ig \langle B|V|B\rangle)$. Thus the coupling $G$  acts on the state prepared at $t=0$ as,
\be
 e^{ig V} Q_L(t) |\mathrm{TFD}\rangle=e^{ig \langle V \rangle_Q}  Q_L(t) |\mathrm{TFD}\rangle~.
 \label{eq:phase}
\ee
To conclude, we see from Eq.~\eqref{eq:phase-size} and \eqref{eq:phase} that  the effect of the non-trivial coupling is to apply an operator size dependent phase to the state $Q_L(t)\tfd$.

\subsubsection{Criterion for a successful teleportation}
\label{sec:success-criterion}
We now ask the question of when is the  teleportation successful according to the circuit \ref{fig:tw-many-body}. As presented in the circuit, having implemented the coupling $G$ at $t=0$, we need to evolve the right circuit with $U^T$ for a time $t$. After this, as shown below and also presented in Ref.  \cite{Brown},  we get the operator $Q^T$ at the right message subsystem. We can do a further decoding operation $D$ to obtain the $Q$. We explain it shortly. First, we redraw the circuit in Fig.~\ref{fig:tw-many-body} with this decoding operation as,  
\begin{equation}
\qquad\qquad\qquad    \includegraphics[width=0.4\linewidth]{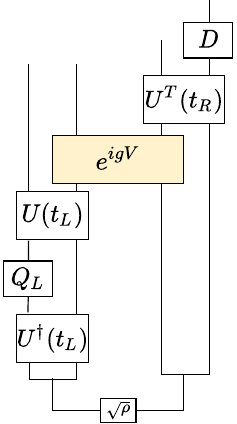}
    \label{eq:with-decoder}
\end{equation}
 It can be noted that (explained below) for the teleportation to be successful, the following must hold \cite{Schuster2021}, 
\begin{equation}
    \includegraphics[width=0.9\linewidth]{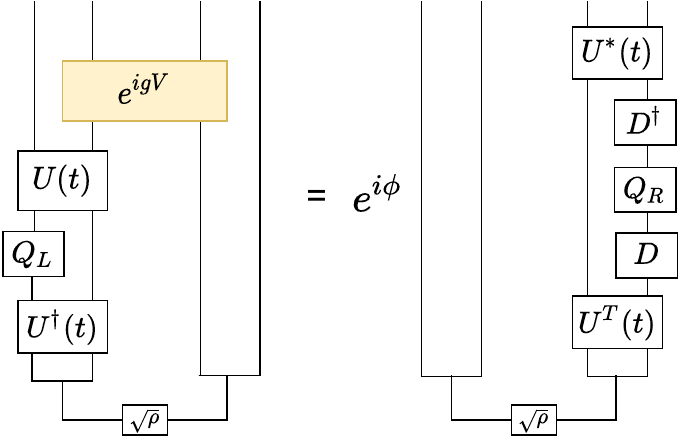}
    \label{eq:success-condition}
\end{equation}
where, the phase $\phi=g\langle V\rangle_Q$ depends on the operator $Q$ and measures its average size. In the large $K$ limit, this overall phase is justified from Eq.~\eqref{eq:phase}. Away from the large $K$ limit, for multi-qubit teleportation, this overall phase is possible only when the effect of the coupling $\exp(igV)$ on $P\tfd$ is same for all $P$s, such that $\exp(igV)(P\tfd)\sim\exp{(i\phi)}(P\tfd)$. Since, the $G$ acts on $Q_L(t)\tfd$, the $\phi$ measures the  size of $Q_L(t)$. Thus, the overall phase as in \eqref{eq:success-condition} is possible, when the size distributions $\eqref{eq:size-dist}$ are \textit{tightly peaked} around the average size distribution $\sum_P |c_P|^2|P|/N$ of the operator, dubbed as \textit{peaked-size teleportation}. A  situation that occurs in a wide range of many-body dynamics (see below subsection \ref{sec:summary-TW}). 

Assuming peak-size teleportation, we analyze the right side of the above equation \eqref{eq:success-condition}. To begin, for the moment, let us set $D=\mathbb 1$, then in the right side we get an operator $Q_R(t)=U^* Q_R U^T=(U^T)^\dagger Q_R U^T$. Note that when the left side evolves with $U$, the right evolves with $U^T$ in Fig~\ref{fig:tw-many-body}. Thus the above is a transfer of an operator on the left $Q$ at time $-t$ to the transpose of the operator on the right, i.e, $Q^T$ at time $t$. This is exactly the  teleportation protocol circuit presented in Ref.~\cite{Brown, Nezami2021}, and they obtained $O^T$ in the right side when $O$ was inserted on the left, as would be the case with the circuits in Fig.~\ref{fig:tw-many-body}. 

Now, the role of the decoder becomes clear. In order to obtain the operator $O$ teleported to the right, we need a decoding operation $D$ such that $D^\dagger O^T D \propto O$. 
The success of the teleportation protocol for generic $U$, as in many-body dynamics, then boils down to finding out when does the above identity \eqref{eq:success-condition} hold? We begin by taking the inner product of the left and right side operators in \eqref{eq:success-condition} as,
\begin{equation}
C_Q = \langle\mathrm{TFD}|\widetilde{Q_R}(t)^\dagger e^{igV}Q_L(t)|\mathrm{TFD}\rangle~,
\label{eq:two-point}
\end{equation}
where, $\widetilde{Q_R}(t)= U^*D^\dagger Q_R D U^T$, as shown in the right side of \eqref{eq:success-condition}. 
So, following Eq.~\eqref{eq:success-condition}, the first condition for the successful teleportation is that \cite{Schuster2021},
\paragraph*{$(i)$ the magnitude of   $C_Q$ is maximal for any operator $Q$.} To ensure that the teleportation succeeds for arbitrary initial state, or equivalently arbitrary sum of operators $Q$. And the second condition is that,
\paragraph* {$(ii)$ the coupling applies the same phase $e^{i\phi}$ to all input states.$\ $} Such will be the case when the size distributions for all sizes are tightly peaked around the average size of the operator. 

We summarize in \ref{sec:summary-TW}, that these two conditions are generically satisfied in many-body models, however, the holographic models follow the wormhole teleportation mechanism.  
In the next subsection we provide an illustration of this form of teleportation in a many-body model described by Hamiltonian \eqref{eqn:H-tim}.

\subsection{Illustration in many-body dynamics}
\label{sec:many-body-tw-illustration}
For illustration of the teleportation protocol in many-body system, we consider the Hamiltonian \eqref{eqn:H-tim} and numerically run the left circuit in Fig.~\ref{fig:tw-many-body} in a spin-$1/2$ system with $N=7$ qubits. We present the results for single-qubit teleportation in  infinite temperature TFD, i.e., EPR state. Preparing an EPR at $t=0$, we do backward time evolution up to $-t$ and then swap the first qubit with a state which has the expectation value $\langle \sigma^z_1\rangle=1$. This can be done by inserting an up state $|0\rangle$, denoted in the computation basis of $\{0,1\}$  by $|0\rangle=(1,0)^T$. Then we evolve forward, perform the coupling, and evolve the right with $U^T$.
\begin{figure}[!ht]
    \centering
    \includegraphics[width=\linewidth]{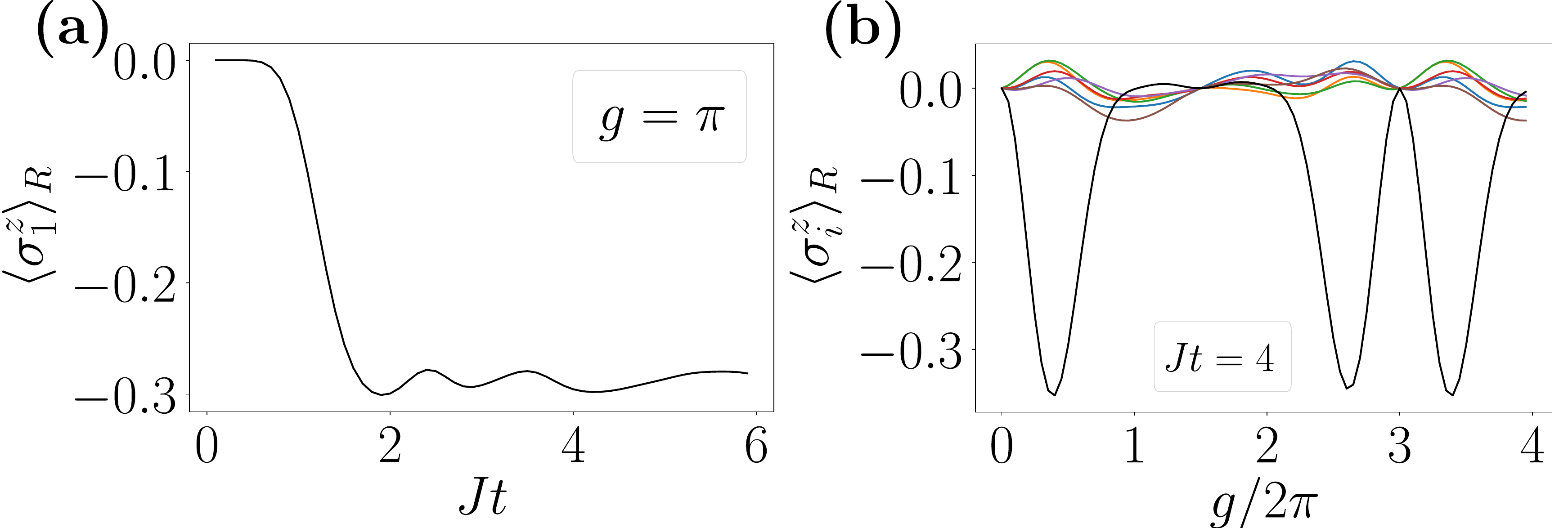}
    \caption{\textit{Illustration of teleportation protocol in Hamiltonian \ref{eqn:H-tim}:} (a) For an input state at qubit 1 on the left, such that  $\langle \sigma^z_1\rangle$=1, following Fig.~\ref{fig:tw-many-body} for $g=\pi$ the expectation value on the right at qubit 1 is presented. (b) For a time $Jt=4$, we present $\langle \sigma^z_i\rangle_R$ measured at the right on each qubit. The teleportation succeeds at qubit 1, shown in black color, while at all other qubits, $\langle \sigma_{i\neq 1}\rangle_R\approx 0$.
    \label{fig:many-body-numerics}}
\end{figure}

In Fig.~\ref{fig:many-body-numerics}(a) on the right we observe the expectation value $\langle \sigma^z_1\rangle_R$  at qubit 1 for the coupling strength $g=\pi$.  With time, the magnitude of $\langle \sigma^z_1\rangle_R$ increases and saturates once the information has reached to all qubits, i.e., at the scale of $t_{\mathrm{scr}}$, this corresponds to $|t_L|=t_R\geq t_{\mathrm{scr}}$. In Fig.~\ref{fig:many-body-numerics}(b), we fix the evolution time at $Jt=4$, and observe the expectation $\langle \sigma^z_i\rangle_R$ on all qubits.  We note that the teleportation is successful (black curve) only at the message qubit, labeled $1$, while at all other qubits, $\langle \sigma^z_{i\neq 1} \rangle_R \approx 0$. The teleported signal  has a maximum magnitude for some values of $g$, and  there is an infidelity in teleportation. For more details on  the dependence on $g$ and  fidelities for spin model we refer to Ref.~\cite{Nezami2021}.

\subsection{Summary and remarks}
\label{sec:summary-TW}
 We derived above the requirements for a successful teleportation. It has been shown by analytical calculations in high temperature SYK \cite{Gao2019}, spin models \cite{Nezami2021}, random unitary circuits \cite{Schuster2021} and using several numerical models that the criterion of success holds for a large class of models and parameters. Such is summarized in Fig.~\ref{fig:summary-yao} taken from \cite{Schuster2021}. In this subsection we summarize their results while also adding some remarks. 
\figyao

\paragraph*{$\bullet$ Holographic and peak-sized teleportation:$\ $} Peak-size teleportation means that the coupling applies the same phase (operator size dependent) to all Pauli strings making up the operator on the left. Therefore, the two sided correlator \eqref{eq:two-point} is $C_Q =G_\beta e^{i\phi}$ where $\phi=g\langle V\rangle_Q\sim \sum_{P}|c_P|^2|P|/N$ (see Eq.\eqref{eq:phase-size}) measures approximately the average size of the operator $Q_L(-t)\sqrt{\rho}$ and,
\begin{eqnarray}
G_\beta&=&\langle \mathrm{TFD}|\widetilde{Q_R}(t)^\dagger Q_L(-t)|\mathrm{TFD}\rangle \nonumber\\&=& \mathrm{Tr}(\widetilde{Q}^*_L\rho^{1/2}Q_L\rho^{1/2})~,
\label{eq:two-point}
\end{eqnarray}
is the  two-point function between the right and left operators. Using the property of TFD state, in the second line, we have rewritten it as the thermal two-point function on one side of the TFD with notation $\rho=\rho_L$. The thermal function decreases as the temperature decreases, thus the two-point function 
\be
   G_\beta\le 1~,
\ee
with the limit saturating for $\beta=0$. 
Since $C_Q$ measures the overlap of the right at $t>0$ and the left at $t<0$, the two-point function $G_\beta$ governs the fidelity of the teleportation, and the fidelity decreases with decreasing temperature in the peaked-size teleportation mechanism (presented in red-pink in the summary Fig.~\ref{fig:summary-yao}(a)). It has been discussed that when the size distribution has a width, resulting from imperfect peaked-size distribution, the fidelity decreases further \cite{Schuster2021}.

In Fig.\ref{fig:summary-yao}(b), the mechanism of the peak-size teleportation (in red) is contrasted with the holographic wormhole teleportation \cite{Maldacena2017} as in the low temperature SYK (depicted in blue). The low temperature SYK teleports with perfect fidelity at time around scrambling time $t_{\mathrm{scr}}$ and zero otherwise. In contrast, in the peak-size teleportation the fidelity is of order $G_\beta$ and shows certain features with time.  However in the low temperature SYK one notes a revival in the fidelity at long times (denoted with $t^*$), with a decreased fidelity $\propto G_\beta$ as in the peak-size teleportation. Thus above this time $t^*$, all scramblers teleport with peak-size mechanism. To conclude, due to the distinct behavior of fidelity of the teleportation with time, it is a strong signature of the holographic or peak-size teleportation. 

\paragraph*{$\bullet$ Connection with the thermal OTOC:$\ $} Recall from Eqs. \eqref{eq:phase-size} and \eqref{eq:phase}, the action of the coupling $G=\exp(igV)$ is to apply a size dependent phase $\exp(ig\langle V\rangle_Q)$, where the $\langle V \rangle_Q$ from \eqref{eq:phase-size} can be expanded to be, 
\begin{align}
    &\langle V\rangle_Q=\nonumber\\
    &\frac{1}{K}\sum_{i=1}^K\langle\mathrm{TFD}|(Q_L^\dagger(t)\otimes\mathbb1)(O_{i, L}\otimes O^*_{i, R})(Q_L(t)\otimes \mathbb 1)|\mathrm{TFD}\rangle \nonumber\\
&=\frac{1}{K}\sum_{i=1}^K \mathrm{Tr}[{\rho}^{1/2} Q_L^\dagger(t)O_{i, L}Q_L(t)\rho^{1/2}O^\dagger_{i, L}] \nonumber\\
&=\frac{1}{K}\sum_{i=1}^K O_{\mathrm{th}}(\beta, t)~.
\label{eq:coupling-size-otoc}
\end{align}
This is the average of the thermal OTOC defined in the previous Section \ref{sec:info-spreading}, Eq.~\ref{eq:otoc-def} for operators $Q$ and $O_i$.

\paragraph*{$\bullet$ Connection with the HPR protocol: $\ $} In the late time, high temperature limit, the teleportation protocol can be shown to be the same as the Hayden-Preskill recovery (denoted by red diamond HPR in Fig.~\ref{fig:summary-yao}(a)) for single-qubit teleportation. For long times $|t_L|=t_R=t>t_{\mathrm{scr}}$, and infinite temperature limit the coupling acts at $t=0$ as, 
\begin{align}
e^{igV}Q_L(t)\epr&=e^{ig\langle V\rangle} Q_L(t)\epr\nonumber\\
&=Q_L(t)\epr=[Q_R(t)]^T\epr~.
\label{eq:late-times}
\end{align}
The absence of the phase factor follows directly from the relation of the $\langle V \rangle$ to the averaged OTOC as in Eq.~\eqref{eq:coupling-size-otoc}. At times $t>t_{\mathrm{scr}}$, the OTOC for infinite temperature states decays to zero, and thus the overall phase above is $1$ whenever a non-trivial $Q_L$ is applied. In the case when $Q_L=\mathbb 1$, then following Eq.~\eqref{eq:coupling-size-otoc}, $\langle V \rangle=1$, and thus Eq.~\eqref{eq:late-times} holds for generic $Q_L$, whenever $g=n \pi$, where $n\in \mathbb Z$.

The coupling $V$, also including identity operations will be, 
\begin{align}
V &=\frac{1}{K}\sum_{i=1}^{K} P_{\mathrm{EPR},i}\nonumber\\
&= \frac{1}{K}\sum_{i=1}^{K}\frac{1}{4}\left(\sum_{P_j}{P_{j,L}P_{j,R}^*}\right)_i
\end{align}
where the outer sum runs over all the carrier qubits, and the inner sum represents EPR pair on $i$th carrier qubits on the two sides. We have used the notation from  Eq.~\eqref{eq:single-epr}, and recall that the $P_j\in\{\mathbb 1, \sigma^x, \sigma^y, \sigma^z\}$. At this point we use the property of late times $t\ge t_{scr}$ when the time evolved operator $Q_L(t)$ have evolved to all available sites. At this time the effect of the above coupling will be the same if we replace the sum over local EPR pairs with an EPR projector on the full carrier subsystem. This is, in the late time, we can equally take, 
\begin{equation}
    V=\mathbf{P}_{\mathrm{EPR}}=\frac{1}{d_D^2}\sum_{\mathbf{P}_D}\mathbf{P}_{D, L}\mathbf{P}^*_{D, R}
\end{equation}
here, we have changed the previous notation $C$ for carrier subsystem with the letter $D$, for comparison with the Fig.~\ref{fig:HP}(c). The sum now runs over the Pauli operators on the full subsystem $D$. With this, we now have, 
\begin{equation}
    e^{igV}=e^{i\pi \mathbf{P}_{\mathrm{EPR}}}= 1-2 \mathbf{P}_{\mathrm{EPR}}
\end{equation}

We wish to show the equivalence between the Yoshida-Kitaev Fig.~\ref{fig:HP}(c) and the many-body teleportation circuit \eqref{eq:with-decoder}. For this purpose, we identify, $G_D=1-2 (\mathbf{P}_{\mathrm{EPR}})_{DD'}$, and $G_A=1-2 (\mathbf{P}_{\mathrm{EPR}})_{A'N'}$. Note that for a single-qubit $d_A=2$ at the subsystem $A$, this becomes, 
\begin{eqnarray}
    G_A &=1-2(\mathbf{P}_{\mathrm{EPR}})_{A'N'}
&=\sigma^y_{A'}(\mathrm{SWAP})\sigma^y_{N'}
\end{eqnarray}
where the SWAP is the swap operator, 
\be
\mathrm{SWAP}=\frac{1}{d_A}\sum_{\mathbf{P}_A}\mathbf{P}_{A,A'}\mathbf{P}_{A, N'}
\ee
Thus comparing with the teleportation figure, the decoder $D=\sigma^y$.  With these operations, on the circuit \eqref{eq:with-decoder}, the output for an input operator $Q_L=O$ will be $Q_R=\sigma^yO^T\sigma^y=O$, $\forall$ $O$ of the form Eq.~\eqref{eq:op-pauli}. Hence, the single-qubit teleportation when we replaced $D$ in \eqref{eq:with-decoder} with the Grover's oracle for a single-qubit succeeds with fidelity $1$. Using the $G_D$ and $G_A$ in Eq.~\ref{eq:with-decoder}, for the case of infinite temperature initial state, and  sliding the left $U^\dagger$ to the right to make $U^*$, we see that the teleportation circuit for single-qubit in $A$ subsystem is the same as the  Fig.~\ref{fig:HP}(c).

\subsection{The size-winding mechanism}
\label{sec:size-winding}
In contrast to the size distribution Eq.~\eqref{eq:size-dist} for the operator $Q_L\rho^{1/2}$, Eq. \eqref{eq:op-pauli}, in Refs.~\cite{Brown, Nezami2021} winding-size distribution is defined as, 
\begin{equation}
\tilde{q}(l)=\sum_{|P|=l}c_P(t)^2
\end{equation}
The important difference is that the $\tilde{q}$ can be complex and the distribution is over the complex plane. For infinite temperatures $\beta=0$, the distribution $\tilde q=q$ since then, due the properties of the EPR state, the operator $O(-t)$ as in Eq.~\eqref{eq:operatorQ} is Hermitian, and thus the coefficients $c_P$ in Eq.~\eqref{eq:op-pauli} are real. By using the properties of the TFD state, as in Eq.~\eqref{eq:shifting-tfd}, we rewrite the action of the operator on the left at time $-t$ as, 
\begin{eqnarray}
Q_L(-t)\tfd&=&\sqrt{d}~Q_L(-t)\sqrt{\rho_L}\epr\nonumber\\
&=& \sum_Pc_P(t) P\epr
\label{eq:left-op}
\end{eqnarray}
and on the right operator $Q^T_R$ at time $t$ as,
\begin{eqnarray}
Q_R^T(t)\tfd&=&\sqrt{d}\sqrt{\rho_L}~[U^*Q^T U^T]^T_L\epr~\nonumber\\&=&\sqrt{d}\sqrt{\rho_L}Q_L(-t)\epr~\nonumber\\
&=& \sum_Pc_P^*(t) P\epr
\label{eq:right-op}
\end{eqnarray}
The success criterion in the Section~\ref{sec:success-criterion} following Ref.~\cite{Schuster2021} is developed by analyzing the overlap of the $Q_R^T(t)\tfd$ with $\exp(ig V) Q_L(-t)\tfd$ (here we take the decoder $D=\mathbb 1$, which means we are interested in operator $Q^T$ on the right). The teleportation succeeds whenever the LR coupling acts similarly on all Pauli strings, and the coupling action generates a phase $\exp(i\phi)$, and that the overlap, i.e. the two point function is maximum for any input operator. 

For holographic systems, it is shown that perfect \textit{size-winding} occurs such that the coefficients in the operator expansion take the form \cite{Brown, Nezami2021},
\begin{equation}
    c_P(t)=e^{i \alpha |P|}r_P(t)\quad ; \quad r_P(t)\in\mathbb R
\end{equation}
and thus the LHS of Eqs.~\eqref{eq:left-op}, \eqref{eq:right-op} differ only by a phase linear in the operator size. The action of the operator on the left has opposite phase winding compared to the action of the operator on the right. As derived in the previous sections, the operator $V$ acts as, 
\begin{equation}
    V(P\epr)=\left(1-\frac{4}{3}\frac{|P|}{N}\right)(P\epr)
\end{equation}
leading to, 
\begin{equation}
    e^{igV}~Q_L(-t)\tfd=\sum_Pe^{i\left(\alpha-\frac{4g}{3N}\right)|P|}r_P(t)\epr
\end{equation}
up to a constant phase, which we have dropped here. For the coupling strength $g=(\alpha\pm n\pi)3N/2$, the action of coupling  at $t=0$ is the same as an operator $Q^T$ on the right at $t$. So the phase factor on the left unwinds under the LR coupling to give the phase as an operator on the right would have. This perfect operator size winding takes place for models with holographic dual, and the teleportation can be seen as the unwinding of the left phase in the complex plane to produce the phase on the right.  For models away from holographic limit, the imperfect size winding is expected, in which case the phases in the expression for operators may not be linear. We refer the readers to Ref.~\cite{Brown, Nezami2021} for proofs and detailed discussion regarding size-winding in holographic and non-holographic teleportation.

As noticed in the previous subsection, the thermal OTOC and the two-point function  encode crucial information about the mechanism of teleportation and the success fidelity.  These are measurable in present day quantum simulators. Successful measurements of infinite temperature OTOC \cite{li2017measuring, wei2018exploring, nie2019detecting, Joshi2020,pegahan2021energy,garttner2017measuring, Landsman2019,braumuller2021probing, mi2021information}
finite temperature OTOC \cite{alaina}, and teleportation protocol \cite{Landsman2019} has been achieved in recent years. 
Thus, we now turn towards the quantum simulations to summarize the current state-of-the-art. We first describe the platforms available and then discuss some important results which make the initial steps towards conducting holographic studies in the lab.

\section{Quantum simulation platforms}
\label{sec:Qsim-platforms}
Realizing the ideas described in this review requires quantum simulators which can controllably prepare desired quantum states, and realize suitable Hamiltonians. In this regard, many-body quantum simulation platforms based on ultracold gases~\cite{bloch2012quantum, gross2017quantum}, trapped ions~\cite{blatt2012quantum,monroe2021programmable}, Rydberg atoms~\cite{gallagher2005rydberg, saffman2010quantum, wu2020concise, browaeys2020many}, superconducting circuits~\cite{houck2012chip, kjaergaard2020superconducting}, nuclear magnetic resonators \cite{vandersypen2005nmr, jones2010quantum, oliveira2011nmr}, and photonic systems~\cite{aspuru2012photonic, chang2014quantum} have demonstrated tremendous potential to simulate useful physical models and phenomena from various fields of physics and beyond. These capabilities are enabled by relatively clean systems, significant degree of control over experimental parameters, strong tunable interactions between the particles, and single-particle addressability in some cases. In this section, we will describe two of these experimental platforms -- based on Rydberg atoms and trapped ions -- that show promise to explore the physics described in this review.

\subsection{Rydberg atoms}

A Rydberg atom is an atom with a highly excited electron, i.e. in an orbital with a large principal quantum number $n$. One of the main advantages of using Rydberg atoms for quantum simulation is their strong dipole moment, which leads to strong inter-atomic interactions~\cite{wu2020concise, browaeys2020many, browaeys2016experimental}. Due to being in a highly excited state, the radius of the electron's orbit in a Rydberg atom is on the order of a few microns, which is thousands of times larger than that of typical ground state atoms. Therefore, the atom is easily polarizable and easily acquires a strong dipole moment (relative to other energy scales in experiments). The resulting strong dipole interactions allows researchers to simulate e.g. quantum many-body Hamiltonians, or realize universal quantum gates that can be used to build a quantum computing architecture. Popular candidate species for Rydberg atoms have been $^{87}$Rb, $^{88}$Sr, and $^{171}$Yb.

An electron is typically excited to one of the atom's Rydberg states via a two-photon transition. Once excited, the atom in the Rydberg state interacts with other atoms in that Rydberg state via a van-der-Waals interaction, $V_{ij} \propto n^{11}/r_{ij}^6$. At typical inter-particle separations in these experiments, $\sim 500-1000$ nm, ground state atoms are nearly non-interacting, and the only interactions occur between Rydberg atoms. Advances in trapping and laser cooling~\cite{mourachko1998many, anderson1998resonant, schlosser2001sub}, and more recent ideas involving atom-by-atom assemblies with trap rearrangements~\cite{endres2016atom, barredo2016atom, kim2016situ, lee2016three, de2019defect, barredo2018synthetic}, have led to successful efforts in near-deterministic creation, trapping, and loading of large numbers of Rydberg atoms in a periodic array in space~\cite{bernien2017probing, scholl2021quantum, ebadi2021quantum}.

A qubit can be encoded in these atoms as two internal atomic states, e.g., two long-lived hyperfine ground states (i.e. a state with a small principle quantum number). The qubit states can be coupled to the Rydberg state via laser pulses.

The essential ingredient offered by Rydberg atoms is that they have strong interactions. Using this, experimentalists can implement a controlled-Z gate,\\ $\exp(-i\pi \ket{e_ie_j}\bra{e_ie_j})$, between spatially nearby qubits $i$ and $j$. A naive way to implement this gate involves directly accruing a phase proportional to the Rydberg-Rydberg interaction strength. This naive scheme, however, is sensitive to the distance between the atoms, and could therefore lead to large errors due to atomic motion. An alternative scheme to realize fast high-fidelity entangling gates between Rydberg atoms, proposed in Ref.~\cite{jaksch2000fast, lukin2001dipole}, uses the phenomenon called Rydberg blockade.

Rydberg blockade arises when the van-der-Waals interaction are so strong, e.g. due to large $n$, that having two Rydberg atoms near each other is energetically too expensive~\footnote{Technically, when one atom is in a Rydberg state, the energy of an adjacent atom's Rydberg state is shifted by an amount equal to the van-der-Waals interaction. The latter atom will not be excited to the Rydberg state when the two-photon Rabi coupling is much smaller than the laser detuning plus van-der-Waals interaction}. This so-called Rydberg blockaded regime has been experimentally observed~\cite{walker2012entanglement, urban2009observation, gaetan2009observation, wilk2010entanglement, comparat2010dipole}. Recently, Rydberg blockade has led to the first experimental evidence for quantum scar states~\cite{bernien2017probing} and topological quantum spin liquids~\cite{semeghini2021probing}.

Essential for simulating the protocols discussed in this review, the Rydberg blockade underpins the implementation of the entangling gate between qubits. The entangling scheme involves coupling ground state qubits to the Rydberg state via laser pulses, and is described in detail in \ref{sec:appen Rydberg}. Entanglement using Rydberg blockade has been widely realized experimentally~\cite{madjarov2020high, isenhower2010demonstration, levine2019parallel, zhang2010deterministic, wilk2010entanglement, maller2015rydberg, walker2012entanglement}. Experiments have demonstrated gate fidelities exceeding $99\%$ for entangling gates, and up to $99.6\%$ for single-qubit gates~\cite{madjarov2020high}. To realize universal quantum computing, it is sufficient to have  controlled-Z gate, together with arbitrary single-qubit rotations which can be implemented via magnetic fields or stimulated Raman transitions.

\subsection{Trapped ions}

Trapped ion chains are one of the most promising platforms for analog and digital quantum simulation. With currently the best gate fidelities for digital quantum gates, they form one of the pillars of today's NISQ devices along with superconducting circuits. Popular candidate species for trapped ions have been $^{171}$Yb$^+$ and $^{40}$Ca$^+$. Qubit states are encoded in two long-lived electronic states of the ions which are either coherently manipulated by narrow linewidth laser fields (optical qubits) or microwave fields (hyperfine qubits). State dependent interactions are mediated by laser fields that interact with the ions' electronic and motional degrees of freedom, eventually providing spin-spin interactions for analog quantum simulators and a universal gate set for digital quantum computers. For simplicity, in this review, we will focus on the optical qubit systems.

A 1D chain of ions is trapped in a Paul trap, which consists of an oscillating quadrupole field that provides, on average, a confining force on the ions\footnote{It is known from Earnshaw's theorem that charged particles cannot be trapped with a static electric field; the oscillating quadrupole field is the simplest geometry which can trap charged particles}.
Due to being electrically charged, the ions experience Coulomb repulsion from each other. This repulsion, together with the confining force provided by the trap, results in a nearly periodic array of trapped ions in space, and has yielded long chains of one dimensional ion chains for quantum simulation and computing~\cite{joshi2021observing, zhang2017observation}.

Although the ions interact via Coulomb interactions, these interactions are independent of the ions' internal state, and therefore do not provide qubit interactions. That is, unlike the Rydberg atoms where van-der-Waals interactions give qubit interactions, the ions do not directly have qubit interactions. Instead, effective qubit interactions are obtained by coupling the ions to motional degrees of freedom, which are the normal modes of the chain, by shining bichromatic laser fields over the ion chain. The normal mode excitations can be found classically by solving the normal mode equations in the limit of large transverse trapping frequency~\cite{kielpinski2000sympathetic, home2011normal}.

There are two main schemes for realizing qubit interactions using these normal modes. The first was developed by Cirac and Zoller~\cite{cirac1995quantum}, which relies on having zero phonons in the ion chain during normal operations, and exciting one phonon during the entangling operation. This scheme therefore requires the system to be cooled to the motional ground state, i.e. the state with zero phonons~\cite{feng2020efficient}. The second scheme, which is more commonly adopted nowadays, was developed by M{\o}lmer and S{\o}rensen~\cite{sorensen2000entanglement}. This scheme does not require cooling the ion chain to its motional ground state. Understanding how both the schemes work requires some understanding of the physics of the ion-laser coupling, which is described in detail in \ref{sec:appen ions}.

Essential for simulating the protocols in this review, the M{\o}lmer-S{\o}rensen scheme implements the entangling operation $\exp(-i \theta \sigma^x_i \sigma^x_j)$ between  two qubits $i$ and $j$. This scheme can, in principle (up to caveats about ion spacing and normal mode frequency spacing), implement this gate between any two qubits $i$ and $j$ in a finite time, and thus achieves all-to-all connectivity between the qubits. Two-qubit M{\o}lmer-S{\o}rensen gates have been widely realized in experiments~\cite{kirchmair2009deterministic, kirchmair2009high, edwards2010quantum, gaebler2016high, tan2015multi, egan2020fault}, with the highest current gate fidelity in the range of $99.9\%$~\cite{clark2021high}.

\section{Quantum simulation of many-body models}
\label{sec:many-body-realization}
The quantum simulation platforms discussed above can realize a universal quantum gate set, and can therefore in principle realize any unitary quantum evolution. A powerful application of quantum simulation, from the perspective of holography, would be to realize holographic models. The SYK model is a particularly simple 0+1 dimensional model which at large $N$ and low energy is dual to  the nearly AdS$_2$ gravity \cite{maldacena2016conformal}. This  model can potentially be realized in quantum simulators, although requiring a large number of quantum gates and ancillary qubits that could only be within reach of simulation in the future. In this section, we will briefly review how to simulate the SYK model with a quantum circuit. Later, in the next section, we discuss  how to prepare the $\tfd$ state. These two ideas, preparing the TFD and realizing the model, can be seamlessly incorporated with the quantum protocols for measuring the OTOC and implementing HPR, and wormhole teleportation protocols. More details follow in Section~\ref{sec:MeasProt}.

The SYK model is a model of interacting Majorana particles~\cite{Sachdev:1992fk, Sachdev:2010um, Kitaev-talks:2015, maldacena2016conformal, Mandal2017, Gaikwad2020},
\begin{equation}
\hat{H} = \frac{1}{4\times 4!} \sum_{p,q,r,s=0}^{N-1} J_{pqrs} \gamma_p \gamma_q \gamma_r \gamma_s,
\end{equation}
where $\gamma_i$ are Majorana operators, and $J_{pqrs}$ are real-valued scalars drawn randomly from a normal distribution with variance $\sigma^2 = 3!J^2/N^3$. For simulating on a quantum circuit, one first writes the SYK model in terms of complex fermions, and then maps it to a spin Hamiltonian via e.g. the Jordan-Wigner transformation. Due to the Jordan-Wigner transformation, a typical term in the Hamiltonian consists of a four-qubit exchange interaction, multiplied by long Jordan-Wigner strings, for example,
\begin{equation}
\hat H_{pqrs} \propto \left( \prod_{m=s}^{r-1} \sigma^z_m \right) \left( \prod_{m=q}^{p-1} \sigma^z_m \right)\sigma^{\alpha_p}_p \sigma^{\alpha_q}_q \sigma^{\alpha_r}_r \sigma^{\alpha_s}_s,
\end{equation}
where $\sigma^{\alpha_i}_i$ is a spin raising or lowering operator on qubit $i$. This term, and similarly for all the other terms in the Hamiltonian, can be realized utilizing local and collective M{\o}lmer-S{\o}rensen gates. Time evolution with the Hamiltonian can be implemented in a Trotterized fashion~\cite{sonner-syk, babbush2019quantum}. 

Apart from the SYK model, there are some recent works on Hamiltonian simulation of certain gauge theories, which are also based on  Trotterization of the Hamiltonian \cite{Buser:2020cvn,Gharibyan:2020bab,Stetina:2020abi,Ciavarella:2021nmj,Davoudi:2021ney,Culver:2021rxo,Ciavarella:2021lel,Liu:2021otn,Klco:2021lap,Honda:2021aum}. It is possible to construct the ground state for these systems, and measure some observables, via a mapping to a qubit system \cite{Buser:2020cvn}. A detailed discussion of these is beyond the scope of this review. We suggest interested readers refer to these references for further details. 

In~\ref{sec:appen Rydberg} and~\ref{sec:appen ions}, we also review the simulation of other quantum many-body spin models that naturally arise in quantum simulation platforms based on trapped ions or Rydberg atoms.

\section{Measurement protocols}
\label{sec:MeasProt}
We can devise an implementation of the quantum protocols described in this review to measure OTOCs, and realize a simulation of teleportation across wormholes, using the quantum simulators described in Section~\ref{sec:Qsim-platforms}. In this section, we will describe concrete quantum circuits to realize these protocols, and highlight a few pioneering experiments that have already accomplished these feats.

\subsection{Protocols for OTOC}\label{subsec:protocol-otoc}
First, we describe two protocols to measure OTOCs. The first protocol measures the thermal OTOC defined in Eq.~(\ref{eq:otoc-def}) using a TFD state. The second protocol obtains the infinite-temperature OTOC from correlating measurements on two sets of qubits that were prepared as a product of correlated qubits in randomized bases. The essential idea of the latter protocol is that an ensemble of correlated qubits initialized in randomized bases realizes a state closely related to EPR pairs, which is the infinite-temperature TFD.

\subsubsection{Thermal OTOC from TFD}
We recall the definition of the  regularized thermal OTOC, $O_{\rm th}(\beta, t)$,  shown in Eq.~(\ref{eq:otoc-def}). 
This OTOC can be naturally interpreted as
\begin{equation}\label{eq:otoc-interpret}
O_{\mathrm{th}}(\beta, t) = \langle \psi \vert V^\dagger(t) \otimes [V(t)]^T \vert \psi \rangle,
\end{equation}
where $\ket{ \psi } = (W \otimes 1) \tfd$. Note that
\begin{align}
[V(t)]^T = &[\exp(i H t) V(0) \exp(-i H t)]^T \nonumber\\
&= [\exp(-i H^* t) V^T(0) \exp(i H^* t)]   
\end{align}
is the Heisenberg operator for $V^T$ at time $t$ when evolved with $-H^*$. The interpretation in Eq.~(\ref{eq:otoc-interpret}) suggests an implementation as follows:
\begin{itemize}
\item Prepare $\tfd$ on $\mathcal{H}_L\otimes \mathcal{H}_R$.
\item Apply $W$ on, say, the left system. This is possible for unitary $W$.
\item Evolve the left and right systems with $H_L$ and $-H_R^*$. The right system should be evolved with $-H_R^*$ for the reason explained above.
\item Measure $\langle V_L^\dagger \otimes V_R^T \rangle$.
\end{itemize}
Next, we describe one method to prepare $\tfd$. Realizing the remaining steps, for example on a digital quantum computer, is straightforward.

Thermofield double states have been prepared for particular models and small system sizes on a trapped-ion based digital quantum computer. The technique used to prepare the TFD is a quantum-classical hybrid technique called the Quantum Approximate Optimization Algorithm (QAOA)~\cite{farhi2014quantum}, which has more recently been called the Quantum Alternating Operator Ansatz~\cite{hadfield2019quantum} (and denoted QAOA as well).

\begin{figure}[!ht]
\centering
\includegraphics[width=1.01\columnwidth]{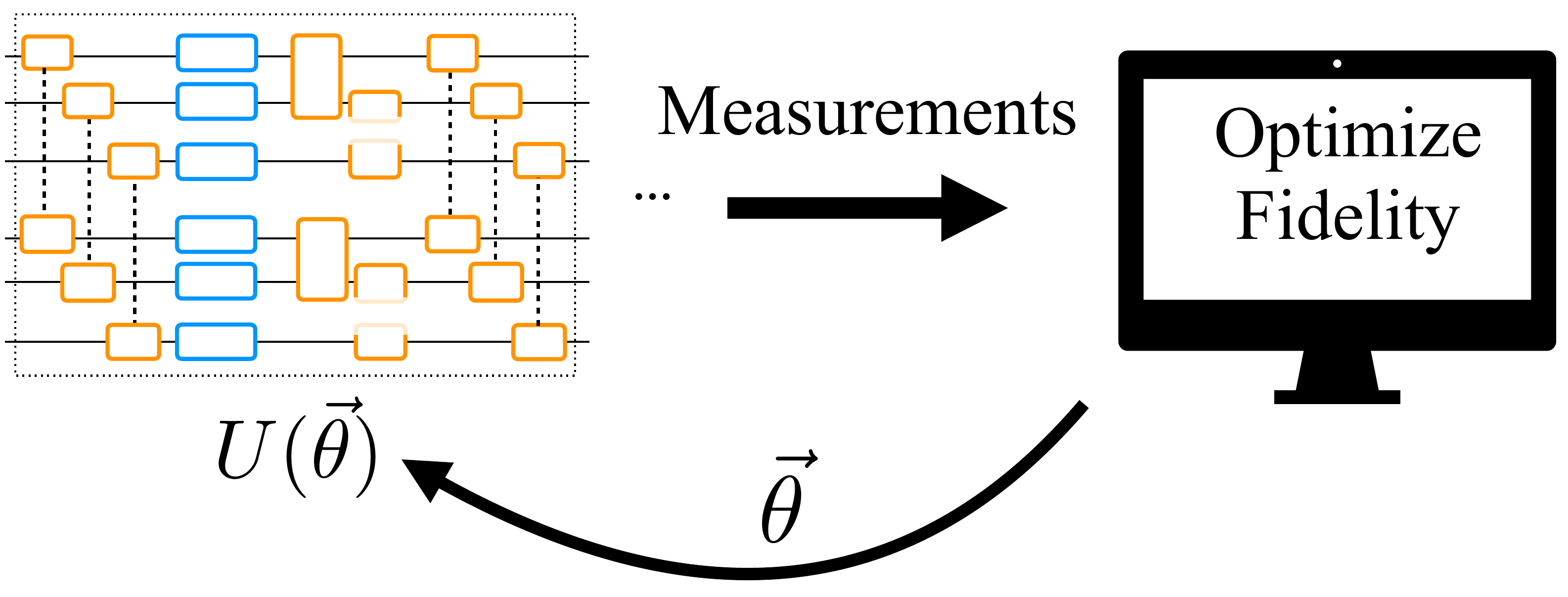}
\caption{\textit{Schematic of the QAOA algorithm} to prepare the TFD for the transverse Ising model. The algorithm consists of a parameterized quantum circuit, $U(\vec\theta) = \cdots U_2(\theta_2)U_1(\theta_1)$, implemented on a quantum computer, where the angles $\vec\theta$ are found by a classical computer. Usually, one finds these angles in a classical-quantum feedback loop, where the classical computer updates $\vec\theta$ based on the output of the quantum computer. Details of the gates used in the quantum circuit are in Fig.~\ref{thermal-otoc-fig1}(a).}
\label{variational}
\end{figure}

QAOA is a variational algorithm~\cite{cerezo2021variational} originally proposed to minimize Hamiltonians. The algorithm is schematically drawn in Fig.~\ref{variational}. It produces a parameterized ansatz wavefunction, $$\ket{\psi(\vec\theta)} = U(\vec\theta)\ket{\psi(0)},$$ where the unitary $U$ is composed of a set of quantum gates $\{U_i\}$, and these quantum gates are parameterized by gate angles $\vec\theta \equiv \{\theta_i\}$, i.e., $U(\vec\theta) = \cdots U_2(\theta_2)U_1(\theta_1)$. The parameters $\vec\theta$ are chosen such that $\ket{\psi(\vec\theta)}$ minimizes the Hamiltonian. In the most common setting, the parameters $\vec\theta$ are chosen in a quantum-classical feedback loop. A classical computer feeds in $\vec\theta$, the quantum computer returns  $\langle \psi(\vec\theta) \vert H \vert \psi(\vec\theta)\rangle$, and the loop continues until $\langle \psi(\vec\theta) \vert H \vert \psi(\vec\theta)\rangle$ is minimized over the space of all $\vec\theta$. The classical computer can use any classical optimization algorithm, e.g. gradient descent, to optimize the necessary $\vec\theta$ to minimize the Hamiltonian. For small system sizes, one can compute the optimal parameters $\vec\theta$ classically, without requiring a classical-quantum feedback loop.

QAOA, and related variational algorithms such as the Variational Quantum Eigensolver, have been used in several applications to minimize target Hamiltonians~\cite{kokail2019self, harrigan2021quantum, pagano2020quantum, o2016scalable, hempel2018quantum, peruzzo2014variational, dumitrescu2018cloud, kandala2017hardware, klco2018quantum}. Finding new applications of variational algorithms is an active area of research~\cite{sundar2019quantum, zhu2020generation, wu19, su21, mar19, kokail2021quantum, wecker2015progress}.

\begin{figure}[!ht]
\centering
\includegraphics[width=0.9\linewidth]{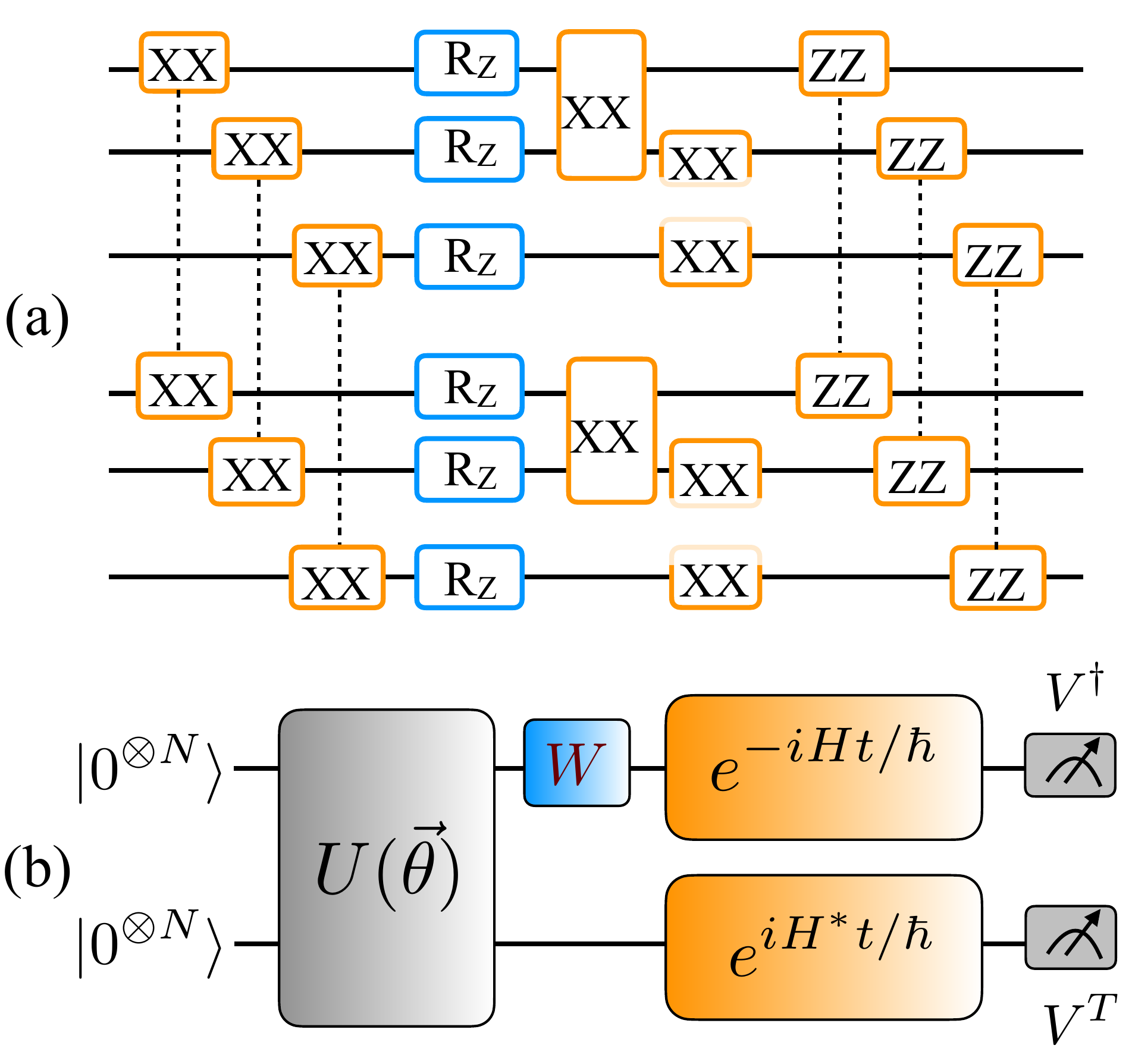}
\caption{(a) Building block of the QAOA circuit that prepares the  thermofield double state for the transverse Ising model. The XX gate is the local M{\o}lmer-S{\o}rensen gate discussed in the main text, the ZZ gate is the analog of the M{\o}lmer-S{\o}rensen for interactions along the $z$ direction, and $R_z$ is a single-qubit rotation around $z$. The gate angles are found classically. (b) Measurement protocol for the thermal OTOC. The block $U(\vec\theta)$ is shown in (a). Depth-$p$ QAOA repeats this block $p$ times, with different angles $\vec\theta_1, \vec\theta_2, \cdots \vec\theta_p$.}
\label{thermal-otoc-fig1}
\end{figure}

Recently, QAOA has been used~\cite{zhu2020generation} for preparing the TFD for the transverse Ising model. One possibility for the basic building block of the quantum circuit for preparing the TFD for this model is shown in Fig.~\ref{thermal-otoc-fig1}(a). Different models require different building blocks for the variational circuit. The variational angles $\vec\theta$ can be chosen such that the fidelity
\begin{equation}
F(\vec\theta) = \left| \langle \psi(\vec\theta) \tfd \right|^2
\end{equation}
is maximized, i.e.
\begin{equation}
\vec\theta_{\rm opt} = {\rm argmax}_{\vec\theta} F(\vec\theta).
\end{equation}
Maximizing the fidelity, however, is restricted to small systems. This is because classically calculating $F(\vec\theta)$ or measuring $F(\vec\theta)$ from the quantum computer are both exponentially difficult tasks.

To mitigate the above challenge, there are alternative proposals to prepare the TFD by maximizing the thermal entropy, or as the ground state of a local parent Hamiltonian for cases where the target Hamiltonian satisfies the eigenstate thermalization hypothesis~\cite{cottrell2019build, maldacena2018eternal}. Specifically for the transverse Ising model ($h=0$ in the Hamiltonian \eqref{eqn:H-tim}), the parent Hamiltonian may take the form
\begin{equation}
H_{\rm parent}(\lambda) = H_A + H_B + H_{AB}(\lambda),
\end{equation}
where $H_A$ and $H_B$ are the transverse Ising Hamiltonian in the A and B chains respectively, and
\begin{equation}
H_{AB} = \lambda J \sum_{i=1}^N (\sigma^y_{iA}\sigma^y_{iB} - \sigma^x_{iA} \sigma^x_{iB}).
\end{equation}
Here, $\lambda$ must be appropriately chosen for each temperature $T$ labeling the TFD. The above parent Hamiltonian exactly produces the TFD at $T=0$ and $T=\infty$. Choosing $\lambda=0$ produces the TFD at $T=0$, which is the product of the transverse Ising model's ground states on the A and B chains. Choosing $\lambda=\infty$ produces the TFD at $T=\infty$, which is a product of EPR pairs. At intermediate temperatures, the ground state of $H_{\rm parent}(\lambda)$ produces the TFD (at $\lambda$-dependent temperature) to a good approximation \cite{sundar-otoc}, where the approximation may be improved by adding more terms to $H_{AB}$. We note that this realization of the TFD model as the ground state of a parent  Hamiltonian is fairly general for chaotic Hamiltonians \cite{cottrell2019build}, see \cite{maldacena2018eternal} for SYK model. The variational ideas to prepare the TFD state are also generic, and can in principle be applied to prepare the TFD state for other models, e.g. the SYK model~\cite{su21}.

After preparing the TFD, measuring the thermal OTOC requires evolving one of the halves forward in time, i.e. with $+H$, and the other backwards in time, i.e. with $-H^*$ (see Fig.~\ref{thermal-otoc-fig1}(b)). Hamiltonian evolution in a digital quantum computer is possible as Trotterized evolution, with sufficiently small Trotter step $dt$. Evolution with $-H^*$ can easily be achieved due to the availability of a universal gate set. The time up to which the system can be evolved is currently limited by gate errors, which restricts high-fidelity quantum simulation to a few Trotter steps.

\subsubsection{Infinite-temperature OTOC from randomized initial states}
The protocol described above can be readily applied to measure the OTOC at $T=\infty$. In particular, $\tfd$ at $T=\infty$ is equal to $\epr$, which can be readily prepared in the lab.

However, there are also other protocols to measure OTOCs at $T=\infty$. One of them, that was proposed by Ref.~\cite{vermersch2019probing} and implemented by Ref.~\cite{Joshi2020} for measuring the OTOC at $T=\infty$, obtains the OTOC from randomized measurements of qubits. This protocol can be extended to large finite $T$ as well, by perturbatively expanding the thermal factor $\exp(-\beta H)$ in powers of $\beta$. Other experiments have used similar ideas with randomized measurements to measure the OTOCs~\cite{garcia2021quantum, mi2021information}.

The infinite-temperature protocol in Ref.~\cite{vermersch2019probing, Joshi2020} works as follows:
\begin{itemize}
\item Prepare two sets of $N$ qubits, one in $\ket{0^{\otimes N}}$ and the other in $\ket{x}$, where $\ket{x}$ is an $N$-qubit product state in the computational basis. We will label the two sets of $N$ qubits as $1\leq i \leq N$ and $N+1\leq i \leq 2N$. None of the operations performed will involve any entanglement between the first $N$ and the second $N$ qubits, therefore we can perform experiments on these in separate experimental runs. Then, each experimental run needs to be performed only on $N$ qubits at a time, which is a significant technical advantage over having $2N$ qubits at one time.
\item Apply $N$ independent single-qubit Haar-random unitaries $u_i$ on qubits $1\leq i\leq N$, and the same $u_i$ on the qubits $N+1 \leq i \leq 2N$.
\item Apply $W$ (assumed to be unitary) on the first $N$ qubits.
\item Evolve both sets of $N$ qubits with the Hamiltonian $+H$. Note that evolution with $-H^*$ is not required.
\item Measure $\langle V^\dagger\rangle$ on qubits $1\leq i \leq N$ and $\langle V^T \rangle$ on qubits $N+1\leq i \leq 2N$. Denote the product of these two measurements as $f_x$.
\item For each $x$, average over the single-qubit Haar-random unitaries $u_i$. Denote the average as $\overline{f_x}$.
\item The weighted sum, $\frac{1}{2^N} \sum_{x=0}^{2^N-1} (-2)^{-|x|} \overline{f_x}$, gives the OTOC, $O_\infty(t) \equiv O_{\mathrm{th}}(\beta=0, t)$. Here, $|x|$ is the Hamming weight of $x$.
\end{itemize}
The crucial step in understanding why this protocol works comes from the realization that the initial state's density matrix, averaged over Haar-random unitaries $\{ u_i \}$ and summed over bit strings $x$ including the weight $(-2)^{-|x|}$, is~\cite{vermersch2019probing}
\begin{align}\label{eq:swap from randomised}
& \frac{1}{N_u 2^N} \sum_{u,x} (-2)^{-|x|} \left( u \otimes u \ket{0^{\otimes N}}\ket{x}\right) \left( \bra{0^{\otimes N}}\bra{x} u^\dagger \otimes u^\dagger \right) \nonumber\\ &\propto {\rm SWAP},
\end{align}
where  ${\rm SWAP} = \sum_{xy} \ket{x}\ket{y}\bra{y}\bra{x}$ swaps the state of the two systems and $u = \bigotimes_{i=1}^N u_i$. For brevity, we will ignore normalization factors for the state that realizes SWAP. The sum in Eq.~(\ref{eq:swap from randomised}) should be understood as averaging over the unitaries first, and then summing over the bit strings $x$, as described in the protocol above.

The SWAP state has the property that
\begin{equation}
\includegraphics[width=0.7\linewidth]{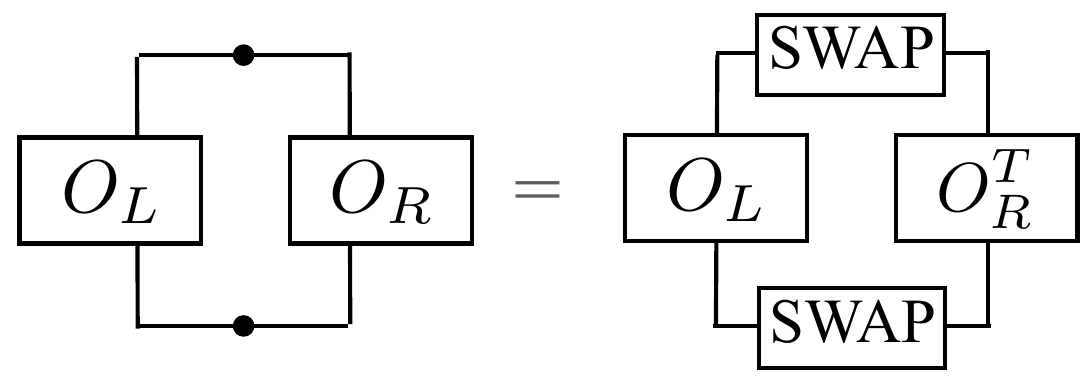}
\end{equation}
and as a corollary,
\begin{equation}\label{eq:swap}
\includegraphics[width=0.9\linewidth]{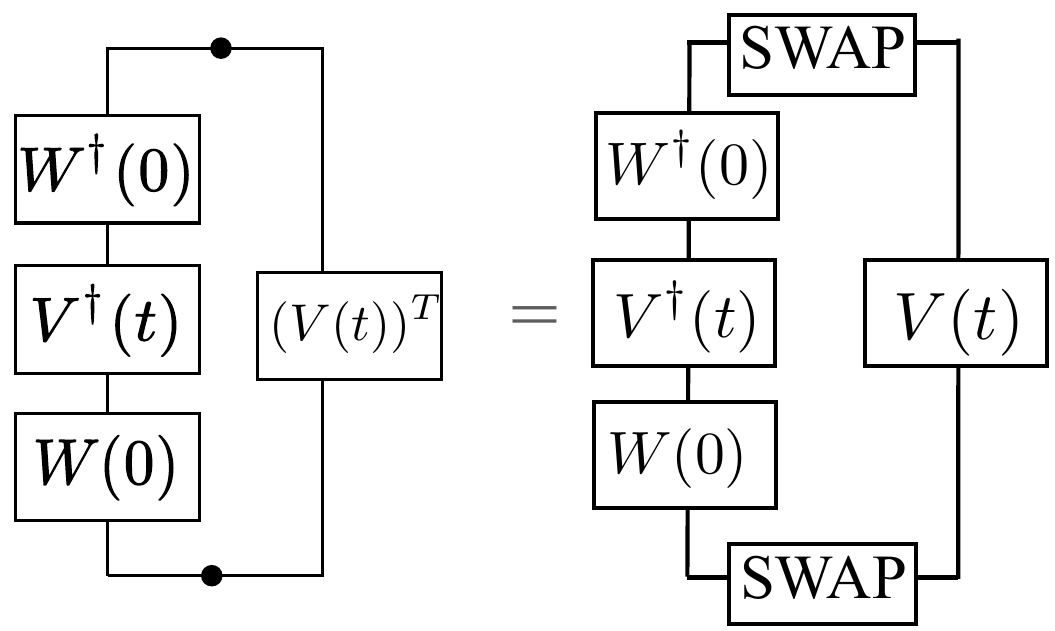}
\end{equation}
Viewing the right hand side of Eq.~(\ref{eq:swap}) as a circuit, the protocol described above is a direct implementation of this circuit, and was implemented in Ref.~\cite{Joshi2020} for the long-ranged Ising model with a transverse field. It is worthwhile to reemphasize the key innovations of this method: Using randomized measurements halved the number of qubits 
as compared to that required by other measurement protocols, and eliminated the need for evolution with $-H^*$. Ref.\cite{vermersch2019probing} also proposed a related alternative protocol with global Haar-random unitaries, instead of local Haar-random unitaries, which proceeds similar to the local protocol, except that the initial states for both sets of $N$ qubits are $\ket{ 0^{\otimes N}}$, and there are no weighting factors $(-2)^{-|x|}$.

\subsection{Simulating teleportation across a wormhole in a trapped ion quantum computer}\label{subsec:verified-qtm-info-scrambling}
There are also experiments that have implemented protocols to simulate the teleportation of one qubit across the analog of a wormhole~\cite{Landsman2019, blok2021quantum}. In one such experiment by Ref.\cite{Landsman2019}, the experimentalists implemented two protocols that simulate the teleportation of one qubit across the analog of an infinite-temperature wormhole -- on EPR states. One of those protocols probabilistically teleports one qubit, and the other deterministically teleports the qubit. Here, we describe the deterministic protocol, and refer the reader to Ref.\cite{Landsman2019} for the probabilistic protocol. 
The protocol is based on the Yoshida-Kitaev version \cite{Yoshida2017} of the Hayden-Preskill protocol \cite{Hayden2007} [see Section \ref{sec:HPR-YK}].

\begin{figure}[t]
\centering
\includegraphics[width=1.05\columnwidth]{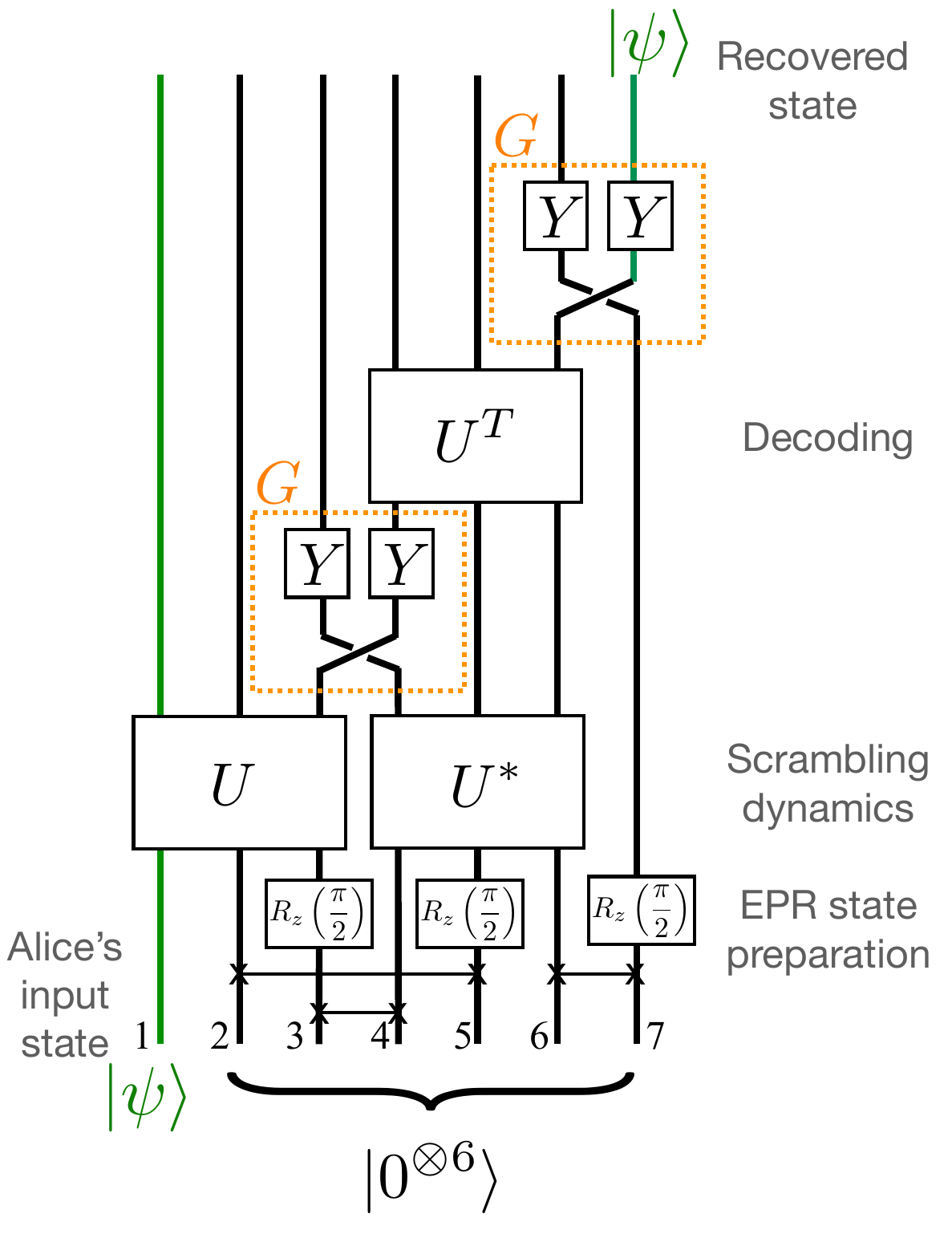}
\caption{\textit{Simulating the many-body teleportation protocol in the lab~\cite{Landsman2019}.} A qubit inserted in the left system is decoded on the right. The quantum circuit is drawn vertically here for convenience and to relate to the protocol in Fig.~\ref{fig:HP}. A horizontal line ending in two crosses is the M{\o}lmer-S{\o}rensen gate, $\exp(-i \frac{\pi}{4}\sigma^x_i \sigma^x_j)$. The single-qubit $R_z$ gate is defined as $R_z(\theta) = \exp(-i\frac{\theta}{2}\sigma^z)$, and $Y$ is the $\sigma^y$ Pauli operator.} 
\label{fig:Landsman}
\end{figure}

The experiment in Ref.~\cite{Landsman2019} implemented a particular instance of Fig.~\ref{fig:HP} with seven qubits, as follows:

\begin{itemize}
\item The experiment begins by initializing qubit 1 in $\ket{\psi}$, which is the state to be teleported, and qubits 2-7 as EPR pairs, with qubits 2 and 5 forming one pair, qubits 3 and 4 forming one pair, and qubits 6 and 7 as one pair. Qubits 2-5 are analogous to the black holes and the past radiation, interpreted as $B$ and $B'$ in Fig.~\ref{fig:HP}, and qubits 6-7 are the ancillary pair for decoding, interpreted as $A'$ and $R'$.

\item Then, they evolved qubits 1-3 with a \textit{maximally scrambling} unitary $U$, i.e. a unitary which evolves all single-qubit Pauli operators into three-qubit Pauli strings. They evolved qubits 4-6 with $U^*$. The probabilistic protocol
measured qubits 3 and 4 and terminated here [see Ref.~\cite{Landsman2019}]. But we will move on to the deterministic protocol.

\item They apply a Grover oracle $G = 1-2\epr \bra{{\rm EPR}}$ on qubits 3 and 4. A circuit compilation trick allows $G$ to be implemented with a SWAP gate (performed classically by relabeling the qubits) followed by single-qubit $Y$ gates [see Fig.~\ref{fig:Landsman}].

\item Then according to Fig.~\ref{fig:Landsman}, one should evolve qubits 4-6 (which are now interpreted as $D'$ and $C'$) with $U^T$, apply $G = 1-2\epr \bra{{\rm EPR}}$ on qubits 6 and 7, and evolve qubits 4-6 with $U^*$. As stated above, $G$ can be implemented with classical relabeling and $Y$ gates. At the end of this, $\ket{\psi}$ will be successfully teleported to qubit 7 (which  is interpreted as $R'$). This concludes the experiment in Ref.~\cite{Landsman2019}. The experiment did not need to implement the final $Y$ and $U^T$ on qubits 4-6, because they don't affect the state of qubit 7.

\end{itemize}

Ref.~\cite{Landsman2019} successfully demonstrated teleportation of various choices of $\ket{\psi}$, specifically $\ket{\psi} = \ket{0}$, $\ket{\psi} = \ket{1}$, $\ket{\psi} = (\ket{0} \pm \ket{1})/\sqrt{2}$, and $\ket{\psi} = (\ket{0} \pm i\ket{1})/\sqrt{2}$, with an average teleportation fidelity of $78\%$. The teleportation fidelity is $<100\%$  due to gate errors in the experiment.  The same protocol can teleport any single-qubit state $\ket{\psi}$.

Future experiments may use the above protocol to teleport multiple qubits with more Grover iterations, as well as simulate teleportation across the analog of a finite-temperature wormhole by appropriately generalizing the circuit, e.g., replacing $\epr$ with $\tfd$ in the initial state, where $\tfd$ could for example be prepared by techniques outlined in Section~\ref{subsec:protocol-otoc}.

\section{Conclusions and Discussions}
\label{sec:conclusion}
In this work, we have presented the recent advances in realizing analogous models of gravity in a lab, in the sense of holography. In particular, our focus has been the wormhole teleportation inspired protocols for teleportation in the many-body systems. The mechanism of the teleportation is governed by the operator size that is inserted just before the coupling is applied. We described the experimental protocols and observations of OTOC and small scale teleportation in state-of-the-art quantum simulators. 

It should be noted that the exact model of gravity in a lab, where one can not only verify the holographic principle but also learn about the gravity from a lab, is still not available. Our summary of the recent advances should be seen as advances in the theoretical translation of the tools and observables in gravity and many-body models available on a lattice, as  well as advances in experimental technology. With the proof-of-principles done for the holographic models that have a semi-classical dual, in the long run one can hope to study the more complex bulk dual, involving stringy corrections, in a quantum lab.

 It is always crucial to study the effects of experimental decoherence and noise sources in implementing protocols.  We have not discussed them in this review, but one should keep in mind the limitations they pose and the rectifications thereof, for example see \cite{sundar-otoc,  benoit-otoc, Landsman2019, Joshi2020, errors-Daley} for possible error sources and corrections. We discussed here that the behavior of the teleportation fidelity with time is a strong signature of the nature of the dynamics, namely generic scramblers or the holographic scrambler. Even better, the teleportation fidelity identifies the real scrambling dynamics and decays due to decoherence \cite{yoshida-yao}. Furthermore, it would be interesting to find out the validity and corrections of the Hayden-Preskill protocol as well as the many-body teleportation protocol in presence of errors \cite{Bao2021}. 

We also note that the operator size distribution is a more refined description of the time-evolved operators than the averaged OTOC that we have presented here. It remains a question as to how and when the size distribution discussed here compares with the usual notion of the operator size \cite{Qi2019a}, and to those amenable in experiments \cite{qi2019measuring}. It is argued in \cite{Susskind:2019ddc,Susskind:2020gnl,Barbon:2020olv,Lin:2019qwu}, that the rate of change of momentum of the particle falling in to the bulk spacetime is dual to the complexity of the dual operator at the boundary. This complexity basically  measures the growth of the size of the operator under time evolution. Some recent progress has been made towards understanding complexity for the dual field theory \cite{Jefferson:2017sdb,Hackl:2018ptj,Khan:2018rzm,Chapman:2017rqy,Bhattacharyya:2018bbv,Ali:2018fcz,Chapman:2018hou,Bhattacharyya:2019kvj,Bhattacharyya:2021cwf,Caputa:2018kdj,Flory:2020dja,Erdmenger:2020sup,Chagnet:2021uvi,Rabinovici:2020ryf}, see \cite{chapman2021quantum} for a recent review. However, it is in its early stage of development. An interesting direction will  be to develop this idea of operator growth using complexity as a possible diagnostic. This will not only enable us to make connection  with certain predictions coming from holography but will also help us to compare with other diagnostics which are measurable via experiments. Another important theoretical direction is to explore the finite temperature generalizations of the many-body teleportation in the spirit similar to \cite{Gao2019,almheiri2019islands, penington2020entanglement}. In recent times, several toy models based on tensor network construction for holography has been proposed \cite{Pastawski:2015qua,Hayden:2016cfa,Czech:2015kbp,Bhattacharyya:2016hbx,Bhattacharyya:2017aly,Yang:2018iki,Chen:2021ipv,error-review-2021}. In this context, it will be interesting to realize the thermofield double state and the teleportation protocol. Perhaps \cite{Peach:2017npp} will provide a good starting point. This will pave the way forward to some of the predictions coming from holography using interesting quantum many-body systems.

At last, for the experimental prospects of connecting theoretical high energy physics with experiments,  we conclude by outlining directions other than the wormhole teleportation. For example some of the open directions are the realization of the SYK model as a simple model of holography \cite{sonner-syk}, simple models of wormholes \cite{Wheeler_PhysRev.97.511, strominger-1955, verlinde2021wormholes}, time-shifted wormholes and the teleportation therein \cite{VanBreukelen2018} and possibilities to use time shifted wormhole teleportation to distinguish states with similar entanglement \cite{nogueira2021geometric}, among many others.

\begin{acknowledgement}
We thank Hannes Pichler, Beno\^it Vermersch, Norman Y. Yao and Peter Zoller for fruitful discussions, and Andreas Elben, Torsten V. Zache for collaboration on related projects. We thank Ana Maria Rey and Murray Holland for a careful reading of the manuscript. We thank Manoj K. Joshi for comments on the section on quantum simulation platforms, Beno\^it Vermersch for comments on randomized measurement protocol for OTOCs and Andreas Elben for various useful comments. 
A.B would like to thank Wissam Chemissany and Nariman Chrakie and speakers of the workshop ``Quantum Information in QFT and AdS/CFT-II " (\url{https://events.iitgn.ac.in/2021/qi/}) for various useful discussion which have made him interested in this particular topic. A.B is supported by Start-Up Research Grant (SRG/2020/001380), Mathematical Research Impact Centric Support Grant (MTR/2021/000490) by the Department of Science and Technology Science and Engineering Research Board (India) and Relevant Research Project grant (58/14/12/2021- BRNS) by the Board Of Research In Nuclear Sciences (BRNS), Department of mic Energy, India.
LKJ acknowledges European Union's Horizon 2020 research and innovation programme under Grant Agreement  No.\ 731473 (QuantERA via QT-FLAG) and the Austrian Science Foundation (FWF, P 32597 N). LKJ also acknowledges virtual hospitality of Indian Institute of Technology Madras, Chennai where parts of this work were presented in May 2020. 
\end{acknowledgement}
\appendix 
\section{Some details about AdS/CFT dictionary} \label{appendA}
 Here we will briefly sketch out some details of the AdS/CFT dictionary. It has two aspects: Kinematical aspects and Dynamical aspects. We briefly review both of them below. For more details, interested readers are referred to \cite{AHARONY2000183,Ammon:2015wua,Bhattacharyya:2015nvf,Zaffaroni:2000vh} and the citations therein. \\ \\
 \textit{Kinematical Aspect:} First, we begin by discussing the generators of a conformal group in $d$ dimensions. For simplicity, we will assume that the underlying CFT is defined on a flat Minkowski background. The conformal transformations consist of the following four transformations, and we quote the corresponding generators below \cite{DiFrancesco:1997nk}.
 \begin{align}
     \begin{split}
    & \textrm{Translation} (P_{i})\rightarrow i\,\partial_{i}\,,\\&
    \textrm{Rotation} (J_{ij}) \rightarrow -i(x_{i}\partial_{j}-x_{j}\partial_{i})\,,\\&
    \textrm{Dilatation}(D)\rightarrow -i\, x^i\partial_i\,,\\&
    \textrm{Special Conformal Transformation}\\& (SCT)(K_{i})\rightarrow i\,(2 x_i x^j\partial_j- x^2\partial_{j}),
          \end{split}
 \end{align}
 where the $i,j$ takes value from $0$ to $d-1,$ where $0$ denotes the time coordinate. $J^{ij}$ includes both the space-rotation and boost. $J^{ij}$ is completely ant-symmetric in $i,j$ indices. So it is evident that Poincare group (consisting of Translations and Rotations) is a subgroup of conformal group. The dilatation generators, scales the coordinates by a constant factor and the special conformal transformation (SCT) can be thought of a translation preceded and followed by an inversion. Now it can be shown that, these generators after suitable identification satisfy a $SO(d,2)$ algebra.\par  \textit{Now the isometry generators of $AdS_{d+1}$ exactly satisfy this algebra and they are in one-to-one correspondence with the generators (global) of conformal group in one lower dimension.} \cite{DiFrancesco:1997nk}. \\ Lets take a concrete example of $AdS_3/CFT_2.$ For $CFT_2$ we have 6 generators corresponding to the \textit{global} conformal transformations. Now let us first write $AdS_3$ in Poincare coordinates\footnote{We can use other coordinates also. For a detailed review please refer to \cite{Ammon:2015wua,AHARONY2000183}}, 
 \begin{equation} \label{poin}
      ds^2= \frac{L^2(dz^2-dt^2+ dx^2)}{z^2},
 \end{equation}
 where the boundary of it is located at $z=0.$ $t$ is Lorentzian time. We below quote the isometry generators by solving the Killing equation and we quote the result below \cite{kundu-wormhole,Bhattacharyya:2015nvf,Bagchi:2012cy,Caceres:2019giy}. 
 \begin{align} \label{gen}
     \begin{split}
         &J_{01}=i\,\Big[\Big(\frac{L^2+z^2+t^2+x^2}{2\,L}\Big)\partial_{t}+\frac{x\, t}{L}\partial_x+\frac{t\, z}{L}\partial_z\Big],\\& J_{02}=i\,\Big[\Big(\frac{-L^2+z^2+t^2+x^2}{2\,L}\Big)\partial_{t}+\frac{x\, t}{L}\partial_x+\frac{t\, z}{L}\partial_z\Big],\\&
        J_{03}=i\,\Big[-x\partial_t-t\partial_x\Big],\\&
        J_{12}=i\,\Big[-z\partial_z-t\,\partial_t-x \partial_x\Big],\\&
        J_{13}=i\,\Big[\Big(\frac{L^2+z^2-t^2-x^2}{2\,L}\Big)\partial_{x}-\frac{x\, t}{L}\partial_t-\frac{x\,z}{L}\partial_z\Big],\\&
         J_{23}=i\,\Big[\Big(\frac{-L^2+z^2-t^2-x^2}{2\,L}\Big)\partial_{x}-\frac{x\, t}{L}\partial_t-\frac{x\,z\partial_z}{L}\Big].
     \end{split}
 \end{align}
We can show that they satisfy
 \begin{align}
     [J_{ab}, J_{cd}]=i\,[\eta_{ac}J_{bd}+\eta_{bd}J_{ac}-\eta_{ad} J_{bc}-\eta_{bc}J_{ad}],
 \end{align}
$SO(2,2)$ algebra and $a,b,c,d \in \{0,1,2,3\}.$ This precisely matches with the algebra of the global conformal generators in $3$-dimensional Minkowski space and  $\eta_{ab}=\textrm{Diag}(-1,-1,1,1)$ a diagonal metric with two time signature \cite{DiFrancesco:1997nk,Bhattacharyya:2015nvf}.\par Also, the conformal boundary of the AdS in this coordinate is located at $z=0.$ If we take the boundary limit on (\ref{gen}), one can easily see the one-to-one correspondence between theses generators and that of those for the global conformal group in one lower dimension, i.e. d=2 for this specific case. For example, in the boundary limit $J_{12}$ in \eqref{gen} corresponds to the Dilatation generator (D) of the boundary CFT \footnote{To take the boundary limit we basically set $z=constant.$ Hence, $\partial_z$ term goes away and in the rest of the terms we take $z \rightarrow 0$ limit.}. One can easily generalize this result for arbitrary dimensions\footnote{For d=2, one can have infinite number of Virasoro generators apart from the one mentioned in this subsection which often termed as local generators. For details one can refer to \cite{DiFrancesco:1997nk}.} \\\\
 \textit{Dynamical Aspect:}
 Now we will discuss the dynamical aspects of the duality. It states the equivalence between CFT and gravitational path-integral, 
 \begin{align}\label{dic}
        Z_{CFT}(\{J_{i}\})=Z_{Gravity},
     \end{align}
where, $$ Z_{Gravity} \sim \int \mathcal{D}(\phi,G_{\mu\nu}, A_{\mu})|_{\{J_i\}_{z=0}}e^{-S_{gravity}+\cdots}.$$ For  $Z_{Gravity},$ we have to evaluate the action $ S_{gravity}$ which consists of bulk fields, metric, scalar field, gauge field etc on-shell, i.e. on the solution of the equation of motion of all these bulk fields. Also, the sources $J_{i}$ of the CFT side can be identified with the boundary values of the bulk scalar field after imposing suitable boundary condition. On the left hand side of (\ref{dic}), we have a functional $Z_{CFT}(\{J_{i}\})$ which depends on  arbitrary (off-shell) sources $J_i$ in $d$ dimensions, and on the right hand side we have the (on-shell) functional $Z_{Gravity}$ in $d+1$ dimensions, involving gravitational action evaluated on the solution of the equations and the fields reduce to the corresponding $J_i$'s at the boundary of the AdS \footnote{In fact, for every source $J_i(x^{i}),$ where $x^{i}$ are the boundary coordinates, we can extend it uniquely inside the bulk after imposing suitable boundary conditions inside the bulk (usually at the center of AdS).  Hence for every source configuration, there exists a corresponding bulk field $\phi_a(x^{i},z),$ where $z$ is the extra bulk radial coordinate \cite{AHARONY2000183,Zaffaroni:2000vh}} . One has to be careful about imposing the boundary limit (for Poincare AdS as shown in (\ref{poin}) it is basically $z\rightarrow 0$ limit) as typically the fields diverges near the AdS boundary \cite{Ammon:2015wua, Zaffaroni:2000vh, AHARONY2000183}. Now utilizing relation (\ref{dic}) one can translate all the field theory correlation functions to the correlation functions of fields in the bulk spacetime.\par
Before we end, we make a few more comments. We need to know how the  CFT operators map to fields in the bulk. In principle, it depends on the details of the two theories (CFT and the gravity theory). String theory provides this map. Roughly, we can observe that the consistent coupling between a  certain field to a certain operator can be often argued using underlying symmetries. Both $J_i$ and the field operator share the same quantum number under the conformal group. This gives some obvious coupling. \begin{equation}
W_{CFT}= S_{CFT}+\int d^d x \Big[g_{i j} T^{i j}+ A_i J^i+ \phi F_{ij}^2+\cdots..\Big].
\end{equation}
So the metric couples to stress tensor, gauge field ($A_{\mu}$) in bulk to current ($J^{i}$) in dual CFT, the scalar field in bulk to some scalar operator at the boundary and so on. Now given the effective action $W_{CFT}$ we can construct the $Z_{CFT}(\{J_{i}\})$ in the usual way. Also, one important point is that mass of the fields in the bulk can be related to the conformal dimensions ($\Delta$) of primary operator of the dual field theory. For example, for a massive scalar field in the bulk with a mass $m,$ we  have,
\begin{equation}
   m^2 L^2= \Delta (\Delta-d),
\end{equation}
where $\Delta$ is the conformal dimension of the dual primary operator for the d-dimensional CFT, and $L$ is the AdS radius. Similar conclusions can be made for spinning fields also. For more details, interested readers are referred to \cite{Ammon:2015wua, AHARONY2000183, Zaffaroni:2000vh}.

\section{Rydberg atoms}\label{sec:appen Rydberg}
\begin{figure}[!ht]
\centering
\includegraphics[width=1.01\columnwidth]{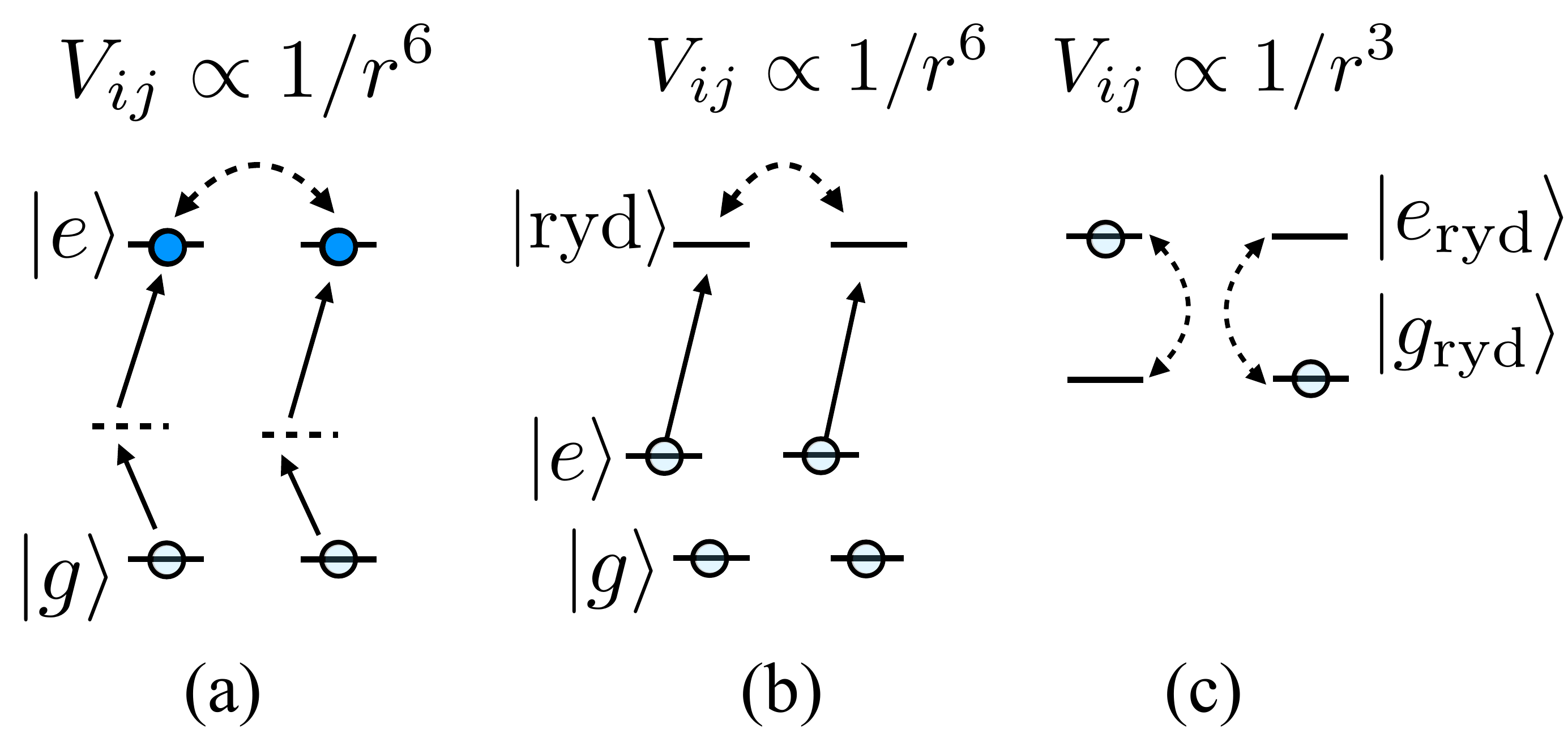}
\caption{\textit{Qubit encodings for Rydberg atoms.} (a) Ground and Rydberg state as qubit states. Atoms are excited from the ground to the Rydberg state via a two-photon transition. Rydberg atoms interact with a van-der-Waals interaction. (b) Two ground states as qubit states. Interactions are induced  by exciting $\vert e\rangle$ to or dressing $\vert e\rangle$ with a Rydberg state. (c) Two Rydberg states as qubit states. They undergo flip-flop interactions due to their dipole moments.}
\label{rydberg}
\end{figure}

In the main text, we stated that Rydberg atoms are one of the platforms for analog and digital quantum simulation. Here, we discuss the physics that can be explored with Rydberg atoms, in more detail.in each of these qubit encodings.

There are several possible ways to encode a qubit in these atoms using any two internal atomic states. These two states could be a long-lived hyperfine ground state (i.e. a state with a small principle quantum number) and a Rydberg state. Or the two qubit states could be two hyperfine ground states, with the Rydberg state acting as an auxiliary state, into which atoms are transferred when strong interactions are needed, or which is admixed to one of the ground states via Rydberg dressing. Or they could even be two Rydberg states. Each of these choices of the qubit states allows different capabilities, and has been used to realize digital quantum computation or analog quantum simulation. Let us now understand the physics that can be explored in each of these qubit encodings.

Let us first consider the case that the qubit states are a ground state and a Rydberg state, as illustrated in Fig.~\ref{rydberg}(a). Any two atoms in the Rydberg state and separated by a distance $r$ interact with each other with strength $V/r^6$, where $V \propto n^{11}$. Additionally, one could drive the atoms from the ground state to the Rydberg state via external lasers and effectively realize, for example, the long-ranged quantum Ising model,
\begin{equation}\label{eqn: Rydberg hamiltonian}
H = \sum_i \Omega \sigma^x_i - \Delta \sigma^z_i + \frac{1}{2} \sum_{ij} \frac{V}{r_{ij}^6} (1-\sigma^z_i)(1-\sigma^z_j).
\end{equation}
Here, $\sigma^\alpha$ are Pauli operators acting on the two qubit states, $\Omega$ is the amplitude of the two-photon transition that excites the atom from the ground to the Rydberg state, and $\Delta$ is the detuning of the two-photon transition from the atomic transition. This is a paradigmatic model in quantum mechanics, and has been realized with Rydberg atoms by various groups~\cite{lienhard2018observing, guardado2018probing, schauss2015crystallization, lee2019coherent, de2018accurate, schauss2012observation, labuhn2016tunable}. Further, the ability to arrange the atoms in arbitrary geometries of tweezers, and the ability to quench various parameters in the above Hamiltonian, leads to a rich playground of physics that is open for exploration.

In the case the qubit states are two hyperfine ground states, as illustrated in Fig.~\ref{rydberg}(b), it is possible to make one of the ground states interacting by dressing it with a Rydberg state. Interactions are induced due to the ad-mixture with the Rydberg state, and obtains a similar model to Eq.~(\ref{eqn: Rydberg hamiltonian}). The advantage to this method is that the atomic lifetimes are longer, and not limited by spontaneous decay from the Rydberg state. This case has also been experimentally realized by various groups~\cite{zeiher2016many, zeiher2017coherent, guardado2021quench, borish2020transverse, jau2016entangling}.

In the case that the qubit states are two different Rydberg states, as illustrated in Fig.~\ref{rydberg}(c), the atomic interactions are slightly different. This is because the dipole-dipole interactions now have a matrix element that resembles a flip-flop interaction, $\vec{d}_i \cdot \vec{d}_j \propto \sigma^+_i \sigma^-_j + {\rm h.c.}$. Additional single-particle terms could be added, via coupling to microwaves that excite atoms from one Rydberg state to the other. This qubit encoding has been used~\cite{de2019observation}, for example, to realize the Su-Schriefer-Heeger model~\cite{su1979solitons, heeger1988solitons}, which is the simplest model presenting topological behavior.

\subsection{Entangling gates on Rydberg atoms}
\begin{figure}[t]
\centering
\includegraphics[width=0.6\columnwidth]{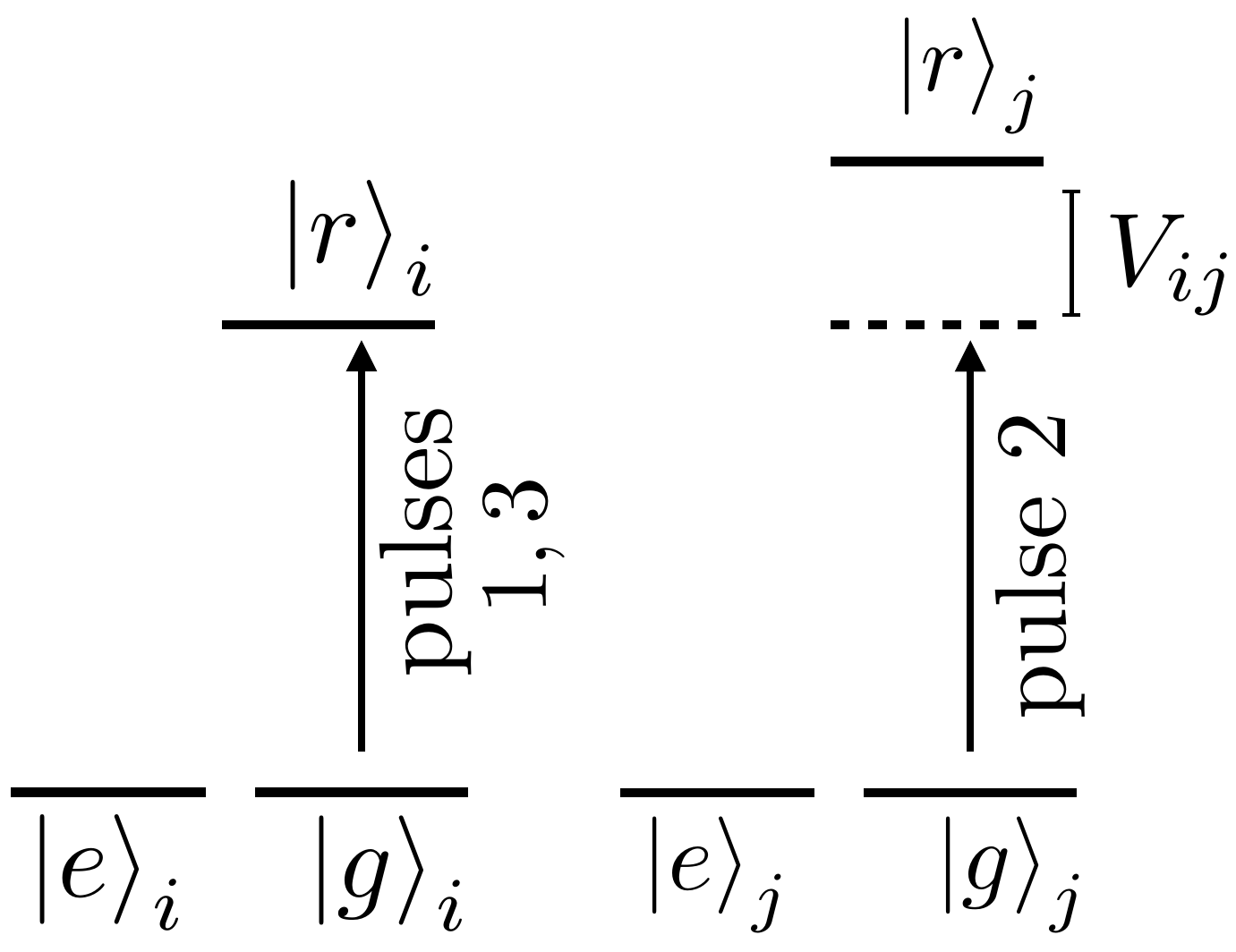}
\caption{\textit{Entangling gate using the Rydberg blockade and three laser pulses.} The first pulse transfers the population in $\vert g_i\rangle$ to a Rydberg state $\vert r_i\rangle$. The second pulse gives $\vert g_j\rangle$ a phase equal to $\pi$ if $\vert r_i\rangle$ is unoccupied, and a phase close to $\pi$ if $\vert r_i$ is occupied. The third pulse transfers the population in $\vert r_i\rangle$ to $\vert g_i\rangle$. In this whole sequence, $\vert e_ie_j\rangle$ is unaffected, and the other three orthogonal states are (approximately) multiplied by $-1$, which realizes a controlled-Z gate.}
\label{blockade}
\end{figure}

The scheme to implement an entangling gate is shown in Fig.~\ref{blockade}. It consists of two hyperfine ground states $\vert g\rangle$ and $\vert e\rangle$ encoding the qubit, and one of the ground state, $\vert g\rangle$ being coupled to a Rydberg state $\vert r\rangle$ via a laser(s). The whole scheme involves three individually addressed laser pulses. First, a laser pulse of duration $t_\pi = \pi/\Omega$ is shone on one atom, then a pulse of duration $2t_\pi$ is shone on the second atom, and finally another laser pulse of duration $t_\pi$ is shone on the first atom. The effect of this sequence can be understood by considering the four initial states, $\vert ee\rangle, \vert ge\rangle, \vert eg\rangle$, and $\vert gg\rangle$, of the two qubits. Since only $\vert g\rangle$ is coupled to the Rydberg state, the whole sequence has no effect on $\vert ee\rangle$. Moreover, for the initial states $\vert ge\rangle$ and $\vert eg\rangle$, the three-pulse sequence is equivalent to applying a single pulse of $2t_\pi$ on $\vert g\rangle$, which only multiplies the state by $-1$. Non-trivial physics happens for the initial state $\vert gg\rangle$. For this case, the first and third laser pulses together multiply the state by $-1$, and the second laser pulse, which is effectively off-resonant due to the $\vert rr\rangle$ being blockaded, gives an additional small phase $\phi \ll \pi$. In total, the only one of four states that does not acquire a sign is $\vert ee\rangle$. This is equivalent to applying a controlled-phase gate. The controlled-phase gate, together with arbitrary single-qubit rotations which can be implemented via magnetic fields or stimulated Raman transitions, are sufficient to realize universal quantum computing. Gate fidelities exceeding $99\%$ for entangling gates, and up to $99.6\%$ for single-qubit gates, have been demonstrated~\cite{madjarov2020high}.

\section{Trapped ions}\label{sec:appen ions}
In the main text, we stated that there are two schemes to implement entanglement between trapped ion qubits. Here, we describe these two schemes, as well as other physics that can be explored.

\begin{figure}[t]
\centering
\includegraphics[width=0.5\columnwidth]{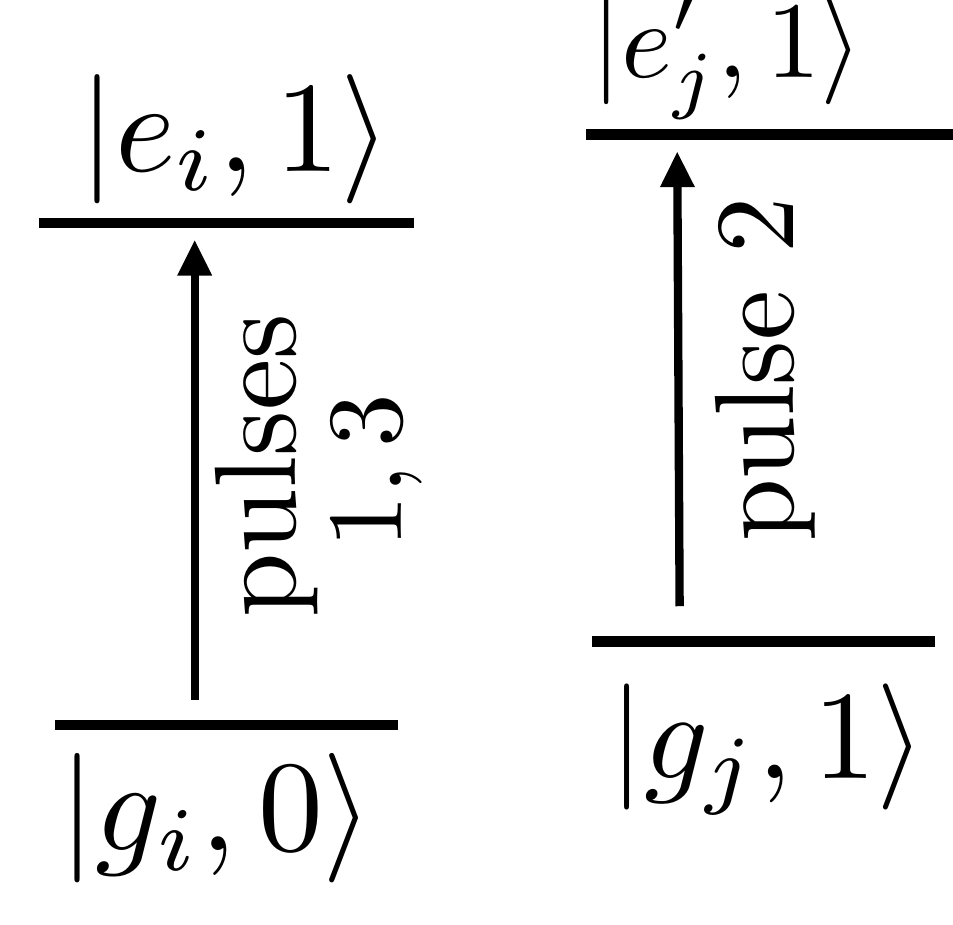}
\caption{\textit{The Cirac-Zoller scheme.} We shine three laser pulses. The first pulse transfers the population in $\vert g_i,0\rangle$ to $\vert e_i,1\rangle$, creating one phonon in the process. The second pulse gives a $(-1)$ sign to $\vert g_j,1\rangle$, thus realizing a controlled-Z gate between the ion and the phonon. The third pulse is identical to the first pulse, destroying the phonon and returning the system to its motional ground state.}
\label{CiracZoller}
\end{figure}

\subsection{The Cirac-Zoller scheme}
The Cirac-Zoller scheme~\cite{cirac1995quantum}, illustrated in Fig.~\ref{CiracZoller}, is a three-step process that realized the controlled-Z gate and requires individual qubit addressability. In the first step, one shines a laser at frequency $\omega = \omega_0 + \omega_t$ on a specific ion, where $\omega_0$ is the energy spacing between the qubit states $\vert g\rangle$ and $\vert e\rangle$, and $\omega_t$ is the frequency of the center-of-mass mode. The Hamiltonian for a single ion coupled to the laser is
\begin{align}
H = & \frac{\hbar\omega_0}{2} \sigma^z + \hbar\omega_t( a^\dagger a + \frac{1}{2}) \nonumber\\ &+ \hbar\Omega \cos(\omega t) (a^\dagger + a)(\sigma^+ + \sigma^-),
\end{align}
where $a (a^\dagger)$ annihilates (creates) a mode excitation, and $\Omega$ is the ion-laser coupling strength. For $\omega_0, \omega_t \gg \Omega$, which is typically the case, we apply the rotating wave approximation, i.e. go to a rotating frame and neglect terms rotating at frequency $O(\omega_0)$ or $O(\omega_t)$ in this frame, and obtain
\begin{equation} \label{eqn: rwa cirac zoller}
  H = \frac{\hbar\Omega}{2} (a^\dagger \sigma^+ + {\rm h.c.}).
\end{equation}
where $a (a^\dagger)$ destroys (creates) a phonon excitation. Assuming the initial state has no phonons, the effect of the above Hamiltonian is that it transfers the qubit's state to the phonons. Concretely, when one shines a laser pulse of duration $t=\pi/\Omega$ on an ion labeled $i$ in the state $(\alpha \vert g\rangle + \beta \vert e\rangle)_i\vert 0\rangle$, where $\vert 0 \rangle$ refers to having zero phonons, the system's state after the pulse is $\vert e\rangle_i (i\alpha \vert 1\rangle + \beta \vert 0\rangle)$. The second step involves shining a second laser pulse, with a similar form to Eq.~(\ref{eqn: rwa cirac zoller}), but couples $\vert g\rangle_j$ on ion $j$ to a \textit{different} excited state $\vert e'\rangle_j$, for a duration $2\pi/\Omega$. The effect of this second step is to selectively give a $(-1)$ sign to $\vert g\rangle_j\vert 1\rangle$, i.e. accomplishes a controlled-Z gate between the phonon and the qubit. The third step is identical to the first step, and decouples the phonon from the qubit $i$. The net effect of the sequence is to selectively give a $(-1)$ sign to $\vert g\rangle_i\vert g\rangle_j \vert 1\rangle$, and do nothing to all other states, which is a controlled-Z gate.

The Cirac-Zoller gate was first experimentally realized to demonstrate entanglement between an ion and a phonon in Ref.~\cite{monroe1995demonstration}, and later between two ions in Ref.~\cite{schmidt2003realization, schmidt2003realize, riebe2006process}.

\subsection{The M{\o}lmer-S{\o}rensen scheme}
\begin{figure}[t]
\centering
\includegraphics[width=0.9\columnwidth]{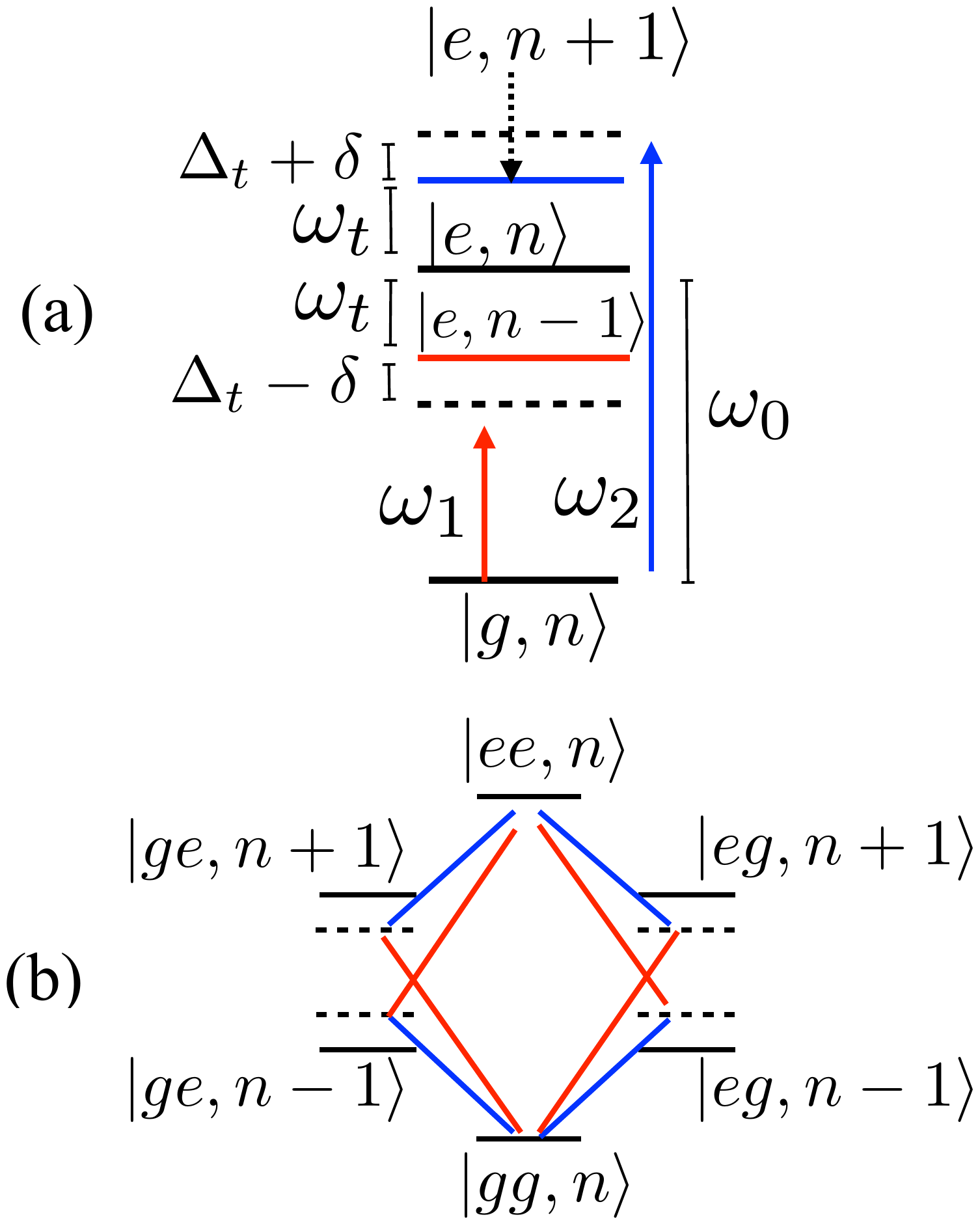}
\caption{\textit{The M{\o}lmer-S{\o}rensen scheme.} (a) We shine two lasers at frequencies $\omega_1$ and $\omega_2$ close to the red and blue motional sidebands at $\omega_0 \pm \omega_t$. Different detuning of the laser frequencies from the motional sidebands give rise to different qubit interactions, as described in the text.  (b) Physical picture for the emergent M{\o}lmer-S{\o}rensen interaction. Ions scatter from $\vert gg,n\rangle$ to $\vert ee,n\rangle$ (also from $\vert ge,n\rangle$ to $\vert eg,n\rangle$) either by emitting and re-absorbing a virtual phonon in the weak coupling case, $\Omega \ll \Delta_t$, or by constructive interference of four paths at special times when the spin and motion decouple in the strong coupling case, $\Delta_t \sim \Omega$.}
\label{MolmerSorensen}
\end{figure}

In the M{\o}lmer-S{\o}rensen scheme, a laser beam containing two frequency components $\omega_1$ and $\omega_2$, as shown in Fig.~\ref{MolmerSorensen}(a) is shone on the ions. The two frequencies are chosen close to the upper and lower motional sidebands, i.e.  $\omega_0 \pm \omega_t \pm \Delta_t +\delta$, where $\omega_0$ is the energy spacing between the qubit states $\vert g\rangle$ and $\vert e\rangle$, and $\omega_t$ is the normal mode frequency of the ions, typically the frequency of the center-of-mass mode.  Depending on the laser frequencies and strengths, this scheme can realize a variety of qubit interactions, including controlled-phase gate between a given pair of qubits, or various long-range Hamiltonians with pairwise interactions between the qubits.

The Hamiltonian for a single ion interacting with two frequency laser field is given as
\begin{align}
H = & \frac{\hbar\omega_0}{2} \sigma^z + \hbar\omega_t( a^\dagger a + \frac{1}{2}) \nonumber\\ &+ \hbar\Omega (\cos\omega_1 t + \cos\omega_2 t)(a^\dagger + a)(\sigma^+ + \sigma^-),
\end{align}
where $a (a^\dagger)$ annihilates (creates) a mode excitation, and $\Omega$ is the ion-laser coupling strength. For $\omega_0, \omega_t \gg \Omega$, which is typically the case, we apply the rotating wave approximation, i.e. go to a rotating frame and neglect terms rotating at frequency $O(\omega_0)$ or $O(\omega_t)$ in this frame, and obtain
\begin{equation} \label{eqn: rwa}
H = \frac{\hbar\Omega}{2} (a^\dagger e^{-i\Delta_t t} + a e^{i\Delta t})(\sigma^+ e^{-i\delta t} + \sigma^- e^{i\delta t}),
\end{equation}
where we denoted $\omega_2 = \omega_0 + \omega_t + \Delta_t + \delta$ and $\omega_1 = \omega_0 - \omega_t - \Delta_t + \delta$. It is to be understood that Eq.~(\ref{eqn: rwa}) is the Hamiltonian for a single ion, with the ion label implicit in the Pauli operators. To include more ions that couple to the lasers, Eq.~(\ref{eqn: rwa}) should be summed over the ion labels.

There are a few cases to consider for the two laser frequencies: the symmetric scheme where $\delta=0$, the asymmetric scheme where $\delta \neq 0$, the case where $\Delta_t \gg \Omega$ and different strengths of $\delta$ within this case, the case where $\Delta_t$ is comparable to $\Omega$, and also the case where more than one normal mode is involved. Each case gives rise to a different qubit interaction, and we will consider them one by one.

Let us begin by considering the weak coupling case, $\Omega \ll \Delta_t$. In this case, the normal modes are only virtually excited, and can be eliminated in second order perturbation theory, giving
\begin{equation}
H = J \sum_{ij} (\sigma^+_i \sigma^+_j  e^{-2i\delta t} + \sigma^+_i \sigma^-_j + {\rm h.c.}),
\end{equation}
where $J \propto \Omega^2/\Delta_t$. The physical picture that explains this emergent interaction is as follows [see also Fig.~\ref{MolmerSorensen}(b)]. The laser drives flip the internal state of an ion labeled $i$, and the ion absorbs (or emits) a virtual normal mode phonon in this process. Another ion labeled $j$ emits (or absorbs) the phonon and flips its internal state due to the laser drive. This virtual exchange of phonons is responsible for mediating long-ranged qubit interactions between the ions. It should be noted that there are similarities between the coupling of an ion's qubit states to the phonon modes and the case of cavity quantum electrodynamics where an atom's internal states are coupled to the electromagnetic modes in the cavity~\cite{mivehvar2021cavity}.

In the limit $\delta = 0$, $H$ reduces to
\begin{equation} \label{eqn: MS}
H = J \sum_{ij} (\sigma^+_i \sigma^+_j + \sigma^+_i \sigma^-_j + {\rm h.c.}),
\end{equation}
which is the global M{\o}lmer-S{\o}rensen interaction. In the limit $\delta \gg J$, the $\sigma^+_i \sigma^+_j$ term is also rapidly rotating and can be averaged to zero, therefore $H$ reduces to 
$H = J \sum_{ij} (\sigma^+_i \sigma^-_j + {\rm h.c.})$.
Finally, when $\delta$ is comparable to $J$, the Hamiltonian after moving to an interaction picture is 
$H = J \sum_{ij} \sigma^x_i \sigma^x_j + B\sum_i \sigma^z_i$. This scheme is widely realized in experiments for quantum simulation [see, e.g.~\cite{kirchmair2009deterministic, lanyon2011universal, kim2009entanglement, edwards2010quantum}].

The disadvantage of the weak coupling case above is that the dynamics are slow, $J \ll \Omega \ll \Delta_t$. The dynamics can be made faster by making $\Delta_t$ comparable to $\Omega$ and setting $\delta=0$. In this case, Eq.~(\ref{eqn: rwa}) is exactly integrable. The exact time evolution operator under $H$ [Eq.~(\ref{eqn: rwa})] has the form
\begin{equation}
U = D(\alpha(t) \sigma^x_{\rm tot}) \exp(i \Phi(t) (\sigma^x_{\rm tot})^2),
\end{equation}
where $D(\alpha) = \exp(\alpha a^\dagger - \alpha^* a)$, and $\sigma^x_{\rm tot} = \sum_j \sigma^x_j$. In this case, the spin degree of freedom and motional degree of freedom are not decoupled in general, except at special times when $\alpha(t) = 0$. This special time occurs at multiples of $t = 2\pi/\Delta_t$. At these special times, the time evolution operator, $U = \exp(i \Phi(t) (\sigma^x_{\rm tot})^2)$, is the same as the one obtained from a M{\o}lmer-S{\o}rensen interaction. One can obtain a physical picture behind this spin-motion decoupling at $t=2\pi/\Delta_t$ by visualizing the center-of-mass mode as a quantum harmonic oscillator. The Hamiltonian [Eq.~(\ref{eqn: rwa})] displaces the quantum harmonic oscillator, and the oscillator undergoes a displacement given by $D(\alpha(t) \sigma^x_{\rm tot})$. It returns to its initial state at $t=2\pi/\Delta_t$ (i.e. when $\alpha(t) = 0$), however, it picks up a spin-dependent geometric phase during each cycle. This spin-dependent phase is exactly equal to the phase given by the M{\o}lmer-S{\o}rensen gate. Figure~\ref{MolmerSorensen}(b) shows another intuitive explanation for the M{\o}lmer-S{\o}rensen interaction. The laser pulses scatter two ions from $\vert gg,n\rangle$ to $\vert ee,n\rangle$ (also from $\vert ge,n\rangle$ to $\vert eg,n\rangle$) via four paths, and the total amplitude of this process is the constructive interference of the four paths.

Local M{\o}lmer-S{\o}rensen interactions, e.g. between exactly two qubits, can be obtained by shining the lasers on only two ions so that the sum in Eq.~(\ref{eqn: MS}) is restricted to those two ions. The gates can be made fast by making $\Delta_t$ comparable to $\Omega$ and applying the laser pulses for a duration that is a multiple of $2\pi/\Delta_t$, as explained above. Two-qubit M{\o}lmer-S{\o}rensen gates have also been widely realized in experiments~\cite{kirchmair2009deterministic, kirchmair2009high, edwards2010quantum, gaebler2016high, tan2015multi, egan2020fault}, with the highest current gate fidelity in the range of $99.9\%$\\\cite{clark2021high}. Together with  single-qubit rotations, they form a universal gate set for digital quantum computation. A major advantage of using trapped ions over superconducting qubits as a quantum computing platform is the global connectivity of the interactions. All ions couple to the center-of-mass mode, which mediates the qubit interactions, therefore one can implement a M{\o}lmer-S{\o}rensen interaction between any pair of ions in a finite time scale regardless of how far apart they are (up to caveats about ion spacing and mode spacing).

Finally, we consider the case that there are other normal modes nearby in frequency to the lasers. This case arises when $\Delta_t$ is comparable to the mode spacing, which can for example be accomplished by parking the lasers close to the radial modes instead of the axial modes. Then Eq.~(\ref{eqn: rwa}) should be modified to include the other modes, $a_m$, as well. After adiabatically eliminating the normal modes for $\Omega \ll \Delta_t$ as above, one again obtains a long-ranged qubit interaction, however the interaction is no longer uniform between all the qubits. Instead, one obtains an approximately power-law decaying interaction, $J_{ij} \sim J/|r_i - r_j|^\alpha$ (with an exponential correction). In the limit of coupling only to the center-of-mass mode, $\alpha=0$ and we recover the infinite-ranged interaction in Eq.~(\ref{eqn: MS}). In the limit that $\Delta_t$ is so large that all the normal modes are nearly at the same frequency relative to the lasers, then $\alpha \approx 3$. For intermediate $\Delta_t$, we have $0 < \alpha < 3$. This case was first realized in~\cite{korenblit2012quantum, kim2009entanglement, richerme2013quantum, islam2013emergence}.

\bibliographystyle{unsrt}
\bibliography{Wormhole-Teleportation, bhuvanesh_sundar_refs}

\end{document}